\documentclass[11pt, a4paper]{report}

\usepackage[english]{babel}
\usepackage{dcolumn} 
\usepackage{amsfonts}
\usepackage{amsmath}    
\usepackage{amssymb}
\usepackage{tensor}
\usepackage{upgreek}
\usepackage{bbm}        
\usepackage{color}
\usepackage{wasysym}
\usepackage{slashed}
\usepackage{multirow, booktabs}
\usepackage{graphicx}
\usepackage{afterpage}
\usepackage{exscale}

\usepackage{geometry}
\usepackage[square, comma, numbers, sort&compress]{natbib}
\renewcommand{\contentsname}{Contents}
\usepackage[ small, labelfont=bf , margin=20pt, tableposition=top]{caption}
\usepackage{fancyhdr}
\renewcommand{\headrulewidth}{0.4pt}

\usepackage[ps2pdf=true,colorlinks, pdfpagelabels]{hyperref}
\usepackage[figure,table]{hypcap} 
\hypersetup{
   bookmarksnumbered,
   pdfstartview={FitV},
   pdfpagemode={UseOutlines},
   pdfauthor={Josef Pradler},
   pdftitle={Electroweak Contributions to Thermal Gravitino Production}
   pdfsubject={Diploma Thesis}
   ,citecolor={blue},
   linkcolor={blue}
}

\usepackage{subfig}

\def\ve{\varepsilon}

\def\a{\alpha}
\def\b{\beta}
\def\c{\chi}
\def\d{\delta}
\def\e{\epsilon}
\def\f{\phi}
\def\g{\gamma}
\def\h{\eta}
\def\i{\iota}
\def\j{\psi}

\def\l{\lambda}
\def\m{\mu}
\def\n{\nu}
\def\o{\omega}
\def\p{\pi}
\def\q{\theta}
\def\r{\rho}
\def\s{\sigma}

\def\x{\xi}
\def\z{\zeta}
\def\D{\Delta}

\def\G{\Gamma}

\def\O{\Omega}
\def\P{\Pi}

\def\S{\Sigma}

\newcommand{\GEV}{\textrm{GeV}} 
\newcommand{\mgrav}{m_{\widetilde{G}}}
\newcommand{\gr}{\ensuremath{\widetilde{G}}}
\newcommand{\mplanck}{\ensuremath{M_{\text{P}}}}
\newcommand{\gammahard}{\ensuremath{\left.{d\Gamma_{\gr}\over
        d^3p}\right|_{\text{hard}}}}
\newcommand{\gammasoft}{\ensuremath{\left.{d\Gamma_{\gr}\over
        d^3p}\right|_{\text{soft}}}}
\newcommand{\BFB}{\ensuremath{\text{BFB}}}
\newcommand{\BBF}{\ensuremath{\text{BBF}}}
\newcommand{\FFF}{\ensuremath{\text{FFF}}}
\newcommand{\fbose}[1]{\ensuremath{f_{\text{B}}(#1)}}
\newcommand{\ffermi}[1]{\ensuremath{f_{\text{F}}(#1)}}
\newcommand{\suthree}{\ensuremath{\text{SU}(3)_{\text{c}}}}
\newcommand{\sutwo}{\ensuremath{\text{SU}(2)_{\text{L}}}}
\newcommand{\uone}{\ensuremath{\text{U}(1)_{\text{Y}}}}
\newcommand{\smgroup}{\ensuremath{\text{SU}(3)_{\text{c}}\times\text{SU}(2)_{\text{L}} \times  \text{U}(1)_{\text{Y}} }}
\newcommand{\TR}{T_{\textrm{R}}}
\newcommand{\yield}{Y_{\gr}}
\newcommand{\coll}{C_{\gr}}

\newcommand{\ogravtp}{\O_{\gr}^{\text{TP}}h^2}
\newcommand{\rhocrit}{\r_{\text{c}}}
\newcommand{\mgut}{M_{\text{GUT}}}
\newcommand{\monetwo}{m_{1/2}}
\newcommand{\omegadm}{\O_{\text{dm}}h^2}
\newcommand{\omegaDMmax}{\O_{\mathrm{dm}}^{\mathrm{max}}}
\newcommand{\stau}{\widetilde{\tau}_1}
\newcommand{\mst}{m_{\widetilde{\tau}_1}}

\newcommand{\gb}[1]{A^{#1}}   
\newcommand{\gf}[1]{\l^{#1}}  
\newcommand{\gfb}[1]{\overline{\l}\vphantom{\l}^{#1}}  
\newcommand{\cb}[1]{\f^{#1}}    
\newcommand{\cf}[2]{\c_{#1}^{#2}}   
\newcommand{\cfb}[2]{\overline{\c}_{#1}^{#2}}   
\newcommand{\gen}[3]{T_{#1,\, #2 #3\,}}  

\newcommand{\gbalpha}[2]{{A}^{(#1)\,#2}}   
\newcommand{\gfalpha}[2]{{\l}^{(#1)\,#2}}  
\newcommand{\gfbalpha}[2]{\overline{\l}\vphantom{\l}^{(#1)\,#2}}  

\newcommand{\order}[1]{\mathcal{O}(#1)} 

\newcommand{\wobs}{W_{\text{o}}}
\newcommand{\whid}{W_{\text{h}}}

\newcommand{\khid}{K_{\text{h}}}
\newcommand{\wobsre}{\widehat{W}_{\text{o}}}

\def\VEV#1{\left\langle #1\right\rangle}        

\def\partder#1#2{{\frac{\partial #1}{\partial #2}}}    
\def\secder#1#2#3{{\frac{\partial^2 #1}{\partial #2 \partial #3}}}  

\begin{document}

\title{
\vspace{-2.5cm}
\begin{flushright}
\huge{\textbf{Electroweak Contributions to\\Thermal Gravitino
    Production}}
\end{flushright} 
\rule[10pt]{\textwidth}{2pt}
\\[2cm]
\begin{tabular}{cl}
\includegraphics[height=2cm, clip=true]{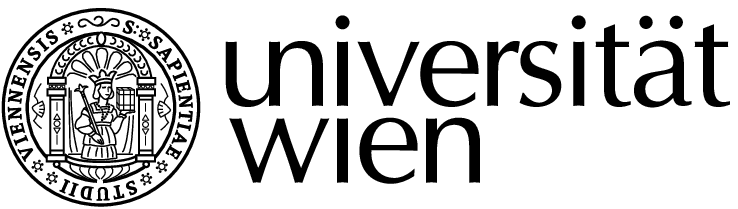} &
\parbox[b][2cm][c]{0.4\textwidth}{\normalsize{Insititut f\"ur
    Theoretische Physik, \\ Universit\"at Wien}} 
\\
\includegraphics[height=2cm, clip=true]{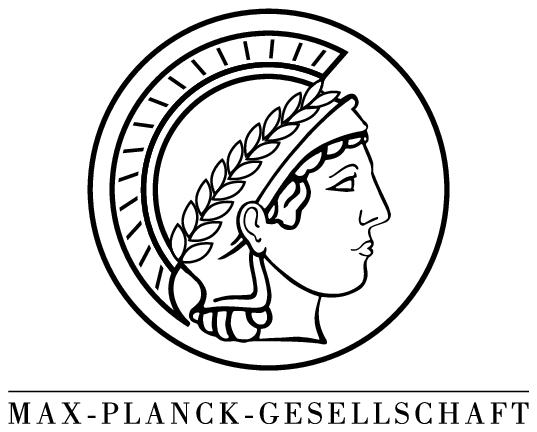} &
 \parbox[b][2cm][c]{0.4\textwidth}{\normalsize{Max-Planck-Institut f\"ur
     Physik, \\ (Werner-Heisenberg-Institut)}} 
\end{tabular}
\\[2cm]
\Large{Diplomarbeit zur Erlangung des Grades \\ Magister der Naturwissenschaften
\\\normalsize{(diploma thesis)}}
\\[4cm]
\begin{flushleft} \normalsize{
\begin{tabular}{ll}
  Verfasser: & Josef Pradler \\
  Matrikelnummer: & 0009878 \\
  Studienrichtung: & Physik \\
  Eingereicht am: & 23.10.2006\\
  Betreuer: & \parbox[t]{6cm}{O.~Univ.-Prof.~Dr.~Alfred Bartl \\
    Dr. Frank Daniel Steffen }
\end{tabular}}
\end{flushleft}
}
\author{}
\date{}
\pdfbookmark[0]{Titlepage}{title} 
\maketitle

    \newenvironment{dedication}
      {\cleardoublepage \thispagestyle{empty} 
        \vspace*{\stretch{1}} \begin{flushright} \em}
      {\end{flushright} \vspace*{\stretch{3}} \clearpage}
    \begin{dedication}
    Dedicated to my parents.
    \end{dedication}

%

%
%
%

%
%
\cleardoublepage
\pdfbookmark[0]{\contentsname}{toc} 
\tableofcontents 
\listoffigures
\listoftables 

\cleardoublepage
\pagenumbering{arabic}
\chapter{Introduction}
\label{cha:introduction}

The Standard Model of elementary particle physics provides a most
successful description of the electroweak and strong interactions
among all presently observed particles. 
However, the observational fact that most of the matter of the
Universe resides in the form of cold non-baryonic dark matter provides
an impressive evidence for physics beyond the Standard
Model~\cite{Spergel:2006hy}.  In fact, the nature and identity of dark
matter is one of the most pressing questions in the natural sciences.

Remarkably, supersymmetry (SUSY) offers an attractive solution of the
dark matter problem. Supersymmetry assigns to each particle a
superpartner whose spin differs by~$1/2$.
One can distinguish between ``normal'' matter,~i.e., Standard Model
fields, and their supersymmetric partners by assigning to each
particle the $R$-parity quantum number $R=(-1)^{3\text{B}+\text{L}+2\text{S}}$,
where $\text{B}$, $\text{L}$, and $\text{S}$ denote the baryon number,
the lepton number, and the spin of the corresponding particle, respectively.
For conserved \mbox{$R$-parity}, the lightest supersymmetric particle (LSP)
is stable. Thus, the LSP is a compelling candidate for dark
matter, provided that it does not have  electromagnetic or
strong interactions.

The gravitino $\gr$ is a particularly attractive candidate for
such an LSP. Any supersymmetric theory containing gravity predicts the
existence of the gravitino, a spin-$3/2$ particle which aquires
a mass from the spontaneous breakdown of supersymmetry. As the
superpartner of the graviton, it is extremely weakly interacting.
Hence, if the gravitino is the LSP, it can be dark matter.

At high temperatures, gravitinos are generated in inelastic scattering
processes with particles that are in thermal equilibrium with the hot
primordial SUSY plasma. 
Assuming that inflation governed the earliest moments of the Universe,
any initial population of gra\-vi\-tinos must be diluted away by the
exponential expansion during the slow-roll phase. We consider the
regeneration of gravitinos that starts after completion of reheating.
The calculation of the production rate of these thermally produced
gravitinos requires a consistent finite-temperature approach. A result
that is independent of arbitrary cutoffs was derived for
supersymmetric quantum chromodynamics (QCD) in
Ref.~\cite{Bolz:2000fu}.  Following this approach, we provide for the
first time the complete Standard Model gauge group result to leading
order in the gauge couplings. For gravitino dark matter scenarios,
this allows us to calculate the relic density of thermally produced
gravitinos, $\O_{\gr}^{\text{TP}}$. The comparison of this density
with the observed dark matter density is crucial to decide whether
gravitinos can explain dark matter.

If the gravitino is not the LSP, it can decay at late times. The late
decays of thermally produced gravitinos can then affect the abundances
of light elements during big bang nucleosynthesis (BBN).  Thus, the
new results will also serve as central input parameters for deriving
cosmological constraints for scenarios with unstable gravitinos.

Thermal gravitino production becomes very efficient if the reheating
temperature $\TR$ of the Universe after inflation is high. Thus,
gravitinos play an important role in models where thermal leptogenesis
explains the matter-antimatter asymmetry of the Universe. At the dawn
of the Large Hadron Collider era, we face the exciting possiblity to
confirm SUSY directly in experiments.  In fact, we show in
this thesis that a conceivable determination of the gravitino mass at
future colliders will allow for a unique test of the viability of
thermal leptogenesis in the laboratory.

This thesis is organized as follows. Chapter~\ref{cha:supergravity}
discusses the general supergravity Lagrangian. We show how to obtain
an effective low-energy theory. This allows us to derive the complete
set of Feynman rules that is necessary for the subsequent
calculations. In Chapter~\ref{cha:electr-therm-grav} we identify all
processes for thermal gravitino production to leading order in the
Standard Model gauge couplings. This yields the complete $\smgroup$
result for the thermal gravitino production rate.
Chapter~\ref{cha:gravitino-cosmology} covers phenomenological
implications of the new result for cosmology. In particular, for
gravitino dark matter scenarios, we provide $\O_{\gr}^{\text{TP}}$ and
derive an upper limit on the reheating temperature of the Universe.
In Chapter~\ref{sec:coll-tests-lept} we propose the collider test of
the viability of thermal leptogenesis. Here, we take into account also
the non-thermal gravitino production from decays of the
next-to-lightest supersymmetric particle (NLSP) into the gravitino
LSP.


\cleardoublepage
\cleardoublepage
\chapter{From Supergravity to Gravitino Phenomenology}
\label{cha:supergravity}

The natural scale of supergravity is the Planck scale.  For our
purposes, however, we need to find a low-energy version which is
reconcilable with phenomenology. On this quest we shall keep track of
the gravitino interactions. This chapter therefore provides a trail
from supergravity to gravitino phenomenology.

After briefly sketching how the gravitino emerges in supersymmetric
theories, we quote 
the general form of the supergravity Lagrangian\footnote{We
  only consider unextended $N=1$ supersymmetry.}  in four spacetime
dimensions and discuss its properties. We explain the conditions for
spontaneous supersymmetry breaking. For a concrete model of
gravity-mediated supersymmetry breaking, we carry out the transition
to the low-energy effective theory relevant for gravitino
phenomenology. We relate the low-energy supergravity Lagrangian with
the Minimal Supersymmetric Standard Model (MSSM) in the high-energy
limit of unbroken electroweak symmetry. Having identified the
couplings of the gravitino to the fields of the MSSM, we provide the
complete set of Feynman rules necessary for the calculations in this
thesis.

\newpage

\section{Local Supersymmetry}
\label{sec:local-supersymmetry}

Supersymmetry is a spacetime symmetry.  It relates the bosonic and
fermionic degrees of freedom in a supermultiplet of particles:
\begin{align}
\label{eq:what-is-susy}
  Q\,|\text{boson}\rangle \simeq |\text{fermion}\rangle, 
  \quad Q\,|\text{fermion}\rangle \simeq |\text{boson}\rangle\;,
\end{align}
The anti-commuting Weyl spinor $Q$, which is the generator of the SUSY
transformation, extends the Poincar\`e algebra to a
graded Lie algebra.
The generators $Q$ fulfill the fundamental commutator
$[\,\cdot\,,\,\cdot\,]$ and anti-commutator $\{\,\cdot\,,\,\cdot\,\}$
relations\footnote{The conventions on spinor notation are found in
  Appendix~\ref{ch:appendix-spinor-alg}}
\begin{subequations}
\begin{align}
  \{ Q_{\alpha} , \overline{Q}_{\dot{\beta}} \} &= 2
  \sigma^\mu_{\alpha \dot{\beta} } P_\mu\; ,
  \label{eq:fundamental-anticommutator-supercharge} \\
  \{ Q_{\alpha} , Q_{\beta} \} & = \{ \overline{Q}_{\dot{\alpha}} ,
  \overline{Q}_{\dot{\beta}} \} =0 \; , \\
  [ P_\m , Q_{\alpha} ] &= [ P_\m , \overline{Q}_{\dot{\beta}} ] = 0\;
  .
\end{align}
\end{subequations}

Let $\varepsilon$ and $\eta$ be two infinitesimal Weyl spinors which
parameterize the supersymmetry transformations. We then can
write~(\ref{eq:fundamental-anticommutator-supercharge}) in terms of a
commutator:
\begin{equation}
  \label{eq:comm-susy-algebra}
  [ \, \vphantom{ \overline{Q}_{ \dot{\beta} } } \varepsilon\, Q ,
  \overline{\eta }\, \overline{Q} \,] = 2 \,\varepsilon \sigma^\mu
  \overline{ \eta } \, P_\mu \; .
\end{equation}
In a globally supersymmetric theory, the parameters $\varepsilon$ and
$\eta$ are spacetime independent, corresponding to a rigid
translation. Gauging supersymmetry corresponds to making the
supersymmetry parameters local,~i.e., dependent on $x$. Then, the
commutator of two \textit{local} supersymmetry transformations yields
translations $\sim \varepsilon(x) \sigma^\mu \overline{ \eta }(x)$
which differ from point to point. Invariance under these general
coordinate transformations is exactly what we expect from a theory of
gravity. Local supersymmetry is therefore referred to as supergravity
(SUGRA)\footnote{Good reviews about supergravity are~\cite{Cerdeno:1998hs}
  and \cite{Nilles:1983ge}. For further reading see references therein.}.

The gravitino is a central element of SUGRA because it is the gauge
field of local supersymmetry transformations. 
Since supergravity is a non-renormalizable theory, operators of mass
dimension five and higher occur. The gravitino thus is an extremely
weakly interacting particle with couplings suppressed by inverse
powers of the reduced Planck mass
\begin{align}
\label{eq:reduced-planck-mass}
  \mplanck& = \frac{1}{ \sqrt{8\pi G_{\text{N}} }} =  2.4\times 10^{18}\,\GEV\; ,
\end{align}
where $G_{\text{N}} $ is Newton's gravitational
constant~\cite{Yao:2006px}.

\clearpage
\section[The Supergravity Lagrangian]{The Supergravity Lagrangian}
\label{sec:n=1-d=4-supergravity}

Our starting point of our analysis is the Lagrangian given in
Appendix~G in the book of Wess and Bagger \cite{Wess:1992cp}. We
rewrite it in terms of four-component spinors and adopt it to our
conventions given in Appendix~\ref{ch:appendix-spinor-alg}. This
involves a change of signature in the Lorentz metric from $(-,+,+,+)$,
used by Wess and Bagger, to the signature more commonly used in
high-energy physics $(+,-,-,-)$. With our conventions, we can carry
out this task in a safe way. We explicitly restore units of $\mplanck$
so that the supergravity Lagrangian in the four-component formalism
then reads:

\begin{eqnarray}\label{eq:Lsugra}
\frac{1}{e}
{\mathcal L} &=&
- \frac{\mplanck^2}{2} R  
+ g_{ij^*} \mathcal D_\m \phi^i
\mathcal D^\m \phi^{*j} - \frac{1}{2} g^2 \left
[(\textrm{Re}f)^{-1}\right ]^{ab} D_{a} D_{b}
\nonumber \\ &&
+ i g_{ij^*} \cfb Lj \g^\m \mathcal D_\m \cf Li +
\ve^{\m\nu\rho\s} \overline{\j}_{L\m} \g_\nu {\mathcal
D}_\rho \j_{L\s}
\nonumber \\ &&
-\frac{1}{4} \textrm{Re}f_{ab} F_{\m\nu}^{a} F^{b\,\m\nu}
+\frac{1}{8} \ve^{\m\nu\rho\s} \textrm{Im}f_{ab} F_{\m\nu}^{a}
F_{\rho\s}^{b}
\nonumber \\ &&
+\frac{i}{2} \textrm{Re}f_{ab} \gfb a \g^\m \mathcal{D}_\m
\gf b -e^{-1}\frac{1}{2} \textrm{Im}f_{ab} \mathcal D_\m \left [
e\gfb a_R \g^\m \gf b_R \right ]
\nonumber \\ &&
+\Bigg[  - \sqrt{2} g \partial_i D_{a} \gfb a \cf Li
+\frac{1}{4} \sqrt{2} g \left [(\textrm{Re}f)^{-1}\right ]^{ab}
\partial_{i} f_{bc} D_{a} \gfb c \cf Li
\nonumber \\ &&
- \frac{i}{16} \sqrt{2} \partial_i f_{ab} \gfb a [\g^\m,\g^\n] \cf
Li F_{\m\n}^{b}
-\frac{1}{2\mplanck} g D_{a} \gfb a_R \g^\m  \j_{\m} 
\nonumber \\ &&
-\frac{i}{2\mplanck} \sqrt{2} g_{ij^*} \mathcal D_{\m} \phi^{*j}
\overline{\j}_\n \g^\m \g^\n \cf Li + \textrm{h.c.}\Bigg]
\nonumber \\ &&
-\frac{i}{8\mplanck} \textrm{Re}f_{ab} \overline\j_\m [\g^m,\g^n] \g^\m
\gf a F_{mn}^{a}
\nonumber \\ &&
- e^{K/{2\mplanck^2}} \Bigg[ \frac{1}{4\mplanck^2} W^* \overline\j_{R\m} 
[ \g^\m , \g^\n ] \j_{L\n}
+ \frac{1}{2\mplanck} \sqrt{2} D_i W \overline{\j}_{\m} \g^{\m} \cf Li 
\nonumber \\ &&
+\frac{1}{2} {\mathcal D}_i D_j W \overline{\c^{\,c}_L}^i \cf Lj
+ \frac{1}{4} g^{ij^*} D_{j^*} W^*
\partial_i f_{ab} \gfb a_R \gf b_L + \textrm{h.c.} \Bigg]
\nonumber \\ &&
- e^{K/{\mplanck^2}} \left [ g^{ij^*} \left ( D_i W \right ) \left (
D_{j^*} W^* \right ) - 3 \frac{|W|^2}{\mplanck^2} \right ] +
\mathcal{O}(\mplanck^{-2})\; .
\end{eqnarray}

\clearpage
Let us briefly introduce the building blocks of this Lagrangian:
\begin{description}
\item[Matter sector] Matter fermions are described in terms of left-handed four-spinors
  \begin{align}
  \label{eq:definition-chiral-matter-fermions}
     \cf Li &= 
     \begin{pmatrix}
         (\chi_{\a})_{\text{W/B}} \\ 0 
     \end{pmatrix}\;,
   \end{align}
   where $ (\chi_{\a})_{\text{W/B}}$ stands for the two-component Weyl
   spinor. Here and in the following, the subscript {\scriptsize W/B}
   indicates the quantities used in the book by Wess and
   Bagger~\cite{Wess:1992cp}.  Since the matter sector of the
   supergravity Lagrangian is built up in terms of left-chiral
   superfields, only left-handed four-spinors $\cf Li$ appear in the
   resulting Lagrangian in the four-component notation.  The
   corresponding scalar superpartners are denoted as $\cb i$.  The
   index $i$ runs over all chiral superfields.
\item[Gauge sector] The gauge multiplet consists of gauge bosons
  $\gb a_\m$ and their superpartners, the gauginos
  \begin{equation}
    \label{eq:definition-gauginos}
    \gf a = \begin{pmatrix}
      -i ( \l^{a}_{\hphantom{a\,}\a} )_{\text{W/B}} \\
      \hphantom{-} i ( \overline{\l}^{a\,\dot\a} )_{\text{W/B}}
     \end{pmatrix},
  \end{equation}
  which are Majorana fields. Both are in the adjoint representation of
  the gauge group, $a,b,\dots=1,\dots, \mathrm{dim\, }\mathcal{G}$.
  The associated field strengths of the gauge fields $\gb a_\m$ are
  written as $F^a_{\m\n}$.  The auxiliary fields $D_{a}$ are
  generalizations of the $D$-terms in the vector supermultiplets of a
  globally supersymmetric theory.
\item[Gravity sector] The graviton $\tensor{e}{_\m^m}$ shows up
  implicitly as the determinant of the vielbein
  $e=\det\tensor{e}{_\m^m}$
  and the curvature scalar $R$. Flat spacetime indices of the local
  Lorentz frame are denoted by $m,n,\dots$; Einstein indices by $\mu, \nu,\dots\;$.
  
  The gravitino is a spin-$3/2$ field which is written in terms of the
  Majorana vector-spinor,
  \begin{equation}
    \label{eq:definition-gravitino}
    \j_\m = \begin{pmatrix} 
      -i ( \j_{\m\,\a} )_{\text{W/B}} 
      \\ \hphantom{-}i ( \overline{\j}_{\m}^{\hphantom{\m\,}\dot{\a}} )_{\text{W/B}}
     \end{pmatrix}\; .
  \end{equation}
  Since it is the superpartner of the graviton, the gravitino is
  massless for exact local supersymmetry. Note, that we include
  factors of $i$ in the definition of the gravitino as well as for the
  gauginos in (\ref{eq:definition-gauginos}).

\item[Gauge kinetic function] The gauge kinetic function $f_{ab}(\f)$
  is a dimensionless analytic function in the scalars $\f$. In the
  superfield approach, it multiplies the kinetic term for the vector
  supermultiplet. In the component version (\ref{eq:Lsugra}) of the
  supergravity Lagrangian, it therefore shows up as a prefactor in the
  kinetic terms of the gauginos and gauge bosons,
  \begin{subequations}
    \begin{align}
      \frac{i}{2}\textrm{Re}f_{ab} \gfb a \g^\m \partial_\m \gf b\;, 
      \label{eq:gauge-fermion-kinetic-term}\\
      -\frac{1}{4} \textrm{Re}f_{ab} F_{\m\nu}^{a} F^{b,\,\m\nu}\;,
      \label{eq:gauge-boson-kinetic-term}
    \end{align}
  \end{subequations}
  respectively.  Because of gauge invariance, it transforms under
  gauge transformations as the symmetric product of adjoint
  representations of the gauge group $\mathcal{G}$.

  Note that derivatives of the dimensionless gauge kinetic function
  $f_{ab}(\f)$ with respect to the scalars~$\phi^i$,
  \begin{equation}
    \partial_{i} f_{ab} \equiv \frac{\partial f_{ab}} { \partial\f^i}\; ,
  \end{equation}
  have negative mass dimension. In the course of spontaneous symmetry
  breaking, it is reasonable to consider models where $
  \partial_{i} f_{ab} = \mathcal{O} (\mplanck^{ -1 })$. Therefore, we
  have not explicitly written out terms in the
  Lagrangian~(\ref{eq:Lsugra}) which are proportional to $\propto
  \mplanck^{-1}\partial_i f_{ab}$ since they are of order $\order{\mplanck^{-2}}$.

\item[Superpotential] The superpotential is an analytic
  function in the chiral scalar fields. It has mass dimension three
  and its general from is restricted by gauge invariance,
  \begin{align}
    \label{eq:general-superpotential}
    W&=\whid(h)+\frac{1}{2}\mu_{i j }(h)\f^i \f^j
    +\frac{1}{6} y_{ijk}(h) \f^i \f^j \f^k + \order{\mplanck^{-1}}\;.
  \end{align}
  Here, we distinguish between the \textit{observable} part of
  the superpotential and the \textit{hidden}
  superpotential~$\whid(h)$. The latter depends only on hidden
  scalar fields $h$, which have no or only very small couplings to
  ordinary matter and gauge fields of the \textit{observable sector}.
  The simplest choice is a superpotential with separated hidden and
  observable sectors
  \begin{equation}\label{eq:seperated-superpotential}
    W = \whid(h) + \wobs(\f).
  \end{equation}
  Let us remark here that a hidden field dependence $\mu_{i j }(h)$
  which mixes observable and hidden sector can be crucial in models
  offering a dynamical solution to the
  \mbox{$\mu$ problem}~\cite{Giudice:1988yz, Brignole:1997dp}.

\item[K\"ahler structure] The K\"ahler potential $K(\f,\f^*)$ is a
  real-valued function in the chiral scalar fields and has mass
  dimension two.  It has the generic form:
  \begin{equation}\label{eq:general-kaehler-potential}
    K=\khid (h,h^*) + \a_{ij}(h,h^*)\f^i\f^{*i} + 
    \left[ Z_{ij}(h,h^*)\f^i\f^j + h.c.\right] + \order{\mplanck^{-1}}\; ,
  \end{equation}
  where $\khid$ denotes the hidden part of the K\"ahler potential that
  is independent of observable fields.

  In fact, the theory is endowed with a rich K\"ahler
  structure. One can think of the scalars $\f^i$ as coordinates
  of a K\"ahler manifold whose metric $g_{ij^*}$ can be expressed by
  the second derivatives of the K\"ahler potential $K(\f,\f^*)$:
  \begin{align}\label{kahlermetric}
    g_{ij^*} = \secder K{\f^i}{\f^{*j}} \equiv \partial_i \partial_{j*}
    K
  \end{align}
  with its inverse $g^{ij^*}$,~i.e., $g^{ij^*}g_{j^*k}=\d^i_k $. The
  K\"ahler connection is given by
  \begin{equation}\label{kahlerconnection}
    \G^k_{ij}=g^{kl^*}\partder{g_{jl^*}}{\f^i}\; .
  \end{equation}

  For example, the K\"ahler metric enters in the chiral sector in
  terms of a prefactor for the kinetic terms of the chiral fermions
  \begin{align}
    \label{eq:kinetic-term-chiral-fermions}
    i g_{ij^*} \cfb Lj \g^\m \partial_\m \cf Li\;.
  \end{align}
  
  The K\"ahler covariant derivatives of the superpotential read
  \begin{subequations}
    \label{eq:kaehler-covariant-derivatives}
  \begin{align}
    D_iW&= W_i + \mplanck^{-2} K_iW\;,\\
    \mathcal D_iD_jW &= W_{ij} + \mplanck^{-2}
    (K_{ij}W+K_iD_jW+K_jD_iW) \nonumber\\ &
    -\G^k_{ij}D_kW+\mathcal{O}(\mplanck^{-3})\; ,
  \end{align}
  \end{subequations}
  where $K_i=\partial_i K$, $W_i =\partial_i W$, $K_{ij}
  =\partial_i\partial_j K$, and $W_{ij} =\partial_i\partial_j W$.
  Beside being invariant under local su\-per\-sym\-metry- and gauge
  transformations, (\ref{eq:Lsugra}) is inert under K\"ahler
  transformations with arbitrary holomorphic functions $F(\f)$
  \begin{subequations}\label{Kahlertrafo}
    \begin{align}
      K(\f,\f^*)&\rightarrow K(\f,\f^*) + F(\f)+F^*(\f^*)\; ,\\
      W&\rightarrow e^{-F(\f)/{\mplanck^2}}W\;,
    \end{align}
  \end{subequations}
provided that the spinor fields undergo $F$-dependend Weyl
rotations
\begin{subequations}
\begin{align}
  \label{eq:weylrotations}
  \cf Li &\rightarrow e^{\frac{i}{2}\textrm{Im}F/{\mplanck^2}\,\g_5}\cf Li  \;,\\
  \gf a &\rightarrow e^{-\frac{i}{2}\textrm{Im}F/{\mplanck^2}\,\g_5}\gf a  \;,\\
  \j_\m &\rightarrow e^{-\frac{i}{2}\textrm{Im}F/{\mplanck^2}\,\g_5}\j_\m\;.
\end{align}
\end{subequations}

It is easily seen  that the metric $g_{ij^*}$
and the K\"ahler connection $\G^k_{ij}$ remain unchanged under a
K\"ahler transformation (\ref{Kahlertrafo}).  More generally, the
isometries of the K\"ahler manifold can be expressed by the real
Killing potentials~$D_{a}(\phi,\phi^*)$. The corresponding
(holomorphic) Killing vector fields are given by
  \begin{subequations}
    \label{eq:killing-vector-fields}
    \begin{align}
      X^{i}_a & = -i g^{ij*} \partial_{j*} D_a \; ,\\
     X^{*j}_a & = i  g^{ij*} \partial_{i} D_a \; . 
    \end{align}
  \end{subequations}
  In the Lagrangian~(\ref{eq:Lsugra}) we used this to express the
  Killing vector fields in terms of the Killing potentials.  In
  principle, the Killing potentials $D_{a}$ can be found by solving
  the corresponding Killing equation. We quote their explicit form
  later in~(\ref{dterm}), where we restrict ourselves to flat K\"ahler
  manifolds.

\end{description}

The physical fields $\gb a_\mu$ and $\gf a$ in the vector multiplet
are written with upper gauge indices $a,b,\dots$ while the Killing
potentials $D_{a}$ have lower indices. To bring them in canonical
form, they are lowered or raised with $\textrm{Re}f_{ab}$
or $ \left [(\textrm{Re}f)^{-1}\right ]^{ab}$ , respectively.

Furthermore, the supergravity Lagrangian above is  given for a
simple gauge group only.  An extension to a non-simple gauge group
\mbox{$\mathcal{G}=\prod_\a \mathcal{G_\a}$} like \mbox{$\smgroup$}
requires to introduce an additional index. We will do so in
Sec.~\ref{sec:MSSM high energy limit}.

The covariant derivatives are defined as\footnote{We drop the
  contributions from the spin-connection ${\omega_{\mu}}^{mn}$ in the
  covariant derivatives of the fermion fields since we will only consider
   flat spactime in this thesis for which
   ${\omega_{\mu}}^{mn}\rightarrow 0$.}
\begin{subequations}
\label{eq:covariant derivatives}
\begin{align}
\mathcal{D}_\mu \phi^i &=
\partial_\mu \phi^i
+ i g g^{ij^*} \partial_{j^*} D_{a} A_\mu^{a},
\label{eq:covariant derivative-phi}
\\
\mathcal{D}_\mu \cf Li &=
\partial_\mu \cf Li 
+\Gamma^i_{jk} \mathcal{D}_\mu \phi^j \cf Lk \nonumber \\ &
+ i g \partial_k \left( g^{ij^*}\partial_{j^*} D_{a}\right) A_\mu^{a}\cf Lk 
+\mathcal{O}(\mplanck^{-2}),
\label{eq:covariant derivative-chi}
\\
\mathcal{D}_\mu \gf a &=
\partial_\mu \gf a
- g f^{abc} \gb b_\m \gf c +\mathcal{O}(\mplanck^{-2}),
\label{eq:covariant derivative-lambda}
\\
\mathcal{D}_\mu \j_\nu &=
\partial_\mu \j_\nu
+ \mathcal{O}(\mplanck^{-2}).
\end{align}
\end{subequations}

The Lagrangian~(\ref{eq:Lsugra}) is invariant under local supersymmetry
transformations, parametrized by means of an anticommuting Majorana
spinor
\begin{align}
  \label{eq:definition-susy-parameter}
 \z(x)=&
\begin{pmatrix}
 ( \z(x)_\a )_{\text{W/B}} \\  ( \overline{\z}(x)^{\dot{\a}} )_{\text{W/B}}
\end{pmatrix} 
\end{align}
of mass \mbox{dimension $-1/2$}, namely,
\begin{subequations}
\label{susytrafos}
\begin{align}
  \d_\z \tensor{e}{_\m^m} &= \frac{1}{\mplanck}(
  \overline \z \g^m \j_{\mu R} -
  \overline \z \g^m \j_{\mu L}  )  \\
  \d_\z \f^i &= \sqrt{2}\overline{\z} \cf Li\\
  \d_\z \cf Li &= -i \sqrt{2} \g^\m \z_R \mathcal{D}_\m \cb i
  -\G^i_{jk}\d_\z \cb j \cf Lk \nonumber \\
  & - \sqrt{2} e^{K/{2\mplanck^2}} g^{ij^*} D_{j^*} W^* \z_L
  +\mathcal{O}(\mplanck^{-1})\\
  \d_\z \gb a_\m &= \overline{\z} \g_\m \gf a_R - \overline{\z} \g_\m \gf a_L  \\
  \d_\z \gf a_L &= \frac{1}{4} F_{\m\n}^{a} [ \g^\m , \g^\n ]    \z_L 
  - i g  \left [(\textrm{Re}f)^{-1}\right
  ]^{ab}   D_{a} \z_L +\mathcal{O}(\mplanck^{-1})\\
  \d_\z \j_\m &= 2 \mplanck \mathcal D_\m \z + \frac{i}{\mplanck}
  e^{K/{2\mplanck^2}} ( W \g_\m \z_R +  W^* \g_\m \z_L) +
  \mathcal{O}(\mplanck^{-1})
\end{align}
\end{subequations}

Let us stress that the Lagrangian (\ref{eq:Lsugra})
depends on two \textit{arbitrary} functions of the chiral scalars,
namely, the K\"ahler function
\begin{equation}\label{kahlerfunction}
G(\f,\f^*)=\frac{K(\f,\f^*)}{\mplanck^2}+\ln{\frac{|W(\f)|^2}{\mplanck^6}}\;,
\end{equation}
and the gauge kinetic function $f_{ab}(\f)$. This is a consequence of
the K\"ahler-Weyl invariance and can be seen by virtue of
(\ref{Kahlertrafo}) with $F(\f)=\ln W(\f)$.  Within the framework of
supergravity, there is no mechanism which tells us one particular form
of $G$ and $f_{ab}$ from the other. On the other hand,
phenomenological considerations will suggest a certain minimal choice
as described below.

\section{Supersymmetry Breaking}
\label{sec:low-energy-sugra}

Particles within the same supermultiplet are degenerate in mass, because
\begin{align}
   [ P^2 , Q_{\alpha} ] &= [ P^2 , \overline{Q}_{\dot{\alpha}} ] =  0\;.
\end{align}
Since we do not yet have experimental evidence for supersymmetry, we
know that it has to be a broken symmetry if realized in nature.

The condition for the spontaneous breakdown of a symmetry is a ground
state that does not respect this symmetry. In the case of local
supersymmetry, we find from the supersymmetry
transformations~(\ref{susytrafos}) that the only Lorentz
invariant ways to achieve this are\footnote{We do not consider the
  possibility of fermion-antifermion condensates which require the
  presence a strong gauge coupling force.}:
\begin{subequations}
\begin{align}
\label{breakingcondition}
  \langle \d_\z \cf Li\rangle & \propto \langle F^i\, \rangle \z \neq
  0 &&  \textrm{F-term breaking}\intertext{and/or}
  \langle \d_\z \gf a\rangle & \propto \langle D_{a} \rangle \z \neq
  0 &&  \textrm{D-term breaking}.
\end{align}
\end{subequations}
Here, $\langle\,\cdot\,\rangle$ denotes the vacuum expectation value
(VEV) and $F^i$ is given by
\begin{equation}
  \label{eq:f-term}
  F^i = e^{ K/ {2\mplanck^2} } g^{ij^*} D_{j^*} W^*.
\end{equation}
This is a generalization of the auxiliary fields
$F^{\,\textrm{glob}}_i=-\partial _{i*} W^*$ which are part of the chiral
supermultiplets in a globally supersymmetric theory.

Many models of spontaneous symmetry breaking have been proposed.
Common to these models is that supersymmetry breaking occurs in the
hidden sector of particles.
Supersymmetry breaking is then communicated to the observable sector
either at tree level or radiatively. In the course of spontaneous SUSY
breaking, the gravitino acquires a mass $\mgrav$. Depending on the
SUSY breaking scheme, $\mgrav$ can range from the eV scale up to
scales beyond the TeV region~\cite{Martin:1997ns}.

\section{Gravity-Mediated Supersymmetry Breaking}
\label{sec:grav-mediated-supersymm-breaking}

An appealing scenario of supersymmetry breaking is
gravity-mediation. We show how
phenomenological considerations can lead to certain minimal choices of
the gauge kinetic function and the K\"ahler potential in such a
scenario.  Furthermore, we illustrate the super-Higgs mechanism and
show how soft supersymmetry-breaking terms arise in the low-energy
limit of supergravity. Finally, we also comment on the way Yukawa
couplings emerge.

Supersymmetry is broken by VEVs $\langle
F^m\, \rangle$ of some scalar fields $h$ which are gauge singlets.
The hidden fields $h$ appear in non-renormalizable terms of mass
dimension higher than four, which are suppressed by powers of $\mplanck$.
We expect soft terms of order
\begin{equation}
  \label{eq:order-of-softterms}
  m_{\text{soft}} \sim \frac{\langle F\, \rangle}{\mplanck}
\end{equation}
to arise. This is intuitive since $ m_{\text{soft}} $ has to vanish
in the limiting case  of unbroken
supersymmetry $ \langle F\, \rangle \rightarrow 0$. In order to obtain $ m_{\text{soft}}$ in the
electroweak range, supersymmetry breaking has to occur on rather high
scales, $\sqrt{\VEV{F}}\sim \order{10^{10}\,\GEV}$.

In the following we will sketch such a scenario and derive a
low-energy limit of the supergravity Lagrangian~(\ref{eq:Lsugra}).
The phenomenological imperative for any concrete
supersymmetry-breaking scenario is that it delivers a supersymmetric
theory equipped with terms that softly break it. The soft terms
obtained from our chosen gravity mediation scenario will have specific
sizes which might differ significantly from other breaking
scenarios. Nevertheless, their structure will be the same. In the end,
we therefore can treat the soft parameters as free parameters (to be
determined experimentally) and forget about the details of the
underlying breaking scenario. What we will have gained is a consistent
set of Feynman rules for the MSSM in the high-energy limit together
with a set of gravitino rules which share the same convention.

Let us start with the gauge kinetic function. The trivial choice
$f_{ab}=\d_{ab}$ would render the kinetic term of the
gauginos~(\ref{eq:gauge-fermion-kinetic-term}) and the kinetic term of
the gauge bosons~(\ref{eq:gauge-boson-kinetic-term}) renormalizable.
This form, however, is phenomenologically disfavored since we cannot
produce gaugino masses at the tree level.\footnote{A tree-level gaugino
  mass can also arise from $D$-term breaking which we do not
  consider in this work.}  The candidate for the gaugino mass term in
(\ref{eq:Lsugra}) is
\begin{equation}\label{gauginomassterm}
\frac{1}{4} e^{K/{2\mplanck^2}} g^{ij^*} D_{j^*} W^*
\partial_i f_{ab} \gfb a_R \gf b_L + h.c.\; ,
\end{equation}
which vanishes for $f_{ab}=\d_{ab}$. We can obtain
a diagonal gaugino mass matrix with the generic ansatz:
\begin{equation}\label{gaugekineticfunction}
f_{ab}=\d_{ab} f(h)\;.
\end{equation} 
After some hidden field $h$ has aquired a VEV $\VEV h$ at some high
scale, we obtain canonically normalized gauge kinetic terms by a
rescaling
\begin{subequations}
\label{eq:rescalings}
\begin{align}
  \hat{\l }^{a} &= \sqrt{ \VEV{\textrm{Re}f }} \gf a 
 \label{eq:rescale-gauginos-fields}\\
  \hat{ A }_\mu^{a}  &= \sqrt{ \VEV{\textrm{Re}f }}  A
  _\mu^{a}.  \label{eq:rescale-gauge-fields}
\end{align}
\end{subequations}
Rewriting (\ref{eq:Lsugra}) in terms of $\hat{\l }^{a}$ and $\hat{
  A }_\mu^{a}$, we see immediately that $ 1/\sqrt{ \VEV{ \textrm{Re}
    f}} $ appears in the covariant derivatives as prefactor of the
gauge coupling $g$. This is remarkable because it suggests that $
\textrm{Re}f $ plays the role of a coupling
constant. The rescaling
\begin{equation}
  \label{eq:rescaling-coupling-const}
  \hat{ g } = 
  \frac{g}{\sqrt{\VEV{\textrm{Re} f }}}
\end{equation}
then completely hides any dependence on $\textrm{Re}f_{ab}$ in
(\ref{eq:Lsugra}). If we further assume that $\VEV h$ is real, all
terms proportional to $\textrm{Im} f_{ab} $ vanish in
(\ref{eq:Lsugra}). For our purposes, this assumption is safe since
only terms of $\order{\mplanck^{-1}}$ with a gravitino in the vertex
will be considered; and an operator \mbox{ $\propto \mplanck^{-1}
  \textrm{Im} f_{ab} \psi_\mu $} is absent in the
Lagrangian~(\ref{eq:Lsugra}).

The term (\ref{gauginomassterm}) contains also the superpotential $W$
and the K\"ahler potential $K$. The general form of the K\"ahler
potential is  specified in~(\ref{eq:general-kaehler-potential}).
There, a diagonal choice $\a_{ij}(h,h^*) = \a(h,h^*) \d_{ij}$ is phenomenologically
favored. A non-diagonal form of $\a_{ij}$ leads to off-diagonal terms
in the the scalar mass matrix in the low-energy limit. This can induce
flavor-changing neutral currents for which strong bounds exist~\cite{Martin:1997ns}.

In the case of the MSSM emerging as a low-energy effective theory, we
can have a contribution like $[Z(h,h^*)H_1H_2+h.c.]$ in the K\"ahler
potential where $H_1$ and $H_2$ denote the the two \mbox{${SU(2)_L}$}
Higgs doublets. The hidden field dependence of $Z$ is closely
connected to the \mbox{$\mu$ problem}~\cite{Giudice:1988yz,Brignole:1997dp},
but is beyond the scope of this thesis. A particularly appealing
choice of the K\"ahler potential is therefore given by
\begin{equation}
K=\khid(h,h^*)+ \sum_i \f^{*i}\f^i + \left[ Z(h,h^*)H_1 H_2 + h.c. \right]
\end{equation}
with $ \a_{ij} = \d_{ij} $ [see
Eqn.~(\ref{eq:general-kaehler-potential})], corresponding to a flat
K\"ahler manifold in the observable sector. With this choice, the
K\"ahler metric~(\ref{kahlermetric}) becomes trivial,
\begin{equation}
\label{eq:trivial kaehler metric}
g_{ij^*}=\d_{ij^*}\; ,
\end{equation}
and leads to canonical kinetic terms in
(\ref{eq:kinetic-term-chiral-fermions}). Moreover, the connection
coefficients~(\ref{kahlerconnection}) vanish: $\G^k_{ij}=0$. The
Killing potentials $D_{a}$ then coincide with the $D$-terms of a  globally
supersymmetric Yang-Mills theory:
\begin{equation}\label{dterm}
D_{a}= \f^{*i}\gen aij \f^{j}.
\end{equation}
The generators of the Lie algebra of the gauge group are denoted by
$T_{a}$.  They are chosen to be hermitian and do obey the commutator
relation
\begin{equation}
  \label{eq:commutator-generators}
\left[ T_{a}, T_{b} \right] = i f^{abc} T_{c}
\end{equation}
with the structure constants $f^{abc}$.

\smallskip \noindent\textbf{Super-Higgs mechanism}\smallskip

In supergravity, an analogon to the Higgs mechanism of
electroweak-symmetry breaking exists.  When supergravity is
spontaneously broken, the corresponding massless Goldstone fermion, or
goldstino, is absorbed by the gravitino which aquires thereby its $\pm
1/2$ helicitiy components.

Mass terms for the gravitino,~i.e., terms in~(\ref{eq:Lsugra})
involving the gravitino field, which are quadratic in the fermionic
fields and do not have derivative couplings are
\begin{equation}
  \label{eq:L2f-gravitino-superHiggs}
  -  \frac{1}{4\mplanck^2} e^{K/{2\mplanck^2}}  W^* \overline\j_{R\m}
  [ \g^\m , \g^\n ] \j_{L\n}  
  - \frac{1}{2\mplanck} e^{K/{2\mplanck^2}} \sqrt{2} D_i W
  \overline{ \j}_{\m} \g^{\m} \cf Li  + \text{h.c.}\,. 
\end{equation}
Since hidden sector fields do not share any gauge interactions, there
are no $D$-terms contributing to supersymmetry breaking via $\VEV
{D_{a}} \neq 0$. Therefore, we do not consider
\begin{equation}\label{eq:Dterm-mixing-goldstino}
    - \frac{1}{2\mplanck} g D_{a} \gfb a_R \g^\m  \j_{\m}  + \text{h.c.}
\end{equation}
as a potential mass term.

Because of their hidden field dependence, the gauge kinetic function, the
superpotential, and K\"ahler potential will get vacuum expectation
values $\VEV{f_{ab}}$, $\VEV{\khid}$, and $\VEV{\whid}$, respectively. From the
first term in (\ref{eq:L2f-gravitino-superHiggs}) we can read off the
gravitino mass after spontaneous symmetry breaking:
\begin{equation}
  \label{eq:gravitino-mass}
  \mgrav =  \frac{1}{\mplanck^2} e^{\VEV{ \khid } /{2\mplanck^2}}  \VEV{ \whid^* }.
\end{equation}
The second term in (\ref{eq:L2f-gravitino-superHiggs}) mixes the
gravitino $\j_{\m}$ with chiral fermions $\cf Li$. Let us define the
spinor\footnote{In case of $D$-term breaking, the definition
  (\ref{eq:goldstino-def}) of the goldstino has to be extended in
  order to remove the mixing in (\ref{eq:Dterm-mixing-goldstino}).}
\begin{equation}
  \label{eq:goldstino-def}
  \eta_L =  D_i W \cf Li
\end{equation}
and apply the supersymmetry transformation (\ref{susytrafos}). We see
that $\eta$ changes by a shift
\begin{equation}
  \label{eq:goldstino-trafo}
  \d_\z \eta_L = - \sqrt{2} e^{K/{\mplanck^2}} g^{ij^*}
  \left ( D_i W \right ) \left ( D_{j^*} W^* \right ) + \dots = 
   - 3 \sqrt{2} \mgrav \z + \dots \;,
\end{equation}
where in the last equality we have assumed a vanishing cosmological
constant as explained below. This implies that
\begin{equation}
  \label{eq:goldstone-fermion}
  \eta =   D_i W \cf Li +  D_{i^*} W^* (\chi_L^{i})^c
\end{equation}
is the Goldstone fermion.  Indeed,
with the choice $\z = \eta \sqrt{2}/(6\mgrav)$, we can choose a
\textit{unitary gauge} where $\eta $ transforms to zero. This
removes the mixing term in (\ref{eq:L2f-gravitino-superHiggs}).

\smallskip \noindent\textbf{Soft terms from the scalar potential}\smallskip

The scalar potential in the supergravity Lagrangian (\ref{eq:Lsugra}) reads
\begin{align}
  \label{eq:sugra-scalar-potential}
  V & = e^{K/{\mplanck^2}} \left [ g^{ij^*} \left ( D_i W \right ) \left (
      D_{j^*} W^* \right ) - 3 \frac{|W|^2}{\mplanck^2} \right ] + \frac{1}{2}
  \hat{g}^2   D_{a} D_{a}
  \nonumber \\
  & =  g_{ij^*}  F^{i} F^{j^*} -3  e^{K/{\mplanck^2}} \frac{|W|^2}{\mplanck^2} + \frac{1}{2}
  \hat{g}^2 D_{a} D_{a} \; ,
\end{align}
where the rescaled coupling~(\ref{eq:rescaling-coupling-const}) is used. 
We have already seen that successful symmetry breaking is
achieved if some of the auxiliary fields $F^i$ (\ref{eq:f-term}) get a
non-zero vacuum expectation value
\begin{equation}
  \label{eq:spont-breaking-cond2}
  \VEV{F^i}= \VEV{e^{ K/ {2\mplanck^2} } g^{ij^*} D_{j^*} W^*} \neq
  0\; .
\end{equation}
While  the scalar potential 
\begin{equation}\label{eq:scalar-potential-global-case}
  V_{\text{glob}} = F F^* + 1/2D^2
\end{equation}
is positive semi-definite in global supersymmetry, the minus sign of
the second term in (\ref{eq:sugra-scalar-potential}) offers the
appealing possibility to break supersymmetry with vanishing
cosmological constant:
\begin{equation}
  \label{eq:vanishing-cosmo-const}
  \VEV V =  \VEV{ g^{ij^*} \left ( D_i W \right ) \left (
      D_{j^*} W^* \right ) - 3 \frac{|W|^2}{\mplanck^2}} = 0\; .
\end{equation}

We therefore can work in the limit of flat Minkowski spacetime and
neglect the interactions of the graviton: $R \rightarrow 0$ and $e
\rightarrow 1$.  Moreover, we drop the distinction between flat and
curved spacetime indices. Because of notational habit, we denote from
now on Minkowski spacetime indices with $\mu, \nu = 0, \dots ,3$.

For the sake of simplicity, we further assume that we do not have any
mixing between the observable and hidden sectors in the superpotential
and the K\"ahler potential. For the superpotential, this corresponds to
the choice (\ref{eq:seperated-superpotential}). For the K\"ahler
potential, we write 
\begin{align}
  K=\khid(h,h^*) + \sum_{i}\phi^i\phi^{*i}
\; .
\end{align}
It is instructive to get a feeling for the scales
involved:
\begin{equation}
  \label{eq:scales-for-derving-softterms}
  \begin{array}{c}
  \VEV h  \sim  \order{\mplanck},\quad
  \VEV {\whid}  \sim  \order{\mplanck^2},\quad
  \VEV {\khid}  \sim  \order{\mplanck^2},\quad
  \VEV {\partial_m \whid}   \sim  \order{\mplanck} .
  \end{array}
\end{equation}
The second and third relations come from the requirement that we want
to obtain a gravitino mass (\ref{eq:gravitino-mass}) in the
electroweak range $\mgrav \apprle \text{few TeV}$.  This requirement
is crucial because $\mgrav$ will govern the size of the soft terms.
The fourth relation is a consequence of the second.

In the following we will use indices $m,n\dots$ for fields $h$ which
we allow to aquire a VEV and indices $i,j\dots$ for observable fields
$\f$. Using the definitions (\ref{eq:kaehler-covariant-derivatives})
for the K\"ahler covariant derivatives, the scalar potential
(\ref{eq:sugra-scalar-potential}) without the $D$-term becomes
\begin{align}
  \label{eq:scalarpotential1}
  V_F = e^{K/{\mplanck^2}} & \left [ \sum_i \left| \partder{\wobs}{\f^i}
       + \frac{
        \f^{*i} }{ \mplanck^2 } (\wobs + \whid ) \right|^2 \right. \nonumber \\
         & \left. + \sum_m \left|  \partder{ \whid }{ h^m } + \frac{1 }{ \mplanck^2
        } \partder{\khid }{h^{m}} (\wobs + \whid ) \right|^2 - 3 \frac{ | \wobs + \whid |^2}{
        \mplanck^2} \right ].
\end{align}
We expand this to
\begin{align}
  \label{eq:scalarpotential2}
  V_F = & e^{\khid / {\mplanck^2} }  \Bigg[ \sum_i \left( \left|\partder{\wobs}{\f^i}
      \right|^2 
      + \frac{  | \whid |^2 }{ \mplanck^4 } \f^{*i} \f^i  
      + \frac{\whid^* }{ \mplanck^2 } \partder{ \wobs}{\f^i}  \f^i 
      +  \frac{\whid }{ \mplanck^2 } \partder{ \wobs^*}{\f^{*i}}  \f^{*i}
    \right) 
      \nonumber \\
     & + \sum_m \left( \frac{ 1 }{ \mplanck^4 } \partder{\khid}{h^{m}}
     \partder{\khid}{h^{*m}} (\wobs \whid^* + \wobs^* \whid ) + \frac{
       \wobs^* }{ \mplanck^2 } \partder{ \khid}{ h^{*m}} \partder{
       \whid}{h^m} + \frac{ \wobs }{ \mplanck^2 } \partder{ \khid }{h^{m} }
     \partder{ \whid^*}{h^{*m}} \right)
   \nonumber \\
   & - 3 \frac{\wobs \whid^*}{\mplanck^2 } - 3 \frac{\wobs^* \whid}{\mplanck^2 }
   \Bigg] + \order{ \mplanck^2 } + \order{ \mplanck^{-1} } .
\end{align}
To get from (\ref{eq:scalarpotential1}) to
(\ref{eq:scalarpotential2}), we have expanded the exponential of the
observable sector in powers of $\mplanck^{-2}$.

Now we can the perform \textit{flat limit} of supergravity
where one sends the Planck mass $\mplanck$ to infinity but holds the
gravitino \mbox{mass $\mgrav$} (\ref{eq:gravitino-mass}) fixed.  Terms
of $\order{ \mplanck^2 }$ in (\ref{eq:scalarpotential2}) contain relative
signs and have to be \textit{fine-tuned} to zero for vanishing
cosmological constant $\VEV V =0$. With a rescaling of the observable
superpotential,
  \begin{equation}
    \label{eq:rescale-superpotential}
    \wobsre = \wobs \frac{\VEV{\whid^*}}{ \VEV{| \whid |}} e^{ \VEV{
        \khid} / {2  \mplanck^2} }\;,
  \end{equation}
this yields
\begin{align}
  \label{eq:scalarpotential3}
  V_F & = \sum_i \Bigg[
    \left|
      \partder{ \wobsre }{\f^i}      
  \right|^2 +
    \mgrav^2 \f^{*i} \f^i + \mgrav \left( \f^i  \partder{ \wobsre
      }{\f^i} + \text{h.c.} \right)  
  \Bigg] - \left(3
  \mgrav \wobsre + \text{h.c.}\right)
  \nonumber \\
  &+ \sum_m \Bigg[ \mgrav \VEV{ \partder{\khid }{h^{m}}} \left( \frac{1 }{ \mplanck^2 }
     \VEV{\partder{\khid }{h^{*m}}}
    +  \frac{1}{ \VEV{\whid} }\VEV{
      \partder{ \whid^* }{ h^{*m} }} \right) \wobsre + \text{h.c.} \Bigg] \;.
\end{align}
 In the first term of~(\ref{eq:scalarpotential3}) we recover the scalar potential of
global supersymmetry (\ref{eq:scalar-potential-global-case}), where
\mbox{$F^{\,\textrm{glob}}_i=-\partial _{i*} W^*$}. (We do not carry
along the $D$-terms which are already shown to
coincide with the globally supersymmetric case in~(\ref{dterm}).)
From (\ref{eq:f-term}), we find
\begin{align}
  \label{eq:f-term-hidden}
  \VEV{F^m} = \mgrav \mplanck^2   \left( \frac{1 }{ \mplanck^2 }
     \VEV{\partder{\khid }{h^{*m}}}
    +  \frac{1}{ \VEV{\whid} }\VEV{
      \partder{ \whid^* }{ h^{*m} }} \right)\;.
\end{align}
Let us identify the remaining parts in (\ref{eq:scalarpotential3})
with the commonly used notation\footnote{The relation $B=A-\mgrav$ is
  just a consequence of the simple breaking scenario which we have
  chosen. In general, one treats $A$ and $B$ as independent parameters.}:
\begin{subequations}
\label{eq:soft-terms1}
  \begin{align}
    m_0 &= \mgrav\;, \label{eq:universal-scalar-masses} \\
    A &= \mgrav \sum_m \VEV{\partder{\khid }{h^{m}}}
    \frac{\VEV{F^m}}{\mgrav \mplanck^2} \; , \\
    B &= \mgrav \sum_m \VEV{\partder{\khid }{h^{m}}}
    \frac{\VEV{F^m}}{\mgrav \mplanck^2}  -\mgrav\;.
  \end{align}
\end{subequations}
In Sec.~\ref{sec:MSSM high energy limit} we will write down the
superpotential of the MSSM. It is then not hard to show that $A$
corresponds to the trilinear scalar coupling and that $B$ is a
bilinear mass parameter;  $m_0$ is a universal scalar mass.  The
values of the above soft-supersymmetry breaking parameters are in the
electroweak range. Since the gravitino mass $\mgrav$ governs the size
of the soft terms, the choices made in
(\ref{eq:scales-for-derving-softterms}) turn out to be
phenomenologically favorable.

\smallskip \noindent\textbf{Gaugino mass term}\smallskip 

To complete the set of soft breaking
parameters~(\ref{eq:soft-terms1}), we discuss now the gaugino mass
term~(\ref{gauginomassterm}). Following the  logic  above, it is
straightforward to perform the flat limit. One then finds in terms
of the rescaled gaugino fields (\ref{eq:rescale-gauginos-fields}):
\begin{align}
  \frac{1}{2} M_{ab} \overline{\hat{\lambda}}\vphantom{\l}^{a}_R  \hat{\lambda}^{b}_L + h.c.,
\end{align}
with the mass matrix [cf.~(\ref{gaugekineticfunction})]: 
\begin{align}
  \label{eq:gaugino mass matrix}
  M_{ab} &= {1 \over 2} \frac{\VEV{F^m}}{\VEV{\textrm{Re}f(h)}} 
  \VEV{\partder{f(h)}{h^m}} \delta_{ab}\; .
\end{align}
From (\ref{eq:f-term-hidden}) together with
(\ref{eq:scales-for-derving-softterms}), one finds that $\VEV{F^m}
\sim \order{\mplanck}$ so that the gaugino masses (\ref{eq:gaugino mass
  matrix}) are in the electroweak range for
\begin{align}
  \label{eq:hidden field dependence gauge kinetic function}
  \VEV{\textrm{Re}f(h)}& = \order{1}  \qquad \textrm{and} \qquad \VEV{\partder{f(h)}{h^m}} \sim
  \order{\mplanck^{-1}}.
\end{align}
For example, a straightforward choice is $ f(h) = {h/ \mplanck}.$

The second relation in~(\ref{eq:hidden field dependence gauge kinetic
  function}) justifies our dropping of terms $\propto \mplanck^{-1}
\partial_{i} f_{ab} $ in the Lagrangian~(\ref{eq:Lsugra}), because these
terms are of~$\order{\mplanck^{-2}}$.

The flat limit,~i.e., $\mplanck \rightarrow \infty$ and $\mgrav$
fixed, amounts to integrating out fields in the Lagrangian which are
suppressed by powers of $\mplanck$. We therefore expect that the
resulting Lagrangian is valid at a high energy scale.  Thus, the soft
parameters (\ref{eq:soft-terms1}) and (\ref{eq:gaugino mass matrix})
should be interpreted as boundary conditions for the renormalization
group (RG) equations~\cite{Drees:2004jm}.

\smallskip \noindent\textbf{Yukawa interactions and chiral fermion mass terms}\smallskip

As far as spontaneous supersymmetry breaking is concerned, there is
another term in (\ref{eq:Lsugra}) to address, namely,
\begin{align}
  \label{eq:yukawas and higgsino mass term}
  - \frac{1}{2} e^{K/{2\mplanck^2}} {\mathcal D}_i D_j W
  \overline{\c^{\,c}_L}^i \cf Lj + \text{h.c.}\;.
\end{align}
The definition for the K\"ahler covariant derivatives is given in
(\ref{eq:kaehler-covariant-derivatives}). Recalling the dimensional
considerations (\ref{eq:scales-for-derving-softterms}), one finds that
the only contribution for the observable sector fields that survives
in the flat limit is \footnote{In fact, a Weyl
  rotation~(\ref{eq:weylrotations}) of the fermion fields with $F =
  -\frac{\VEV{\khid}}{2} - \mplanck^2\ln{
    \frac{\VEV{\whid^*}}{\VEV{|\whid|}} }$ is necessary.}
\begin{align}
  \label{eq:yukakwas and higgsino mass term final}
  - \frac{1}{2} \secder{\wobsre}{\phi^{i}}{\phi^{j}}\,
  \overline{\c^{\,c}_L}^i \cf Lj + \text{h.c.}\;.
\end{align}
Here we made use of the rescaling of the observable superpotential
(\ref{eq:rescale-superpotential}). This rescaling can be absorbed into
the parameters of the
superpotential~(\ref{eq:general-superpotential}),
\begin{subequations}
  \label{eq:rescaling of yukawas and mu}
 \begin{align}
   \hat{\mu}_{ij} & =  \mu_{ij} \frac{\VEV{\whid^*}}{ \VEV{| \whid |}} e^{ \VEV{
       \khid} / {2  \mplanck^2} }\;, \\
   \hat{y}_{ijk} & = y_{ijk} \frac{\VEV{\whid^*}}{ \VEV{| \whid |}} e^{ \VEV{
       \khid} / {2  \mplanck^2} }\;.
  \end{align}
\end{subequations}
We see that the bilinear part of the superpotential
(\ref{eq:general-superpotential}) produces mass terms for chiral
fermion fields with mass parameter $\hat{\mu}_{ij}$ while Yukawa
couplings $\hat{y}_{ijk}$ arise from the trilinear part of the
observable superpotential. 

\section{The Free Gravitino Field}
\label{sec:free-grav-lagr}

In the preceding section we have seen that the gravitino acquires a
mass $\mgrav$ through the super-Higgs mechanism. From
(\ref{eq:Lsugra}) together with (\ref{eq:gravitino-mass}), we find the
Lagrangian for the free gravitino field
\begin{align}
  \label{eq:free gravitino raw version}
  \mathcal{L}_{\j}^{\text{free}} &=  \ve^{\m\nu\rho\s} \overline{\j}_{L\m} \g_\nu
  \partial_\rho \j_{L\s} - \frac{1}{4} \mgrav 
  \Big(   \overline\j_{R\m} 
[ \g^\m , \g^\n ] \j_{L\n} + \text{h.c.}  \Big)\; 
\end{align}
which can be rewritten as 
\begin{align}
  \label{eq:free gravitino final version}
  \mathcal{L}_{\j}^{\text{free}} &= -\frac{1}{2} \ve^{\m\nu\rho\s}
  \overline{\j}_{\m} \g_5 \g_\nu
  \partial_\rho \j_{\s} - \frac{1}{4} \mgrav 
   \overline\j_{\m} 
[ \g^\m , \g^\n ] \j_{\n} + \text{tot.div.}\; .
\end{align}
Variation of (\ref{eq:free gravitino final version}) yields the
Rarita-Schwinger equation \cite{Rarita:1941mf}
\begin{align}
  \label{eq:rarita-schwinger equation}
  -\frac{1}{2}\e^{\mu\nu\rho\s}\g_5\g_\nu\partial_\rho\psi_\s
  -\frac{1}{4} m_{\gr}[\g^\mu,\g^\nu]\psi_\nu &= 0 \;.
\end{align}
Since the gravitino satisfies  the constraints
\begin{subequations}
    \label{eq:gravitino constraints}
  \begin{align}
    \g^\mu \j_\m &= 0 \;,  \label{eq:gamma constraint for the gravitino}  \\
    \partial^\mu \j_\m & = 0\;,
  \end{align}
\end{subequations}
the Rarita-Schwinger equation (\ref{eq:rarita-schwinger equation})
can be shown to reduce to the Dirac equation for each vector component $\mu$ of the
gravitino,
\begin{align}
  \label{eq:dirac eqn for gravitino}
  \left( i \slashed\partial - \mgrav \right) \j_\m = 0 \;.
\end{align}
The polarization tensor for a gravitino with four-momentum~$P$ is
given by \cite{Bolz:2000fu}
\begin{align}
\label{eq:full-gravitino-polsum}
  \P_{\m\n} (P) &= \sum_{s} \psi^{(s)}_\mu(P)\overline{\psi}^{(s)}_\nu(P) \nonumber\\
  &= -(\slashed{P}+\mgrav)\left(g_{\mu\nu} -\frac{P_\mu
      P_\nu}{\mgrav^2}\right) -\frac{1}{3} \left(\g^\mu +
    \frac{P_\mu}{\mgrav}\right) (\slashed{P}-\mgrav) \left(\g^\nu +
    \frac{P_\nu}{\mgrav}\right)\;,
\end{align}
where the sum is performed over the four gravitino helicities $s=\pm
3/2,\,\pm1/2$. The polarization tensor obeys
\begin{subequations}
  \begin{align}
    \g^\mu \P_{\m\n}(P) = 0\; , \\
    P^\mu \P_{\m\n}(P)  = 0\; , \\ 
    (\slashed{P}-m)  \P_{\m\n}(P)  = 0 \;.
  \end{align}
\end{subequations}
For energies much higher than the gravitino mass, it can be shown that
the  $\P_{\m\n}$ splits into two parts~\cite{Bolz:2000fu},
\begin{align}
  \label{eq:polarization tensor in the high energy limit}
  \P_{\m\n}(P) \simeq - \slashed{P} g_{\mu\nu} 
       + {2\over 3} \slashed{P} {P_\mu P_\nu\over m_{\gr}^2}  \;.
\end{align}
The first term represents the sum over the helicity $\pm 3/2$ states of
the gravitino whereas the second part corresponds to the sum over the
$\pm 1/2$ helicities of the goldstino~(\ref{eq:goldstino-def}).

\section{The MSSM in the High-Energy Limit}
\label{sec:MSSM high energy limit}

In the calculation of the gravitino production rate we will assume a
primordial plasma with the particle content of the Minimal
Supersymmetric Standard Model (MSSM). Since we will consider thermal
production far above the electroweak symmetry-breaking scale
$\mathcal{O}(100\,\GEV)$, we can work with an unbroken
electroweak symmetry group \mbox{$\sutwo \times \uone$}.

We now collect our previous results and extend the above considerations
to the standard model gauge group, i.e., we introduce
an additional index~$\a$ that keeps track of the different factors 
\begin{align}
  \label{eq:smgroup}
  \mathcal{G} & =\prod_{\a=1}^3 \mathcal{G}_\a  = \uone \times \sutwo
  \times \suthree
\end{align}
so that henceforth we use the assignment
\begin{align}
  \label{eq:assignment-alpha}
  \alpha  = 1 \;\; \text{for} \;\;\uone\; , \qquad
  \alpha  = 2  \;\;\text{for} \; \; \sutwo\; , \qquad
  \alpha  = 3  \;\;\text{for} \; \;  \suthree\; .
\end{align}
Accordingly, the gauge couplings $g_\a$ are given by
\begin{align}
  \label{eq:gauge-couplings}
  g_1 \equiv g' \;, \qquad  g_2 \equiv g \;, \qquad  g_3 \equiv g_{\text{s}} \; ,
\end{align}
with the $\uone$ hypercharge coupling $g'$, and  the weak and strong
 coupling constants $g$ and $ g_{\text{s}}$,
 respectively.

In terms of the rescaled quantities (\ref{eq:rescalings}) and
(\ref{eq:rescaling-coupling-const}), we find from (\ref{eq:Lsugra}) for
the $\smgroup$ gauge interactions
\begin{align}
  \mathcal{L}_{\text{gauge}} &= \sum_{\a=1}^3 \mathcal{L}^{(\a) }_{\text{gauge}}
\end{align}
with 
\begin{eqnarray}
  \label{eq:Lsusy-gauge}
  {\mathcal L}^{(\a) }_{\text{gauge}} &=& 
  + \mathcal D_\m^{(\a)} \phi^i \mathcal D^{(\a)\m} \phi^{*i} 
  - \frac{1}{2} {g}_\a^2 \left( \f^{*i}\gen aij^{(\a)} \f^{j} \right )^2
  \nonumber \\ &&
  + i \cfb Li \g^\m \mathcal{D}^{(\a)}_\m \cf Li 
  -\frac{1}{4} {F}_{\m\nu}^{(\a)\,a} {F}\vphantom{F}^{(\a)\, b,\,\m\nu} 
  +\frac{i}{2} 
  \gfbalpha \a a
  \g^\m \mathcal{D}^{(\a)}_\m 
  \gfalpha \a a
  \nonumber \\ &&
  - \sqrt{2} {g}_\a  \gfbalpha \a a  \f^{*i}\gen aij^{(\a)}   \cf Lj  
  - \sqrt{2} {g}_\a \cfb Li   \gen aij^{(\a)}  \f^{j}  \gfalpha \a a \; .
\end{eqnarray}
Note that we have \textit{dropped} the hats which were introduced
originally to indicate the rescalings,~i.e., we set
\begin{align}
  \label{eq:dropping-the-hats}
  \hat{g} \rightarrow g\; , \qquad \hat{A} \rightarrow A\; , \qquad
  \hat{\l} \rightarrow \l \; .
\end{align}

\begin{table}[tb]
\label{tab:gauge-fields-mssm}
\caption[MSSM gauge fields]{Gauge fields of the MSSM} 
\begin{center}
\begin{tabular}{lc@{\qquad}cr@{}l}
\toprule
Name
&  Gauge bosons $A_\mu^{(\a)\,a} $ & Gauginos $\l^{(\a)\,a}$ & 
 \multicolumn{2}{c}{$\big(\suthree,\sutwo\big)_{Y}$} \\ 
\midrule
\vspace{-0.4cm} \\ 
        B-boson, bino 
   & $A_\mu^{(1)\,a}  = B_\m \,\d^{a1} $&
   $\l^{(1)\,a} = \widetilde{B}\, \d^{a1}  $
 & $\quad\quad(\,\mathbf{1}\,,\mathbf{1}\, ) $ &  $\vphantom{1}_{0}$
\vspace{-0.2cm} \\ 
\\
        W-bosons, winos 
   & $A_\mu^{(2)\,a}  = W^a_\m $&
   $\l^{(2)\,a} = \widetilde{W}^a  $
 & $\quad\quad(\,\mathbf{1}\,,\mathbf{3}\, ) $ &  $\vphantom{1}_{0}$
\vspace{-0.2cm} \\ 
\\
        gluon, gluino
   & $A_\mu^{(3)\,a}  = G^a_\m $&
   $\l^{(3)\,a} = \widetilde{g}^a  $
 & $\quad\quad(\,\mathbf{8}\,,\mathbf{1}\, ) $ &  $\vphantom{1}_{0}$
\vspace{0.2cm}
 \\
\bottomrule
\end{tabular}
\end{center}
\end{table}

With the $D$-term (\ref{dterm}) and the vanishing K\"ahler connection for
trivial K\"ahler manifolds, the covariant derivatives
(\ref{eq:covariant derivatives}) in the Lagrangian
(\ref{eq:Lsusy-gauge}) become
\begin{subequations}
   \label{eq:covariant derivatives final}
  \begin{align}
    \mathcal{D}^{(\a)}_\mu \phi^i &=
    \partial_\mu \phi^i + i {g}_\a \gbalpha {\a}{a}_\m \gen aij^{(\a)} \phi^j\;
    ,\\
    \mathcal{D}^{(\a)}_\mu \cf Li &=
    \partial_\mu \cf Li + i {g}_\a \gbalpha {\a}{a}_\m \gen aij^{(\a)} \cf  Lj \; ,
    \\
    \mathcal{D}^{(\a)}_\mu \gfalpha{\a}{a} &= 
    \partial_\mu \gfalpha{\a}{a} - {g}_\a f^{(\a)\,abc} \gbalpha {\a}{b}_\m \gfalpha \a c \; .
  \end{align}
\end{subequations}
The field strength tensor ${F}_{\m\n}^{(\a)\,a}$ reads
\begin{align}
  \label{eq:field strength tensor}
  {F}_{\m\n}^{(\a)\,a} & = \partial_\m {A}_\n^{(\a)\,a} -
  \partial_\n {A}_\m^{(\a)\,a} -{g} f^{(\a)\,abc} {A}_\m^{(\a)\,b} {A}_\n^{(\a)\,c}\; .
\end{align}
\begin{table}[t]
\label{tab:matter-fields-mssm}
\caption[MSSM matter fields]{Matter fields of the MSSM} 
\begin{center}
\begin{tabular}{ll@{\qquad}lr@{}l}
\toprule
Name
& \quad Bosons $\phi^i$ & Fermions $\chi_L^i$ & 
 \multicolumn{2}{c}{$\big(\suthree,\sutwo\big)_{Y}$} \\ 
\midrule
\vspace{-0.4cm} \\ 
\multirow{3}{*}[0.3cm]{%
  \begin{minipage}{3cm}
        Sleptons, leptons \\ $I = 1,2,3$
  \end{minipage}
} 
   & $\widetilde{L}^I = 
  \begin{pmatrix}
     \widetilde{\n}^I_L\\ \widetilde{e}^{\,-\,I}_L
  \end{pmatrix}$ &
 ${L}^I = 
  \begin{pmatrix}
     {\n}^I_L\\ {e}^{-\,I}_L
  \end{pmatrix}$
 & $\quad\quad(\,\mathbf{1}\,,\mathbf{2}\, ) $ &  $\vphantom{1}_{-1}$
 \vspace{0.1cm}
 \\
 & $ \widetilde{E}^{*I} = {\widetilde{e}}^{\,-\,*\,I}_R $ &
  $ E^{c\,I} ={e}^{-\,c\,I}_R $
 & $(\,{\mathbf{1}}\,,\mathbf{1}\, ) $ &  $\vphantom{1}_{+2}$
\vspace{-0.2cm} \\ 
\\
\multirow{4}{*}[0.56cm]{%
  \begin{minipage}{3cm}
        Squarks, quarks \\ $I = 1,2,3$ \smallskip \\ ($\times$ 3 colors)
  \end{minipage}}
 & $\widetilde{Q}^I = 
  \begin{pmatrix}
     \widetilde{u}^I_L\\ \widetilde{d}^I_L
  \end{pmatrix}$ &
 ${Q}^I = 
  \begin{pmatrix}
     {u}^I_L\\ {d}^I_L
  \end{pmatrix}$ 
 & $(\,{\mathbf{3}}\,,\mathbf{2}\, ) $ &
 $\vphantom{1}_{+\frac{1}{3}}$ 
\vspace{0.1cm} \\
  & $ \widetilde{U}^{*I} = \widetilde{u}^{*I}_R$ &
  $  U^{c\,I} = {u}^{c\,I}_R $
& $(\,\overline{\mathbf{3}}\,,\mathbf{1}\, ) $ &
$\vphantom{1}_{-\frac{4}{3}}$  
\vspace{0.1cm}\\
  & $ \widetilde{D}^{*I} = \widetilde{d}^{*I}_R$ &
  $  D^{c\,I} = {d}^{c\,I}_R $
& $(\,\overline{\mathbf{3}}\,,\mathbf{1}\, ) $ &
$\vphantom{1}_{+\frac{2}{3}}$ 
\vspace{-0.2cm} \\ 
\\
\multirow{3}{*}[0.56cm]{%
  \begin{minipage}{3cm}
        Higgs, higgsinos
  \end{minipage}}
 &
 ${H}_d = 
  \begin{pmatrix}
     H_d^0 \\ H_d^-  
   \end{pmatrix}$ &
 $\widetilde{H}_d =
  \begin{pmatrix}
     \widetilde{H}^0_d\\ \widetilde{H}^-_d
  \end{pmatrix}$ 
& $(\,{\mathbf{1}}\,,\mathbf{2}\, ) $ &  $\vphantom{1}_{-1}$ 
\vspace{0.2cm}
\\
 &
 ${H}_u = 
  \begin{pmatrix}
     H_u^+ \\ H_u^0  
   \end{pmatrix}$ &
 $\widetilde{H}_u =
  \begin{pmatrix}
     \widetilde{H}^+_u\\ \widetilde{H}^0_u
  \end{pmatrix}$ 
& $(\,{\mathbf{1}}\,,\mathbf{2}\, ) $ &  $\vphantom{1}_{+1}$  
\vspace{0.2cm}
 \\
\bottomrule
\end{tabular}
\vspace{-0.6cm}
\end{center}
\end{table}
The corresponding gauge fields and their superpartners are listed in
Table~\ref{tab:gauge-fields-mssm}.  Matter fields with gauge couplings
are in the fundamental (anti-fundamental) representation of the
corresponding gauge group, namely, $\mathbf{2}\,
(\mathbf{\overline{2}}=\mathbf{2})$ for $\sutwo$ and
$\mathbf{3}\,(\overline{\mathbf{3}})$ for $\suthree$. The $\sutwo$
doublet structure for the MSSM matter fields is reviewed in
Table~\ref{tab:matter-fields-mssm}. The strongly
interacting particles gather in color triplets \textbf{3},~i.e.,
squarks and quarks indicated in Table~\ref{tab:matter-fields-mssm}
carry an additional color index. Gauge singlets are denoted by
$\mathbf{1}$ and $(\cdot,\cdot)_{0}$ for the strong/weak and $\uone$
interactions, respectively. Note that the normalization for the
hypercharges is such that the electric charge $Q$ is given by $ Q =
T_3 + Y/2$ where $T_3$ denotes the weak isospin eigenvalue $=\pm 1/2$ for
upper/lower entries in the doublets of
Table~\ref{tab:matter-fields-mssm}, respectively; $T_3 \equiv 0$ for
$\sutwo$ singlets~$(\,\cdot\,,\mathbf{1})_Y$. The family index $I$ in
Table~\ref{tab:matter-fields-mssm} refers to one out of three
generations of,
\begin{align*}
  \text{leptons}\:& \left\{ \begin{array}{c@{=\,(\,}l@{,\,}c@{,}c@{\,)}l} 
       {\n}^{\,I}_L & {\n}_{e^-} &{\nu}_{\mu^-} & {\nu}_{\tau^-}&
            \smallskip\\
      {e}^{\,-\,I}_L &
      {e}^{\,-}_L &{\mu}^{\,-}_L & {\tau}^{\,-}_L&
      \smallskip \\
      {e}^{\,-\,c\,I}_R &
    {e}^{\,-\,c}_R & {\mu}^{\,-\,c}_R &
        {\tau}^{\,-\,c}_R&
      \end{array} \right.
    \;,& \text{sleptons}\:& \left\{ \begin{array}{c@{=\,(\,}l@{,\,}c@{,}c@{\,)}l} 
        {\widetilde{\n}}^{\,I}_L & {\widetilde{\n}}_{e^-} &
        {\widetilde{\n}}_{\mu^-} & {\widetilde{\n}}_{\tau^-}&
        \smallskip\\
        {\widetilde{e}}^{\,-\,I}_L &
        {\widetilde{e}}^{\,-}_L &{\widetilde{\mu}}^{\,-}_L & {\widetilde{\tau}}^{\,-}_L&
        \smallskip \\
        {\widetilde{e}}^{\,-\,*\,I}_R &
        {\widetilde{e}}^{\,-\,*}_R & {\widetilde{\mu}}^{\,-\,*}_R &
        {\widetilde{\tau}}^{\,-\,*}_R&
    \end{array} \right. \; ,
\end{align*}
\begin{align*}
  \text{quarks}\:& \left\{ \begin{array}{c@{=\,(\,}l@{\,,\,}c@{\,,\,}c@{\,)}l} 
      {u}^{I}_L &
      {u}_L &{c}_L & {t}_L&
      \smallskip \\
      {u}^{\,c\,I}_R &
    {u}^{\,c}_R & {c}^{\,c}_R &
        {t}^{\,c}_R&
        \smallskip \\
         {d}^{I}_L &
      {d}_L &{s}_L & {b}_L&
      \smallskip \\
      {d}^{\,c\,I}_R &
    {d}^{\,c}_R & {s}^{\,c}_R &
        {b}^{\,c}_R&
      \end{array} \right. 
 \;,& \text{squarks}\:& \left\{ \begin{array}{c@{=\,(\,}l@{\,,\,}c@{\,,\,}c@{\,)}l}  
       {\widetilde u}^{I}_L &
      {\widetilde u}_L &{\widetilde c}_L & {\widetilde t}_L&
      \smallskip \\
      {\widetilde u}^{\,*\,I}_R &
    {\widetilde u}^{\,*}_R & {\widetilde c}^{\,*}_R &
        {\widetilde t}^{\,*}_R&
        \smallskip \\
         {\widetilde d}^{I}_L &
      {\widetilde d}_L &{\widetilde s}_L & {\widetilde b}_L&
      \smallskip \\
      {\widetilde d}^{\,*\,I}_R &
    {\widetilde d}^{\,*}_R & {\widetilde s}^{\,*}_R &
        {\widetilde b}^{\,*}_R&
      \end{array} \right.  \; ,    
\end{align*}
for $I=(1,2,3)$, respectively.

All matter fields are written in terms of left-handed four-spinors
since they stem from left-chiral supermultiplets. For example, a right
handed tau lepton $ {\tau}^{\,-}_R$ with hypercharge~$-2$ is
written in terms of its charge conjugate ${\tau}^{\,-\,c}_R$ with
hypercharge~$+2$.

The generators in (\ref{eq:Lsusy-gauge}) for the standard model gauge
group read
\begin{subequations}
\label{eq:generators gauge group}
  \begin{alignat}{2}
    \gen aij^{(1)}  &= \frac{1}{2} Y_i \d_{ij} \d_{a1} &&\quad \text{for } 
    \uone  \label{eq:gen u1}
    \; , \\
    \gen aij^{(2)}  &= \frac{1}{2} \s_{a,\,ij} &&\quad  \text{for } 
    \sutwo  \label{eq:gen su2}
    \; ,   \\
    \gen aij^{(3)}  &= \frac{1}{2} \l_{a,\,ij} &&\quad \text{for } 
    \suthree  \label{eq:gen su3}
    \; ,
\end{alignat}
\end{subequations}
where $Y_i$ is the hypercharge of the
corresponding particle $\phi^i$ or $\cf Li$ (see
Table~\ref{tab:matter-fields-mssm}). The Pauli sigma matrices $\s_a \;
(a=1,2,3)$ are given in (\ref{sigmas}) in the Appendix and $\l_a\; (a=1,\dots,8)$
denote the Gell-Mann matrices. The chosen basis (\ref{eq:generators
  gauge group}) for the generators implies that the structure constants
$f^{(\a)\,abc}$ for the non-abelian groups are totally antisymmetric.
Since $\uone$ is abelian, the commutator relation (\ref{eq:commutator-generators}) is
trivial,~i.e., 
\begin{alignat}{2}
  f^{(1)\,abc} & \equiv 0 &&\quad \text{ for } 
    \uone.
\end{alignat}

The superpotential of the MSSM is given by
\begin{align}
  \label{eq:superpotential-MSSM}
  \wobsre = W_{\text{MSSM}} &= \widetilde{U}^{*}\, \mathbf{y}_u
  \widetilde{Q} \cdot {H}_u - \widetilde{D}^{*}\, \mathbf{y}_d
  \widetilde{Q} \cdot {H}_d -
  \widetilde{E}^{*}\, \mathbf{y}_e \widetilde{L} \cdot {H}_d
  + \mu H_u \cdot H_d \; .
\end{align}
The doublet structure is tied together as $ \widetilde{Q} \cdot {H}_u
= \varepsilon^{ij} \widetilde{Q}_i {H}_{u\,j}$, with $\varepsilon^{ij}$
given in (\ref{eq:App-epsilons-undotted}). Furthermore, $
\widetilde{U}^{*}\, \mathbf{y}_u \widetilde{Q}$ is meant to be a
matrix multiplication in family space, $ \widetilde{U}^{*}\,
\mathbf{y}_u \widetilde{Q} = \widetilde{U}^{*\,I}\, {y}^{\,IJ}_u
\widetilde{Q}^J$.

The couplings $\mathbf{y}_{u,d,e}$ and the bilinear parameter $\mu$ in
(\ref{eq:superpotential-MSSM}) are understood to be the rescaled
quantities on the left-hand sides in (\ref{eq:rescaling of
  yukawas and mu}). We see from (\ref{eq:yukakwas and higgsino mass
  term final}) that $\mathbf{y}_{u,d,e}$ yield Yukawa coupings $\sim
y\,\phi\, \overline{\c^{\,c}_L} \chi_L $ and that $\mu$ gives mass to
the higgsinos. Neglecting all Yukawa couplings except the one
for the top quark,
\begin{alignat}{3}
  \label{eq:yukawa-approx}
  \mathbf{y}_e &\simeq 0 \; ,   
    &\quad\quad  \mathbf{y}_d &\simeq 0 \; ,&\quad\quad 
   \left( {y}^{IJ}_u \right) & \simeq
   \begin{pmatrix}0&0&0\\0&0&0\\0&0&y_t\end{pmatrix} \;,
\end{alignat}
the superpotential (\ref{eq:superpotential-MSSM}) reduces to
\begin{align}
   \label{eq:superpotential-with-top-yukawa-only}
    W_{\text{MSSM}} & \simeq y_t \,  {\widetilde t}^{\,*}_R \,  {\widetilde
      t}_L H_u^0 -y_t\, {\widetilde t}^{\,*}_R\, {\widetilde b}_L
    H_u^+ +
    \mu \left( H_u^+ H_d^- -  H_u^0 H_d^0 \right)\; .
\end{align}
In this thesis, however, we will not consider contributions to the
thermal gravitino production rate coming from top-Yukawa interactions;
we leave such an analysis for future work.

\section{Gravitino Interactions}
\label{sec:grav-inter}

Let us now collect the terms in the supergravity Lagrangian
(\ref{eq:Lsugra}) which describe the gravitino interactions with
ordinary matter fields. In the calculation of the gravitino production
rate, we will only consider external gravitinos which are subject to
the constraint (\ref{eq:gamma constraint for the
  gravitino}). Therefore,  the relevant
interaction Lagrangian extracted  from (\ref{eq:Lsugra}) is
\begin{align}
  \label{eq:gravitino interaction lagrangian}
  \mathcal{L}^{(\a) }_{\j,\,\text{int}} &= -\frac{i}{\sqrt{2}\mplanck} \left[
  \mathcal D^{(\a) }_{\m} \phi^{*i} \overline{\j}_\n \g^\m \g^\n \cf Li
  -\mathcal D^{(\a) }_{\m} \phi^{i} \cfb Li \g^\n \g^\m  \j_\n
\right]
 \nonumber \\
  &-  \frac{i}{8\mplanck}\overline\j_\m [\g^\r,\g^\s] \g^\m
  {\l}^{(\a)\, a} {F}_{\r\s}^{(\a)\, a}\; .
\end{align}
Here we have used the K\"ahler metric (\ref{eq:trivial kaehler
  metric}) and the definition for the rescaled
fields~(\ref{eq:rescalings}), but dropping the hats as in
(\ref{eq:Lsusy-gauge}).  Note that each operator in (\ref{eq:gravitino
  interaction lagrangian}) is suppressed by $\mplanck^{-1}$.

\section{Effective Theory for light Gravitinos}
\label{sec:goldst-interactions}

In the discussion of the super-Higgs mechanism, we have seen that the
goldstino degrees of freedom become the helicity $\pm 1/2$ components
of the gravitino. Indeed, these longitudinal helicity $\pm 1/2$
components become dominant if the energies involved are much higher
than the gravitino mass $\mgrav$. The dynamics of the Goldstone
fermion is given by the derivative coupling of the goldstino to the
supercurrent. The correct effective Lagrangian for light gravitinos
with goldstino-matter couplings in \textit{non-derivative} form has
been found in \cite{Lee:1998aw}. For a single external goldstino the
non-derivative form is equivalent to the derivative form to all
orders in perturbation theory and reads
\begin{align}
  \label{eq:goldstino-lagrangian}
  \mathcal{L}^{(\a) }_{\j,\,\text{light}} &= i\,\frac{m_{\phi^i}^2 -
    m_{\chi^i}^2}{\sqrt{3}\mplanck \mgrav} \left( \overline{\psi} \cf Li
    \f^{*i} - \cfb Li \psi \f^i \right) - \frac{M_\a}{4 \sqrt{6}
    \mplanck \mgrav } \overline{\psi} [ \g^\m,\g^\n ] \gfalpha{\a}{a}
  {F}_{\m\n}^{(\a)\, a} \nonumber \\ &- i \,\frac{g_\a M_\a}{ \sqrt{6}\mplanck
    \mgrav}
 \f^{*i} \gen aij^{(\a)} \f^j \overline{\psi} \g_5 \gfalpha{\a}{a}\; .
\end{align}
where the Majorana goldstino field is denoted by $\psi$.\footnote{The
  definitions for the gaugino fields in \cite{Lee:1998aw} already
  contain the factors of $i$ of our
  convention~(\ref{eq:definition-gauginos}). For the goldstino
  $\psi$, we have included them in the transition to the
  four-component formalism.}  Note that all vertices are proportional
to supersymmetry-breaking mass terms. The coupling in the first term
is proportional to the \textit{squared} masses $m_{\phi^i}^2$ and
$m_{\chi^i}^2$ of the corresponding matter fields $\phi^i$ and $\cf
Li$. The couplings in the remaining terms are linear in the
gaugino masses~$M_\a$. Therefore, at high energies and temperatures
$T$, contributions involving the goldstino-fermion-scalar vertex are
suppressed  relative to
the gaugino contributions due to the higher mass dimension of the
coupling.

\section{Feynman Rules}
\label{sec:feynman-rules}

We are now in a position to provide all Feynman rules necessary for
the calculation of the thermal gravitino production rate. The gauginos
$\l^a$, the gravitino $\j_\m$, and the goldstino~$\psi$ are Majorana
fermions.  Since these fields are self-conjugate, they yield different
Wick contractions from those of Dirac fields. We therefore use the
method proposed in \cite{Denner:1992vz} and introduce a continuous
fermion flow,~i.e., an arbitrary orientation of each fermion line.
Proceeding against the fermion flow then allows one to form chains of
Dirac matrices such that the relative sign of interfering diagrams can
be obtained in the same manner as one does for Dirac fermions. We
therefore have two analytical expressions for each vertex
corresponding to the two possible orientations of the fermion flow.

\begin{figure}[htb]
\begin{minipage}{\textwidth}%
\hfill\includegraphics[width=0.45\textwidth]{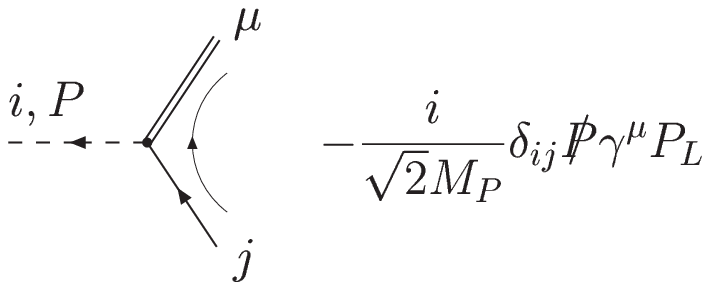}%
\includegraphics[width=0.45\textwidth]{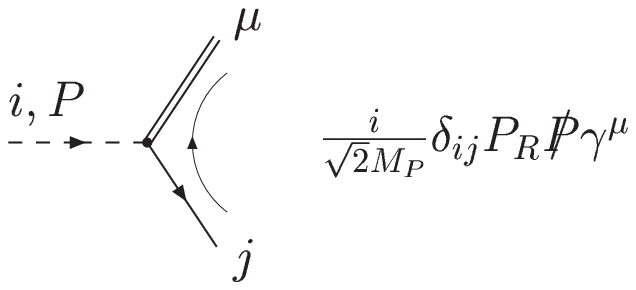}\hspace*{\fill}

\hfill\includegraphics[width=0.45\textwidth]{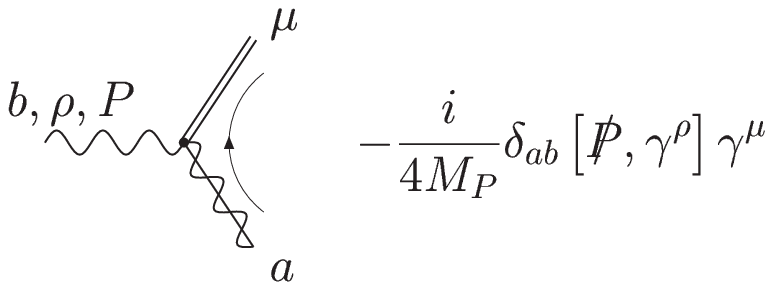}%
\includegraphics[width=0.45\textwidth]{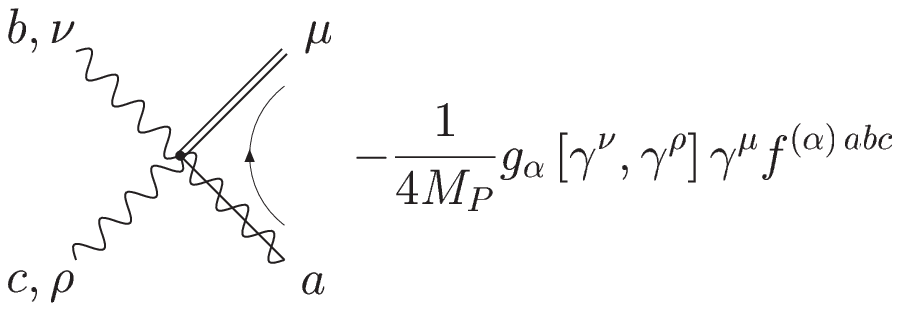}\hspace*{\fill}

\hfill\includegraphics[width=0.45\textwidth]{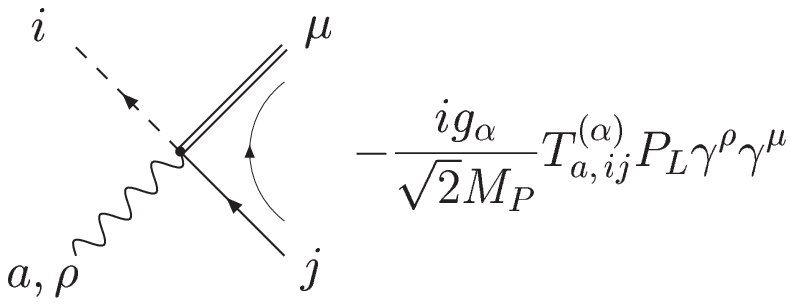}%
\includegraphics[width=0.45\textwidth]{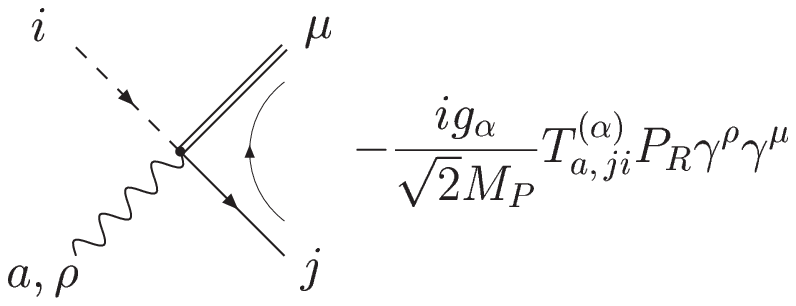}\hspace*{\fill}
\end{minipage}%
\caption[Gravitino Feynman rules]{Feynman rules for the gravitino from
  (\ref{eq:gravitino interaction lagrangian}) for one direction of the
  fermion flow.  The wiggled lines represent gauge fields
  $\gbalpha{\a}{a}$ while the corresponding superpartners
  $\gfalpha{\a}{a}$ are depicted by wiggled lines with additional
  straight solid lines. Scalars $\phi^i$, chiral fermions $\cf Li$,
  and the gravitino are represented respectively by dashed, solid, and
  double-solid lines. All momenta flow into the vertex.}
\label{fig:Feynman-rules-Gravitino}
\end{figure}

For one direction of the fermion flow, the gravitino Feynman rules
derived from (\ref{eq:gravitino interaction lagrangian}) are given in
Fig.~\ref{fig:Feynman-rules-Gravitino}. The gravitino is represented
as a double solid line, scalars $\phi^i$ are given by dashed lines and
chiral fermions $\cf Li$ are given by solid lines. Gauge bosons
$\gbalpha{\a}{a}$ are shown as wiggled lines and the corresponding
gauginos are depicted as wiggled lines with additional straight solid
lines. All momenta are understood to flow into the vertex.

\begin{figure}[htb]
\begin{minipage}{\textwidth}%
\hfill\includegraphics[width=0.45\textwidth]{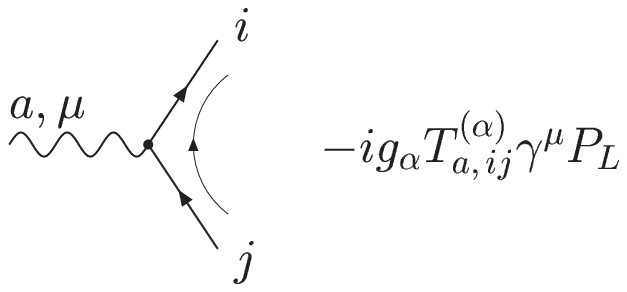}%
\includegraphics[width=0.45\textwidth]{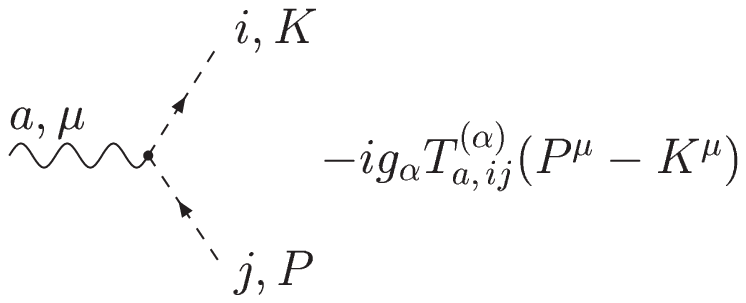}\hspace*{\fill}

\hfill\includegraphics[width=0.45\textwidth]{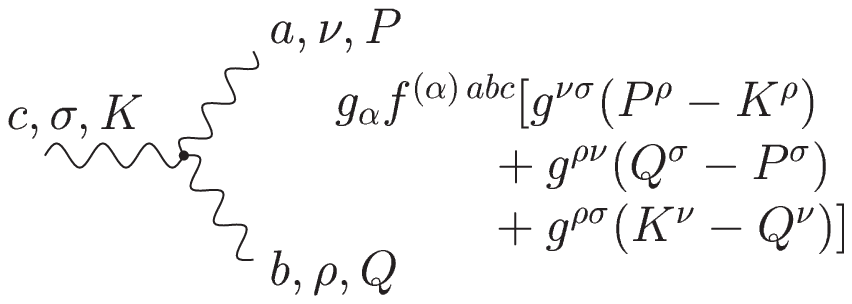}%
\includegraphics[width=0.45\textwidth]{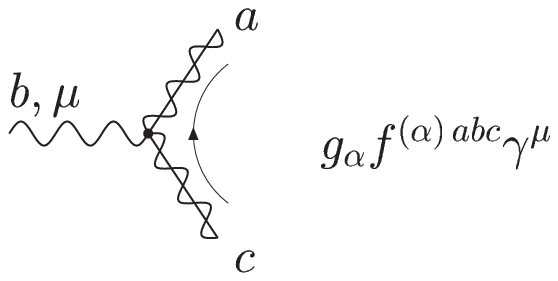}\hspace*{\fill}

\hfill\includegraphics[width=0.45\textwidth]{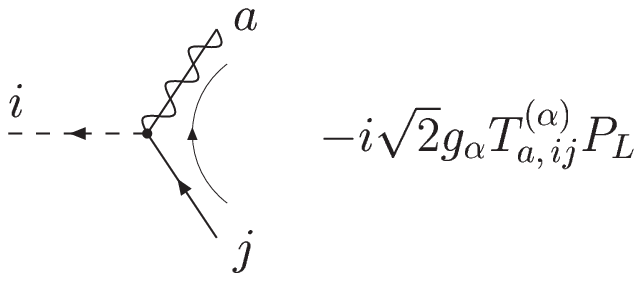}%
\includegraphics[width=0.45\textwidth]{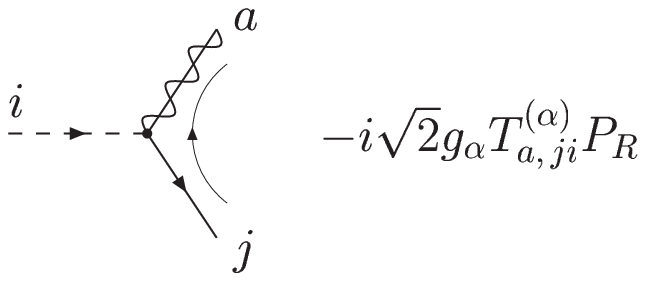}\hspace*{\fill}

\end{minipage}
\caption[Feynman rules for the gauge interactions]{Relevant Feynman
  rules for the supersymmetric gauge interactions derived from
  (\ref{eq:Lsusy-gauge}). As in Fig.
  \ref{fig:Feynman-rules-Gravitino}, it is understood that the gauge
  bosons and gauginos are associated with the corresponding gauge
  group $(\alpha)$. All momenta are ingoing. }
\label{fig:Feynman-rules-gauge}
\end{figure}

Figure~\ref{fig:Feynman-rules-gauge} shows the relevant Feynman rules
for the gauge interactions of the gauge group $\mathcal{G}_\a$ derived
from~(\ref{eq:Lsusy-gauge}) for one direction of the fermion flow.
For $\uone$, or $\a=1$, there is no self-coupling
of the gauge bosons $A^{(1)}_\m= B_\m$. Therefore, the triple gauge
boson vertex and its supersymmetric counterpart in the second line of
Fig.~\ref{fig:Feynman-rules-gauge} are absent.

\begin{figure}[hbt]
\begin{minipage}{\textwidth}%
\hfill\includegraphics[width=0.45\textwidth]{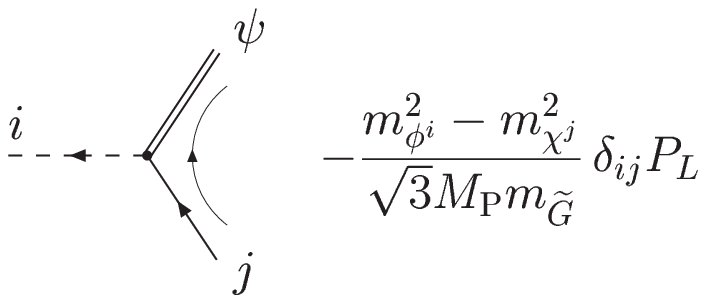}%
\includegraphics[width=0.45\textwidth]{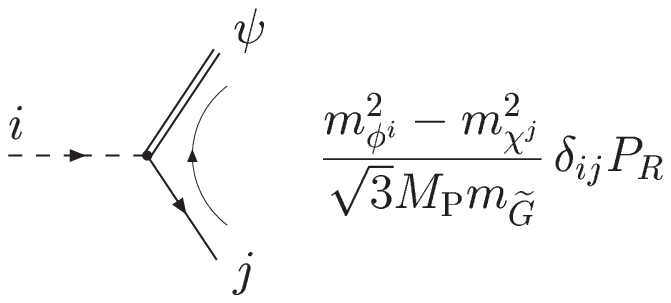}\hspace*{\fill}

\hfill\includegraphics[width=0.45\textwidth]{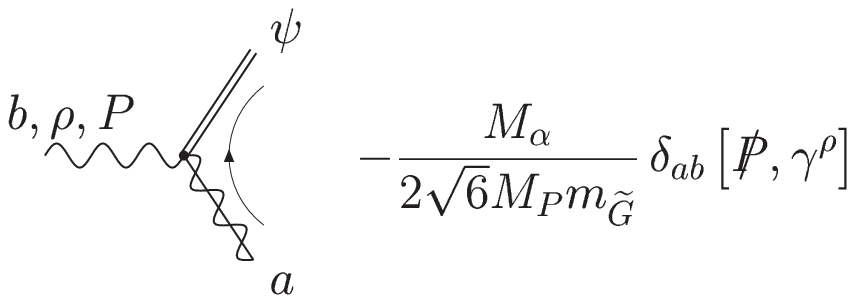}%
\includegraphics[width=0.45\textwidth]{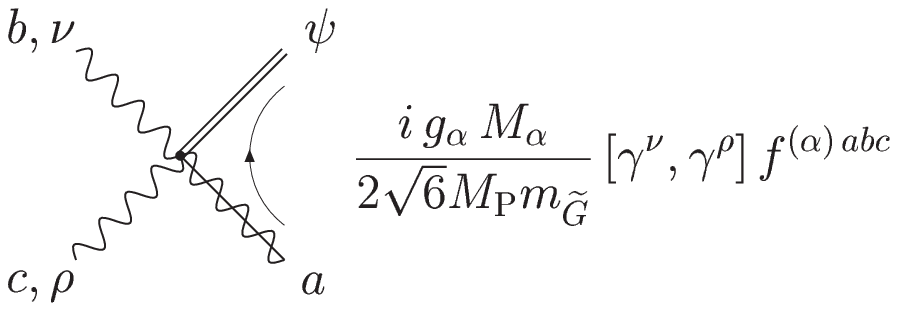}\hspace*{\fill}

\hfill
\includegraphics[width=0.45\textwidth]{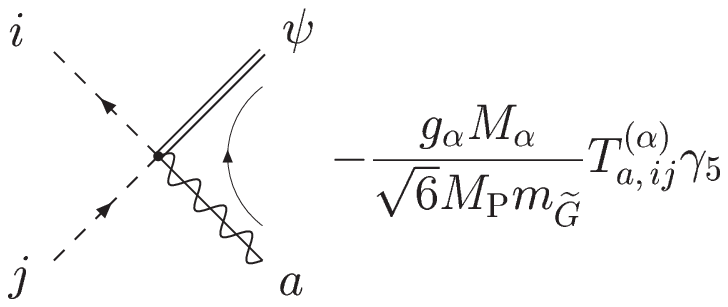}\hspace*{0.5\textwidth}
\end{minipage}%
\caption[Feynman rules for light Gravitinos]{Feynman rules for the
  effective theory of light gravitinos
  from~(\ref{eq:goldstino-lagrangian}) for one direction of the
  fermion flow. The double solid line now denotes the Majorana
  spinor $\psi$. Again, all momenta are understood to be ingoing.}
\label{fig:Feynman-rules-Goldstino}
\end{figure}

For a light gravitino, its interactions are dominated by goldstino
dynamics. The corresponding Feynman rules derived from the effective
Lagrangian~(\ref{eq:goldstino-lagrangian}) are shown in
Fig.~\ref{fig:Feynman-rules-Goldstino}. The double solid lines now
represents the Majorana field $\psi$.

The complete set of the relevant Feynman rules including vertices with
both directions of the fermion flow are given in
Appendix~\ref{cha:feynman-rules}.


\cleardoublepage
\chapter{Thermal Gravitino Production}
\label{cha:electr-therm-grav}

Thermal gravitino production in a consistent thermal field theory
approach has been worked out for supersymmetric quantum chromodynamics
 in Ref.~\cite{Bolz:2000fu}. Taking
into account also the electroweak processes, we extend
the calculation to the full standard model gauge group $\smgroup$ to
leading order in the gauge couplings. We also correct an error 
in the $\suthree$ result of Ref.~\cite{Bolz:2000fu}.

\section{The Braaten--Yuan Prescription}
\label{sec:braaten-yuan-prescription}

In the previous chapter we have seen that gravitino interactions are
suppressed by inverse powers of $\mplanck$. Thus, the dominant
contributions to gravitino production and annihilation processes to
leading order in the gauge couplings are inelastic $2 \rightarrow 2$
reactions with one external gravitino. For one of these $2
\rightarrow 2$ scattering processes, the net gravitino production rate
at finite temperature reads {\cite{Braaten:1991dd}}
\begin{align}
  \label{eq:hard-prod-rate}
   \frac{d\G_{\gr}}{d^3p} & = { 1\over 2 (2\pi)^3 E} \int {d\Omega_p \over 4\pi}
  \int \left[\prod_{i=1}^3 \frac{ d^3 p_i }{ (2\pi)^3 2E_i} \right]
  (2\pi)^4
  \delta^4( P_1+ P_2- P_3 - P)  \nonumber \\
  &\times \left\{ f_{\text{1}}(E_1) f_{\text{2}}(E_2) [1 \pm
    f_{\text{3}}(E_3)][1 - f_{\gr}(E)]|M(1+2 \rightarrow 3 +
    \gr)|^2
  \right.   \nonumber \\
  & \left.\qquad - [1 \pm f_{\text{1}}(E_1)] [1 \pm f_{\text{2}}(E_2)]
    f_{\text{3}}(E_3) f_{\gr}(E) |M(3+ \gr \rightarrow 1+2) |^2
  \right\} ,
\end{align}
where $E$ and $P$ are the energy and four-momentum of the
gravitino, respectively.  The corresponding squared matrix element $|M|^2$ is
weighted with the phase space distribution functions $f_i(E_i)$ of the
particles involved in the scattering; $\pm$ applies for final-state
bosons/fermions and corresponds to Bose enhancement/Pauli blocking,
respectively. In expression~(\ref{eq:hard-prod-rate}), an average
$d\O_p /4\p$ over the directions of the gravitino momentum is taken.
The squared matrix elements $|M|^2$ are assumed to be summed over
initial and final polarizations and to be weighted with the
appropriate multiplicities .

At high temperatures, all particles except the gravitino are in
thermal equilibrium so that $f_1$, $f_2$, and $f_3$ are given by the
equilibrium distributions,
\begin{subequations}
  \begin{align}
    \label{eq:bose-fermi-distributions}
    \fbose{E_i} &= {1 \over e^{E_i/T} - 1} \qquad \text{for bosons}, \\
    \ffermi{E_i} &= {1 \over e^{E_i/T} + 1} \qquad \text{for fermions},
  \end{align}
\end{subequations}
where $T$ denotes the temperature of the thermal bath.

Assuming that inflation governed the earliest moments of the Universe,
any initial population of gravitinos must be diluted away by the
exponential expansion during the slow-roll phase. We consider the
thermal production (or regeneration) of gravitinos that starts after
completion of reheating at the temperature~$\TR$.  Accordingly, the
gravitino phase space density $f_{\gr}$ is much smaller than the
equilibrium distribution $f_{\text{F}}$.  We thus can set
\mbox{$\left(1-f_{\gr}\right) \simeq 1$} and neglect gravitino
disappearance processes $ 3+ \gr \rightarrow 1+2 $. The production
rate~(\ref{eq:hard-prod-rate}) then becomes
\begin{align}
  \label{eq:hard-prod-rate-V2}
   \frac{d\G_{\gr}}{d^3p} & = { 1\over 2 (2\pi)^3 E} \int {d\Omega_p \over 4\pi}
  \int \left[\prod_{i=1}^3 \frac{ d^3 p_i }{ (2\pi)^3 2E_i} \right]
  (2\pi)^4
  \delta^4( P_1+ P_2- P_3 - P)  \nonumber \\
  &\times  f_{\text{F/B}}(E_1) f_{\text{F/B}}(E_2) [1 \pm
    f_{\text{F/B}}(E_3)]|M(1+2 \rightarrow 3 +
    \gr)|^2\; .
\end{align}

Note that a naive application of perturbation theory can lead to
logarithmically singular contributions to the production
rate. If a massless gauge boson with three-momentum $|\mathbf{k}|=k$ is exchanged
in the $t$- or $u$-channel, the corresponding squared matrix
element is divergent for \mbox{$k \rightarrow 0$}. A rigorous method to
deal with such situations is the Braaten--Yuan
prescription~\cite{Braaten:1991dd} which requires the weak coupling
limit, $g\ll 1$: One introduces an intermediate
momentum scale $k^*$ such that $gT \ll k^* \ll T$. This scale seperates
\textit{soft} gauge bosons with momentum transfers of order $gT$ from
\textit{hard} ones with momentum transfers of order $T$.  
The gravitino production rate is then given by the sum
\begin{align}
\label{eq:sum-soft-plus-hard}
\frac{d\G_{\gr}}{d^3p} & =\gammahard +\gammasoft \; .
\end{align}

The hard part is obtained conveniently by computing the squared matrix
elements $|M((1+2 \rightarrow 3 + \gr) |^2$ in standard
zero-temperature perturbation theory. Whenever a massless gauge
boson is exchanged in the $t$- or $u$-channel, $k^*$ will be
introduced as an infrared momentum cutoff.  The resulting
production rate will then be of the form
\begin{align}
  \label{eq:hard-form}
  \gammahard &= \left.\frac{d\G_{\gr}}{d^3p}\right|_{k^*<k} =
  A_{\text{hard}} + B \ln{ \left( \frac{T}{k^*} \right)}\; .
\end{align}

\smallskip

In the region of soft momentum transfer, $k<k^*$, one employs the
finite-temperature version of the optical theorem. At $T=0$, the
imaginary part of the self-energy is related to the decay width of an
unstable particle. In a thermal plasma, however, even stable particles
can disappear and be produced in inelastic scattering off particles in
the thermal background.  Accordingly, the discontinuity in the
self-energy gives the rate at which a non-equilibrium distribution of
the corresponding particle approaches thermal
equilibrium~\cite{Weldon:1983jn}. The soft part of the thermal
production (or regeneration) rate of gravitinos can thus be expressed in
terms of the thermal gravitino self-energy $\S_{\gr} (P)$,
\begin{align}
  \label{eq:soft-production-rate-gravitino}
 \gammasoft & = - \frac{1}{(2 \p)^3} \frac{\ffermi{E}}{E}  \,
  \text{Im}\,\S_{\gr}(E+i\ve,\mathbf{p})|_{k<k^*} \; .
\end{align}

In the naive consideration of the production rate, the singular
behavior stems from long-range forces. The physical cutoff is provided
by the screening of the interactions due to the cooperative motion of
particles in the thermal bath~\cite{Braaten:1991dd}.  Instead of using
bare gauge boson propagators, hard thermal loop (HTL) resummed gauge
boson propagators have to be used in the region of soft momentum
transfer, $k<k^*$ \cite{Braaten:1989mz}.  For gravitino
energies $E\apprge T$, there is no need to consider HTL-resummed
vertices. The soft part of the thermal gravitino production
rate~(\ref{eq:soft-production-rate-gravitino}) will then be of the
form
\begin{align}
 \gammasoft &= A_{\text{soft}} + B \ln{\left( \frac{k^*}{m_{\text{th}}}  \right)  }\;
\end{align}
with $m_{\text{th}} $ denoting the  thermal mass of the corresponding
gauge boson.

Indeed, in the sum~(\ref{eq:sum-soft-plus-hard}) the logarithmic dependence on
$k^*$ cancels out and yields a finite result. The Braaten--Yuan
prescription~\cite{Braaten:1991dd} together with the HTL-resummation
technique~\cite{Braaten:1989mz} allows us to calculate the gravitino
production rate   in a gauge-invariant way.

\section{Hard Contribution}
\label{sec:hard-contribution}

\begin{figure}[!htb]
\label{fig:2to2-scatterings}
\begin{itemize}
\item \textbf{A} (BBF): $A^{(\alpha)\,a} + A^{(\alpha)\,b} \rightarrow
  \lambda^{(\alpha)\,c} + \gr  $ \quad  ($\nexists$ for $\uone$)
  \begin{center}
    \vspace{-0.5\baselineskip}
    \includegraphics[width=0.9\textwidth]{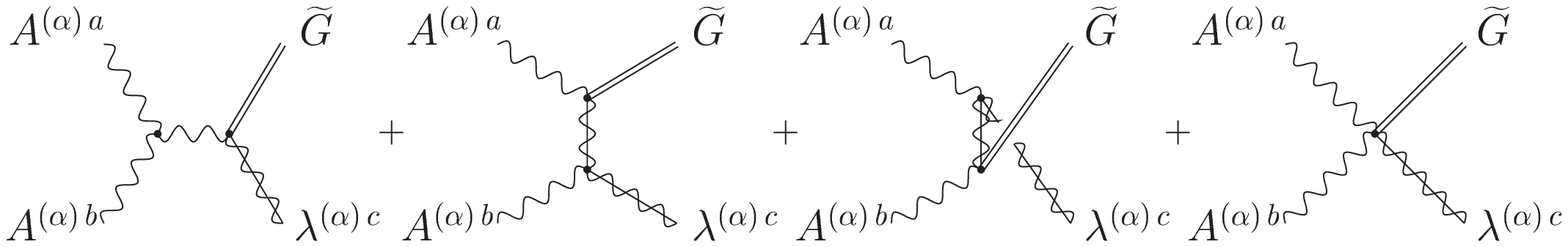}
    \vspace{-0.5\baselineskip}
  \end{center}
\item \textbf{B} (BFB): $A^{(\alpha)\,a} + \lambda^{(\alpha)\,b}  \rightarrow
  A^{(\alpha)\,c} + \gr  $ \quad (crossing of A, $\nexists$ for $\uone$)
\item \textbf{C} (BBF): $\phi^i + A^{(\alpha)\,b} \rightarrow
  \cf Lj + \gr  $  
  \begin{center}
    \vspace{-0.5\baselineskip}
    \includegraphics[width=0.9\textwidth]{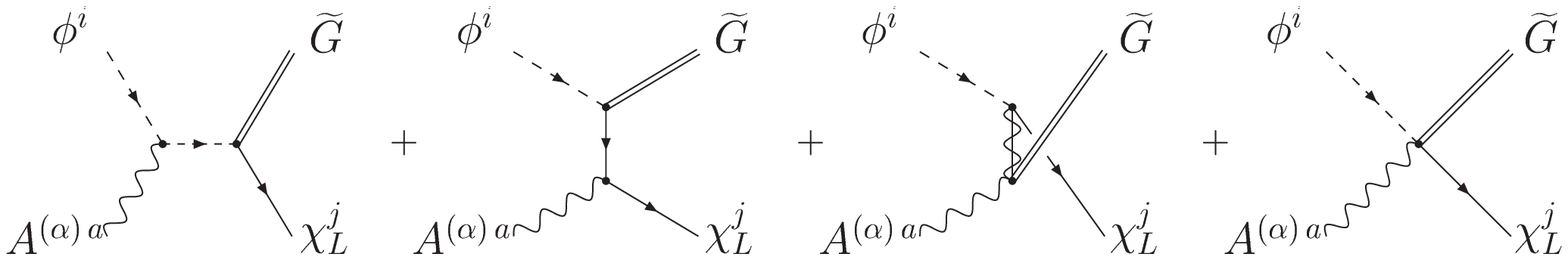}
    \vspace{-0.5\baselineskip}
  \end{center}
\item \textbf{D} (BFB): $ A^{(\alpha)\,a} + \cf Li \rightarrow
  \phi^j + \gr  $ \quad (crossing of C)  
\item \textbf{E} (BFB): $\phi^{*i}  + \cf Lj   \rightarrow
  A^{(\alpha)\,a}  +  \gr  $ \quad  (crossing of C) 
\item \textbf{F} (FFF): $\gfalpha{\a}{a} + \gfalpha{\a}{b}  \rightarrow
  \gfalpha{\a}{c} + \gr  $ \quad  ($\nexists$ for $\uone$)
  \begin{center}
 \vspace{-0.5\baselineskip}
    \includegraphics[width=0.9\textwidth]{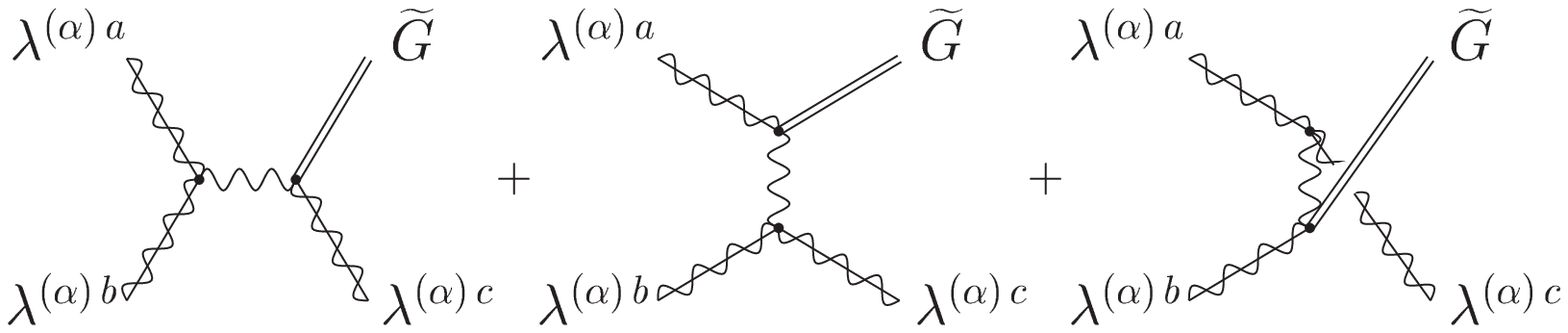}
 \vspace{-0.5\baselineskip}
  \end{center}
\item \textbf{G} (FFF): $\cf Li + \gfalpha{\a}{a} \rightarrow
  \cf Lj + \gr  $  
  \begin{center}
\vspace{-0.5\baselineskip}
    \includegraphics[width=0.9\textwidth]{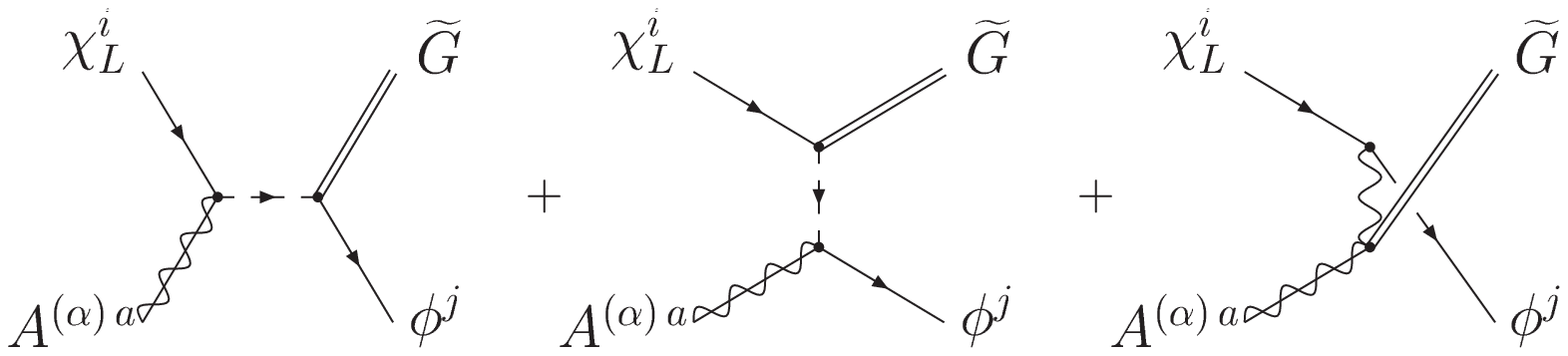}
\vspace{-0.5\baselineskip}
  \end{center}
\item \textbf{H} (BFB): $\phi^i + \gfalpha{\a}{a} \rightarrow
  \phi^j + \gr  $  
  \begin{center}
\vspace{-0.5\baselineskip}
    \includegraphics[width=0.9\textwidth]{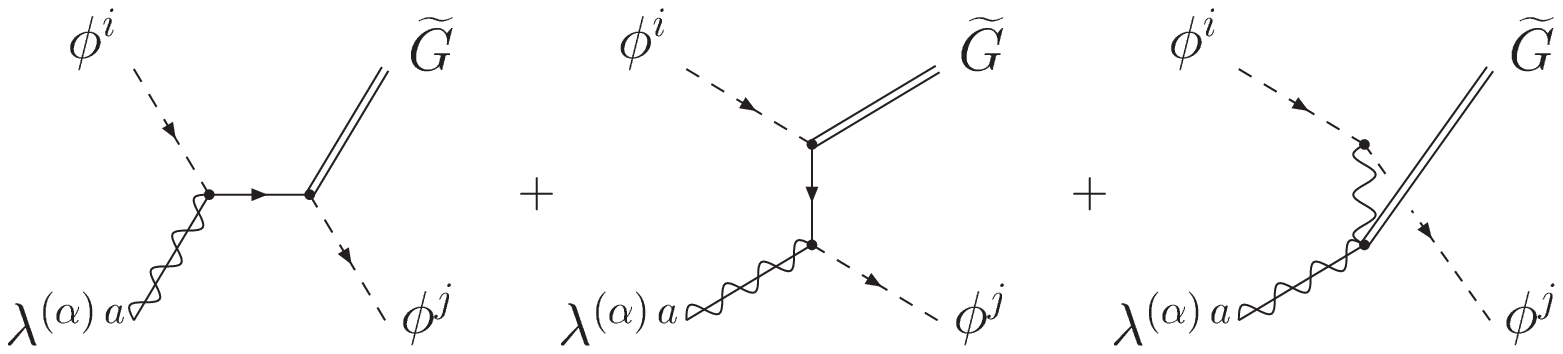}
\vspace{-0.5\baselineskip}
  \end{center}
\item \textbf{I} (FFF): $\cf Li +  \c^{\,c\,j}_L  \rightarrow
  \gfalpha{\a}{a} + \gr  $ \quad (crossing of G)
\item \textbf{J} (BBF): $\phi^i + \phi^{*j} \rightarrow
  \gfalpha{\a}{a} + \gr  $ \quad (crossing of H)  
\end{itemize}
\caption[Leading scattering processes]{The $2 \rightarrow 2$ scattering processes for gravitino
  production. Processes A, B, and F are not
present for $\uone$ since $f^{(1)\,abc} \equiv 0 $.  }
\end{figure}

From the Feynman rules for the gravitino and the supersymmetric gauge
interactions presented respectively in
Figs.~\ref{fig:Feynman-rules-Gravitino} and
\ref{fig:Feynman-rules-gauge}, we find the leading processes for
thermal gravitino production.  Figure~\ref{fig:2to2-scatterings} shows
these processes for one factor $\mathcal{G}_\a$ (\ref{eq:smgroup}) of
the standard model gauge group $\smgroup$.

\afterpage{\clearpage}

Note that the processes A, B, and F are not present for $\a =
1$,~i.e., for $\uone$, since there is no self-coupling of gauge bosons
for an abelian gauge group; thereby also the supersymmetrized version
of the relevant three-gauge boson vertex is absent. The matter
fields $\phi^i$ and $\cf Li$ coupling to the gauge bosons
$\gbalpha{\a}{a}$ of the corresponding group $\mathcal{G}_\a$ can be
read from Table~\ref{tab:matter-fields-mssm}.

The squared matrix elements for the processes shown in 
Fig.~\ref{fig:2to2-scatterings} are given in Table
\ref{tab:squared-matrix-elements} in terms of the Mandelstam variables
\begin{subequations}
  \label{eq:mandelstam}
\begin{align}
  s & = ( P_1 + P_2 )^2\; , \\
  t & = ( P_1 - P_3 )^2\; ,
\end{align}
\end{subequations}
where the four-momenta $P_1$, $P_2$, and $P_3$ are associated with the
particles in the order in which they are written down in the
column ``Process i'' of Table~\ref{tab:squared-matrix-elements}. The
gaugino masses are written as $M_\a$.

\begin{table}[tb]
\label{tab:squared-matrix-elements}
\caption[Squared Matrix elements]{Squared matrix elements for the $2
  \rightarrow 2$ scatterings in terms of the Mandelstam variables $s$
  and $t$; $M_\a$ denote the gaugino masses. Processes  A, B
  and F are not present for $\uone$. Sums over initial and
  final spins have been performed.} 
\begin{center}
\begin{tabular}{cc@{\qquad}r@{$\;\rightarrow\;$}l@{\qquad}l}
\toprule
Label $i$ & Class &
 \multicolumn{2}{c}{ Process $i$ }
 & $ | \mathcal{M}_i|^2 \Big/ g_\a^2 \,\mplanck^{-2}
\left( 1 + \frac{M_\a^2}{3 \mgrav^2}  \right)$   \\ 
\midrule
\vspace{-0.4cm} \\ 
A & BBF  &  $A^{(\alpha)\,a} + A^{(\alpha)\,b} $ & $
  \lambda^{(\alpha)\,c} + \gr  $
  & $ 4 \left(s + 2t + 2 \frac{t^2}{s} \right) |f^{(\alpha)\, abc}  |^2  $
\vspace{-0.2cm} \\ 
\\
B & BFB  & $A^{(\alpha)\,a} + \lambda^{(\alpha)\,b} $ & $
  A^{(\alpha)\,c} + \gr  $
  &$ -4 \left(t + 2s + 2 \frac{s^2}{t} \right) |f^{(\alpha)\, abc}  |^2  $
\vspace{-0.2cm} \\ 
\\
C  & BBF &  $\phi^i + A^{(\alpha)\,b}  $ & $
  \cf Lj + \gr  $  
  & $ 2 s | \gen aij^{(\alpha)} |^2  $
\vspace{-0.2cm} \\ 
\\
D & BFB &   $ A^{(\alpha)\,a} + \cf Li  $ & $
  \phi^j + \gr  $
  &  $ - 2 t | \gen aij^{(\alpha)} |^2  $
\vspace{-0.2cm} \\ 
\\
E & BFB  &   $\phi^{*i}  + \cf Lj   $ & $
  A^{(\alpha)\,a}  +  \gr  $
  &  $ - 2 t | \gen aij^{(\alpha)} |^2  $
\vspace{-0.2cm} \\ 
\\
F  & FFF &   $\gfalpha{\a}{a} + \gfalpha{\a}{b}  $ & $
  \gfalpha{\a}{c} + \gr  $
  &  $ - 8  \frac{ (s^2 + s\,t + t^2)^2 }{s \,t\,(s+t)} |f^{(\alpha)\, abc}  |^2  $
\vspace{-0.2cm} \\ 
\\
G  & FFF &  $\cf Li + \gfalpha{\a}{a} $ & $
  \cf Lj + \gr  $  
  & $ -4\left( s + \frac{s^2}{t}  \right) | \gen aij^{(\alpha)} |^2  $
\vspace{-0.2cm} \\ 
\\
H  & BFB &   $\phi^i + \gfalpha{\a}{a}  $ & $
  \phi^j + \gr  $ 
  & $ -2 \left( t + 2s + 2 \frac{s^2}{t} \right) | \gen aij^{(\alpha)} |^2  $
\vspace{-0.2cm} \\ 
\\
I  & FFF &   $\cf Li +  \c^{\,c\,j}_L   $ & $
  \gfalpha{\a}{a} + \gr  $
  & $ -4\left( t + \frac{t^2}{s}  \right) | \gen aij^{(\alpha)} |^2  $
\vspace{-0.2cm} \\ 
\\
J  & BBF &  $\phi^i + \phi^{*j} $ & $
  \gfalpha{\a}{a} + \gr  $
  & $ 2 \left(s + 2t + 2 \frac{t^2}{s}  \right)  | \gen aij^{(\alpha)} |^2  $
\vspace{-0.2cm} \\ 
\\
\bottomrule
\end{tabular}
\end{center}
\end{table}

The Dirac traces occurring in the evaluation of the squared matrix
elements have been performed using the computer program
\texttt{FORM}~\cite{Vermaseren:2000nd}. Polarizations sums over final
\textit{and} initial states have been carried out (no
averaging).\footnote{This is correct since there is no factor included
  in the definition for the hard production
  rate~(\ref{eq:hard-prod-rate}), which counts the internal degrees of
  freedom.} In process A, there are two gauge bosons with four-momenta
$P_1$ and $P_2$ in the initial state. The simple replacement of the
corresponding polarization sum $\sum_{\text{pol}} \ve^\m_a \ve^{*\n}_b
\rightarrow -g^{\m\n}\d_{ab}$ would amount to the inclusion of
longitudinal polarizations. Since we work in the high-energy limit of
unbroken electroweak symmetry, not only the gluons but also the
electroweak gauge bosons are massless.  Therefore, we have used the
proper polarization sum
\begin{align}
  \label{eq:pol-sum-for-two-gauge-bosons}
  \sum_{\text{pol.}} \ve_a^\m(P_{1/2})\ve_b^{*\n}(P_{1/2})  & = \left[ -g^{\m\n} +
  \frac{2}{s} \left( P_1^\m P_2^\n + P_1^\n P_2^\m\right) \right] \d_{ab}   \; ,
\end{align}
in which the unphysical modes are already
subtracted~\cite{Cutler:1977qm}.  
The Feynman rules used in the evaluation of the Feynman diagrams
are given in Appendix~\ref{cha:feynman-rules}.

The squared amplitudes shown in  Table~\ref{tab:squared-matrix-elements} have
to be weighted with appropriate multiplicities:
\begin{description}
\item[Processes A, B, F] The sum over the gauge-group indices
  $a,\,b,\,c = 1,\dots ,\dim{\mathcal{G}_\a}$ has to be performed:
  \begin{align}
    \label{eq:sum-over-structure constants}
    \sum_{a,b,c} | f^{(\a)\,abc}  |^2 &= N_{\a}  (N_{\a}^{\,2} - 1)
    \quad \text{for}\quad \text{SU}(N_\a)\; .
  \end{align}
  For processes A and F, an additional factor
  of $1/2$ occurs because of identical particles in the initial state.
  Note that the structure constants vanish for $\uone$.
\item[Processes C, D, E, G, H] In the corresponding initial state, there can be
  \begin{subequations}
     \label{eq:nsu2 and nsu3}
    \begin{align}
      2\, n_2 &\equiv 28 ,\label{eq:nsu2}
      \intertext{ distinct isospin-doublets for $\sutwo$ or}
      2 \, n_3 &\equiv 24  ,\label{eq:nsu3}
    \end{align}
  \end{subequations}
  color-triplets for $\suthree$;
  cf.~Table~\ref{tab:matter-fields-mssm}. The factor $2$ takes into
  account the corresponding conjugate multiplet as the initial state.
  The sum over the isospin/color degrees of freedom reads
  \begin{align}
    \label{eq:sum-over-generators}
    \sum_{i,j,a} |\gen aij^{(\a)}|^2 &= \frac{1}{2} ( N_\a^{\,2} - 1 )
    \quad \text{for}\quad \text{SU}(N_\a)\; .
  \end{align}
  For $\uone$, we have $ \gen aij^{(1)} = \d_{ij} \d_{a1} Y_i^2/4$ with
  $Y_i$ given in Table~\ref{tab:matter-fields-mssm}. All matter fields
  of the MSSM carry non-vanishing hypercharge. Thus, the correct
  mul\-ti\-pli\-ci\-ty factor for the $\uone$ contributions is obtained by
  summing over all squared hypercharges of either scalars~$\phi^i$
  \textit{or} fermions~$\cf Li$ listed in
  Table~\ref{tab:matter-fields-mssm}. We find
  \begin{align}
    \label{eq:n1}
    2\, n_1 & \equiv 2\, \sum_{\phi^i/\cf Li} \frac{Y_i^2}{4} =22 \; , 
  \end{align}
  where the factor of 2 accounts for the corresponding
  antiparticles.
\item[Processes I, J] Since I and J describe the same physical
  processes after replacing the incoming fields by their conjugates,
  the multiplicities (\ref{eq:nsu2}), (\ref{eq:nsu3}),
  and~(\ref{eq:n1}) apply without the factor of 2.
\end{description}

The diagrams of Fig.~\ref{fig:2to2-scatterings} fall into three
separate classes depending on the number of bosons and fermions
involved in the inital and final state. The BBF processes A, C, and J
have two bosons in the initial state and one fermion in addition to
the gravitino in the final state.  Correspondingly, B, D, E, and H are
BFB processes and F, G, and I are FFF processes. The hard part of the
thermal gravitino production rate for the full standard model gauge
group $\smgroup$ can then be written as
\begin{align}
\label{eq:hard-production-rate}
 \gammahard & =  {
  1\over 2 (2\pi)^3 E}  \int {d\Omega_p \over 4\pi} \int
\left[\prod_{i=1}^3 \frac{ d^3 p_i }{ (2\pi)^3 2E_i} \right] (2\pi)^4 \delta^4( P_1+ P_2- P- P_3) \\
&\times  \sum_{\alpha = 1}^3 \left( f_{\BFB} |M^{ ( \alpha ) }_{\BFB}|^2 +  f_{\BBF} |M^{ ( \alpha ) }_{\BBF}|^2 
  + f_{\FFF} |M^{ ( \alpha ) }_{\FFF}|^2\right)\Theta(|{\mathbf p}_1-{\mathbf p}_3|-k^*) \nonumber
\end{align}
with the shorthand notation
\begin{subequations}
\begin{align}
\label{eq:distributions-product}
f_{\BFB} &= \fbose{E_1}  \ffermi{E_2}\left[ 1 +\fbose{E_3} \right],\\
f_{\BBF} &= \fbose{E_1}f_B(E_2)\left[1 - \ffermi{E_3}\right], \\
f_{\FFF} &= \ffermi{E_1} \ffermi{E_2} \left[1 - \ffermi{E_3}\right].
\end{align} 
\end{subequations} 

\newpage
\smallskip\noindent\textbf{Contributions from $\mathbf{\sutwo}$
  ($\boldsymbol{\alpha} \mathbf{ = 2}$) and
  $\mathbf{\suthree}$ ($\boldsymbol{\alpha} \mathbf{ = 3}$) }\smallskip

For the weighted sums, we find
\begin{subequations}
   \label{eq:M-BFB-BBF-FFF-version1}
  \begin{flalign}
    |M^{(\a)}_{\BFB}|^2 &= \left( 1 + \frac{M_\a^2}{3 \mgrav^2}
    \right) \frac{4 g_\a^2 (N_\a^2-1)}{\mplanck^2} \left[ \left(
        -t-2s- 2 \frac{s^2}{t} \right) \left(N_\a + \frac{n_\a}{2}
      \right)-  t\, n_\a \right],
    \label{eq:M-BFB-version1} \\
    |M^{(\a)}_{\BBF}|^2 &= \left( 1 + \frac{M_\a^2}{3 \mgrav^2}
    \right) \frac{2 g_\a^2 (N_\a^2-1)}{\mplanck^2} \left[ \left( s +
        2t + 2\frac{2t^2}{s} \right) \left(N_\a + \frac{n_\a}{2}
      \right) +  s\,n_\a \right],
    \label{eq:M-BBF-version1} \\
    |M^{(\a)}_{\FFF}|^2 &= \left( 1 + \frac{M_\a^2}{3 \mgrav^2}
    \right) \frac{4 g_\a^2 (N_\a^2-1)}{\mplanck^2} \left[    
      - \frac{ (s^2 + s\,t + t^2)^2 }{s \,t\,(s+t)} N_\a 
      - \left( t + 2s + \frac{t^2}{s} + 2 \frac{s^2}{t} \right)\! \frac{n_\a}{2}
    \right]\! .
    \label{eq:M-FFF-version1} 
  \end{flalign}
\end{subequations}
Equation (\ref{eq:M-FFF-version1}) can be rewritten as
\begin{align}
  \label{eq:M-FFF-version2}
    |M^{(\a)}_{\FFF}|^2 = \left( 1 + \frac{M_\a^2}{3 \mgrav^2}
    \right) \frac{4 g_\a^2 (N_\a^2-1)}{\mplanck^2} & \left[    
      \left( - t -2s - \frac{s^2}{t} + \frac{s^2}{s+t} - \frac{t^2}{s}
      \right)  \left(N_\a + \frac{n_\a}{2} \right) \right.  \nonumber \\ &
    \left.  - \left(  \frac{s^2}{t} + \frac{ s^2 }{ s + t } \right) \frac{n_\a}{2}
    \right].
\end{align}
Since $ s +t +u = \sum m_i^2$, where  $m_i$ are the masses
of the external particles, and \mbox{$u = ( P_1 - P )^2 $}, we can write $ s + 2t = t
- u$ in the high-energy limit, $T\gg m_i$.  Therefore,
\begin{align}
  \pm\frac{s^2}{t}+\frac{s^2}{s+t} &=\pm\frac{s^2}{t}-\frac{s^2}{u}\; .
\end{align}
Since the difference $t-u$ and $1/t-1/u$ is odd under exchange of
$P_1$ and $P_2$, the integral over such terms will be zero, as long as
the remaining integrand and the measure is even under this
transformation. Therefore, there is no contribution to the integral
from $ s + 2t $ in (\ref{eq:M-BBF-version1}) and we can further
substitute $s$ by $-2t$ in the last term of
(\ref{eq:M-BBF-version1}). In Eq.~(\ref{eq:M-FFF-version2}) we
use the following replacements
\begin{subequations}
\begin{align}
\frac{s^2}{t}+\frac{s^2}{s+t} &\rightarrow 0\; , \\
  -\frac{s^2}{t}+\frac{s^2}{s+t} &\rightarrow -\frac{2s^2}{t}\; .
\end{align}
\end{subequations}
For the gauge groups $\sutwo\;(\a = 2)$ and $\suthree\;(\a = 3)$, we
can therefore write the weighted sums of the squared matrix elements
of Table~\ref{tab:squared-matrix-elements} in terms of three distinct
matrix elements,
\begin{subequations}
   \label{eq:M-BFB-BBF-FFF-version2}
  \begin{align}
     |M^{(\a)}_{\BFB}|^2 &= \left( 1 + \frac{M_\a^2}{3 \mgrav^2}
    \right) \frac{4 g_\a^2 (N_\a^2-1)}{\mplanck^2} \left[
      |M_1|^2 \left( N_\a + \frac{n_\a}{2} \right) - |M_2|^2 n_\a
    \right] \; ,
   \label{eq:M-BFB-version2}  \\
    |M^{(\a)}_{\BBF}|^2 &= \left( 1 + \frac{M_\a^2}{3 \mgrav^2}
    \right) \frac{2 g_\a^2 (N_\a^2-1)}{\mplanck^2} \left[
   |M_3|^2\left(N_\a + \frac{n_\a}{2} \right)
      - |M_2|^2 n_\a \right]\; , 
   \label{eq:M-BBF-version2} \\
   |M^{(\a)}_{\FFF}|^2 &= \left( 1 + \frac{M_\a^2}{3 \mgrav^2}
    \right) \frac{4 g_\a^2 (N_\a^2-1)}{\mplanck^2} \left[ 
      \vphantom{ \frac{n\a}{2} }
      |M_1|^2  - |M_3|^2 \right] \left( N_\a + \frac{n_\a}{2} \right) \; ,
  \end{align}
\end{subequations}
with $n_\a$ as defined in~(\ref{eq:nsu2 and nsu3}) and
\begin{subequations}
\label{eq:matrixelementsM1toM3}
\begin{align}
|M_1|^2 &= -t-2s-\frac{2s^2}{t}\;, \label{eq:matrixelementM1}\\
|M_2|^2 &= t\;, \\
|M_3|^2 &= \frac{t^2}{s}\; .
\end{align}
\end{subequations}
Recall from (\ref{eq:sum-over-structure constants}) and
(\ref{eq:sum-over-generators}) that $N_\a$ denotes the dimension of
the fundamental representation of the corresponding gauge group,~i.e.,
$N_2 = 2$ and $N_3 = 3$.

\smallskip\noindent\textbf{Contribution from $\mathbf{\uone}$  ($\boldsymbol{\alpha} \mathbf{ = 1}$)}\smallskip

In analogy to the considerations which lead
to~(\ref{eq:M-BFB-BBF-FFF-version2}), we obtain for the contributions from the
$\uone$ interactions 
\begin{subequations}
   \label{eq:M-BFB-BBF-FFF-U1}
  \begin{align}
     |M^{(1)}_{\BFB}|^2 &= \left( 1 + \frac{M_1^2}{3 \mgrav^2}
    \right) \frac{4 g_1^{2} n_1}{\mplanck^2} \left(
      |M_1|^2 - 2 |M_2|^2 
    \right) \; ,
   \label{eq:M-BFB-U1}  \\
    |M^{(1)}_{\BBF}|^2 &= \left( 1 + \frac{M_1^2}{3 \mgrav^2}
    \right) \frac{2 g_1^2 n_1}{\mplanck^2} \left(
   |M_3|^2   - 2 |M_2|^2 \right)\; , 
   \label{eq:M-BBF-U1} \\
   |M^{(1)}_{\FFF}|^2 &= \left( 1 + \frac{M_1^2}{3 \mgrav^2}
    \right) \frac{4 g_1^2 n_1}{\mplanck^2} \left(
            |M_1|^2  - |M_3|^2 \right) \; ,
  \end{align}
\end{subequations}
with $n_1$ as defined in (\ref{eq:n1}) and $g_1 = g'$. Note that $M_1$ in the
prefactor of (\ref{eq:M-BFB-BBF-FFF-U1}) denotes the Bino mass while the squared
matrix elements $|M_i|^2$ are the ones of
(\ref{eq:matrixelementsM1toM3}).

For the calculation of the gravitino production
rate~(\ref{eq:hard-production-rate}), various integrations are
necessary. Since the structure of the integrands is identical for all
factors $\mathcal{G}_\a$ of the standard model gauge group, we
introduce the following notation
\begin{subequations}
\label{eq:hard-prod-rate-BFB-FFF-BBF-with-coefficients}
\begin{align}
  \label{eq:hard-prod-rate-BFB}
  \gammahard^{(\BFB)} & =   \sum_{\alpha = 1}^3 
  \frac{g_\a^2}{\mplanck^2} \left( 1+\frac{M^2_{ \alpha
      }}{3\mgrav^2}\right)   \sum_{i=1}^3
  \left[ c_{ \BFB,\, i }^{(\alpha)}\, I_{ \BFB}^{|M_i|^2}
    \right],
\\
  \label{eq:hard-prod-rate-FFF}
  \gammahard^{(\FFF)} & =   \sum_{\alpha = 1}^3 
  \frac{g_\a^2}{\mplanck^2} \left( 1+\frac{M^2_{ \alpha
      }}{3\mgrav^2}\right)   \sum_{i=1}^3
  \left[ c_{ \FFF,\, i }^{(\alpha)}\, I_{ \FFF}^{|M_i|^2}
    \right],
\\
  \label{eq:hard-prod-rate-BBF}
  \gammahard^{(\BBF)} & =   \sum_{\alpha = 1}^3 
  \frac{g_\a^2}{\mplanck^2} \left( 1+\frac{M^2_{ \alpha
      }}{3\mgrav^2}\right)   \sum_{i=1}^3
  \left[ c_{ \BBF,\, i }^{(\alpha)}\, I_{ \BBF}^{|M_i|^2}
    \right].
\end{align}
\end{subequations}

The  coefficients $c_{\BFB,\,i }^{(\alpha)}$ are given in
Table~\ref{tab:coefficientsBFB}. The other factors are given by
\begin{subequations}
  \begin{align}
    \label{eq:relations coefficients}
   c_{\BBF,\,1 }^{(\alpha)} & =  c_{\FFF,\,2 }^{(\alpha)} =
   c_{\BFB,\,3 }^{(\alpha)}\; , \\
   c_{\BBF,\,2 }^{(\alpha)} & =  c_{\BFB,\,2 }^{(\alpha)}\; , \\
   c_{\BBF,\,3 }^{(\alpha)} & =  c_{\FFF,\,1 }^{(\alpha)} =
   - c_{\FFF,\,3 }^{(\alpha)} =   c_{\BFB,\,1 }^{(\alpha)}  \; .
  \end{align}
\end{subequations}

\begin{table}[tb]
\label{tab:coefficientsBFB}
\caption[Multiplicity coefficients]{Multiplicity coefficients
  for the squared matrix elements in~(\ref{eq:hard-prod-rate-BFB-FFF-BBF-with-coefficients})} 
\begin{center}
\begin{tabular}{c@{\hspace{1cm}}c@{\hspace{1cm}}c@{\hspace{1cm}}c}
\toprule
$c_{\BFB,\,i }^{(\alpha)}$  & $\a = 1 $ &
 $\a = 2$ & $\a = 3 $   \\ 
\midrule
\vspace{-0.4cm} \\ 
$i = 1\;$ & $ 4 n_1 $ &    $ 4 (N_2^2 -1) \left(N_2 + \frac{n_2}{2} \right) $  &    $ 4 (N_3^2 -1) \left(N_3 + \frac{n_3}{2} \right) $
\vspace{-0.2cm} \\ 
\\
$i = 2\; $ & $ - 8 n_1  $   & $- 4 (N_2^2 -1) n_2$  &  $ - 4 (N_3^2 -1) n_3$  
\vspace{-0.2cm} \\ 
\\
$i = 3\; $   & 0  &  0  & 0
\vspace{-0.2cm} \\ 
\\
\bottomrule
\end{tabular}
\end{center}
\end{table}

The calculation of the integrals $ I_{ \BFB}^{|M_i|^2}$, $ I_{
  \FFF}^{|M_i|^2}$, and $ I_{ \BBF}^{|M_i|^2}$ is presented in detail
in Appendix~\ref{cha:Appendix-hard-production-rate}. Here are the
results:
\begin{flalign}
 \label{eq:I-BFB-FFF-M1-FINAL}
 I_{\left\{ {\BFB \atop \FFF}\right\} }^{|M_1|^2} & = \left\{ {1 \atop
     1/2 } \right\} \frac{T^3 \ffermi{E}}{192 \p^4} \left[ \ln{\left(
       \frac{2 T }{ k^* } \right)} + {17 \over 6} - \gamma + {\zeta
     '(2) \over \zeta(2)}
   - \left\{ {  \ln{2} \atop 0 } \right\}   \right]  \nonumber \\
 & \quad + \frac{1}{256\pi^6 } \int_0^\infty dE_3 \int_0^{E+E_3}
 dE_1  \ln{\left(\frac{|E_1-E_3|}{E_3}\right)} \nonumber \\
 & \quad \times \Bigg\{ - \Theta(E_1-E_3) \frac{d}{dE_1}\left[
   f_{\left\{ {\BFB \atop \FFF}\right\} }\frac{E_2^2}{E^2}
   \left(E_1^2+E_3^2\right)\right]
 \nonumber \\
 &\qquad \quad + \Theta(E_3-E_1) \frac{d}{dE_1}\left[ f_{\left\{ {\BFB \atop
         \FFF}\right\} } \left(E_1^2+E_3^2\right)\right] \nonumber
 &  \\
 & \qquad\quad + \Theta(E - E_1) {d\over dE_1} \left[ f_{\left\{ {\BFB
         \atop \FFF}\right\} } \left( \frac{ E_1^2 E_2^2}{E^2} - E_3^2
   \right) \right] \Bigg\}\; ,
\end{flalign}
\begin{flalign}
 \label{eq:I-BFB-BBF-M2-FINAL}
 I_{\left\{ {\BFB \atop \BBF}\right\} }^{|M_2|^2} &=
 \frac{1}{256\pi^6} \int_{0}^{\infty} dE_3 \int_{0}^{E+E_3} dE_2\,
 f_{\left\{ {\BFB \atop \BBF}\right\} }
 \nonumber \\
 & \times \Bigg\{  \Theta(E - E_3) \frac{ E_3^2}{E^2} \left( {E_3 \over
     3} - E_1 \right) + \Theta(E_3 - E)
 \left( {E \over 3} - E_2  \right) \nonumber \\
 &
 \qquad+ \Theta(E_3-E_2) \Theta(E - E_3) \frac{
   E_2-E_3 }{ 3 E^2} \left[ (E_2-E_3)(E_2+2 E_3) - 3(E_2+E_3)E \right]
 \nonumber \\  & \qquad
 -  \Theta(E_2-E) \Theta(E - E_3) \frac{(E_2 - E)^3}{3 E^2} \nonumber \\
 &\qquad + \Theta(E-E_2) \Theta(E_3 - E)
 \frac{(E_2 - E)^3}{3 E^2}
 & \nonumber \\
 &\qquad - \Theta(E_2-E_3) \Theta(E_3 - E) \frac{
   E_2-E_3 }{3 E^2} \left[ (E_2-E_3)(E_2+2 E_3) - 3(E_2+E_3)E \right]
 \Bigg\}\;,
\end{flalign}
\begin{flalign}
   \label{eq:I-FFF-BBF-M3-FINAL}
  I_{\left\{ {\FFF \atop \BBF}\right\} }^{|M_3|^2} & =  \frac{1}{256 \p^6} \int_{0}^{\infty}   dE_3  \int_{0}^{E+E_3}  dE_2
   \, f_{\left\{ {\FFF \atop \BBF}\right\} } 
    \nonumber 
    \\ & 
    \times \Bigg\{
        \Theta(E-E_3) \frac{1}{E^2} \frac{E_1^2
     E_3^2}{E+E_3} 
    \nonumber \\
    & \qquad + \Theta(E_3 - E)  \frac{ E_2^2}{E+E_3}
    \nonumber \\
    & \qquad - \Theta(E - E_3) \Theta(E_3 - E_2) \frac{E_2 -E_3}{E^2} 
   \left[  E_2(E_3 -E) - E_3( E_3+ E)   \right] \nonumber \\
    & \qquad + \Theta(E_3 - E) \Theta(E_2 - E_3) \frac{E_2 -E_3}{E^2} 
  \left[  E_2(E_3 -E) - E_3( E_3+ E)   \right] \Bigg\} \; ,&
\end{flalign}
where $\g = 0.57722\dots$ is the Euler-Mascheroni constant, $\z(x)$ is
Riemann's Zeta function with $\zeta (2) = \pi^2 /6 $ and $\zeta
'(2)/\zeta (2)= -0.56996\dots$\;. In the above expressions, $E_1+E_2 =
E_3 + E$ is understood.

Only the part $\propto 1/t$ in the squared matrix element $|M_1|^2 $
given in~(\ref{eq:matrixelementM1}) exhibits a singular behavior for
$t\rightarrow 0$.  Therefore, the integrals contributing to the
logarithmic dependence on~$k^*$ are~$I_{\left\{ {\BFB \atop
      \FFF}\right\} }^{|M_1|^2}$. The logarithm is extracted
analytically using integration by parts.  Details are given in
Appendix~\ref{cha:Appendix-hard-production-rate}; see
Eq.~(\ref{eq:logdep}) and subsequent steps.  In the calculation of
$I_{\left\{ {\BFB \atop \BBF}\right\} }^{|M_2|^2}$ and $I_{\left\{
    {\FFF \atop \BBF}\right\} }^{|M_3|^2}$, the cutoff $k^*$ is
set to zero from the very beginning.

The hard part of the  gravitino production rate (\ref{eq:hard-production-rate}) is
then given by the sum
\begin{align}
  \label{eq:prodrate-BFB-FFF}
  \gammahard & =   \gammahard^{(\BFB)} +  \gammahard^{(\FFF)}+  \gammahard^{(\BBF)} .
\end{align}

We correct an error in the $\suthree$ result of
Ref.~\cite{Bolz:2000fu}.\footnote{For a direct comparison with
  Ref.~\cite{Bolz:2000fu}, it should be stressed that our definition for
  the production rate differs by a factor of $\ffermi{E}/(2\p)^3$.}
We do not find the term
\begin{align}
  T^3(N+n_f)[\mathrm{Li}_2(-e^{-E/T})-\pi^2/6]\;
\end{align}
given as part of $I_{BFB}$ in (C.14) of
Ref.~\cite{Bolz:2000fu}. Although such a contribution appears as a
surface term in our calculation, it is canceled by another surface
term. The crucial spots to look at are Eqs.~(\ref{eq:g122-rest-final})
and~(\ref{eq:g2-surface-final}) in our
Appendix~\ref{cha:Appendix-hard-production-rate}. The authors of
\cite{Bolz:2000fu} agree with our finding and will publish an erratum.
The phenomenological implications will be discussed in
chapters~\ref{cha:gravitino-cosmology} and~\ref{sec:coll-tests-lept}.

\section{Soft Contribution}
\label{sec:soft-contribution}

As mentioned in Sec.~\ref{sec:braaten-yuan-prescription}, hard
thermal loop (HTL) resummed propagators have to be used for soft gauge
boson momentum transfers. Since diagrams of nominally higher order in
the loop expansions can contribute to same order in the coup\-ling
constant $g$ at high temperatures, this corresponds to an improved
perturbation theory where a certain subset of diagrams, the HTL
self-energies, are resummed~\cite{Pisarski:1988vd}.

\begin{figure}[bt]
\centering
\subfloat[Gauge boson contributions for $\suthree$ and $\sutwo$]{
    \label{fig:HTL-gauge-bososn-contr}
     \includegraphics[width=0.4\textwidth]{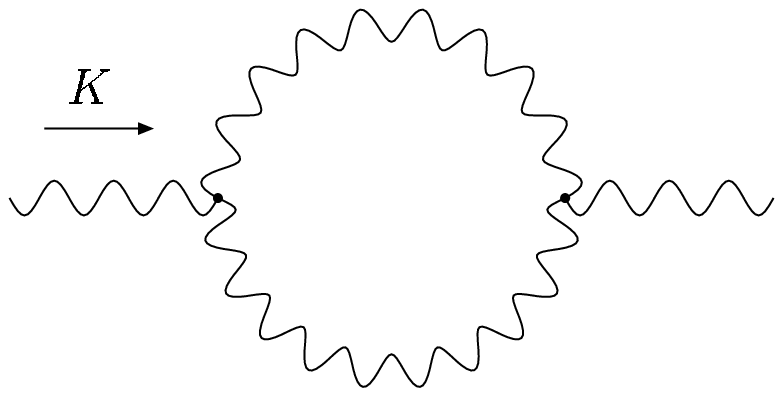}  \hspace{20pt}%
     \includegraphics[width=0.4\textwidth]{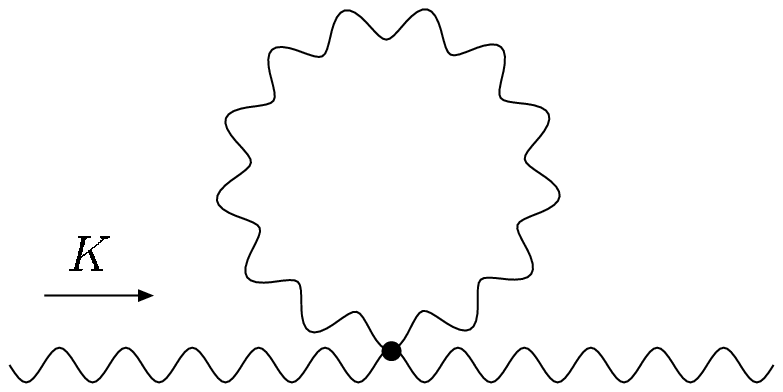}%
   } \\ 
   \subfloat[Gaugino contributions]{
    \label{fig:HTL-gaugino-contr}
    \includegraphics[width=0.4\textwidth]{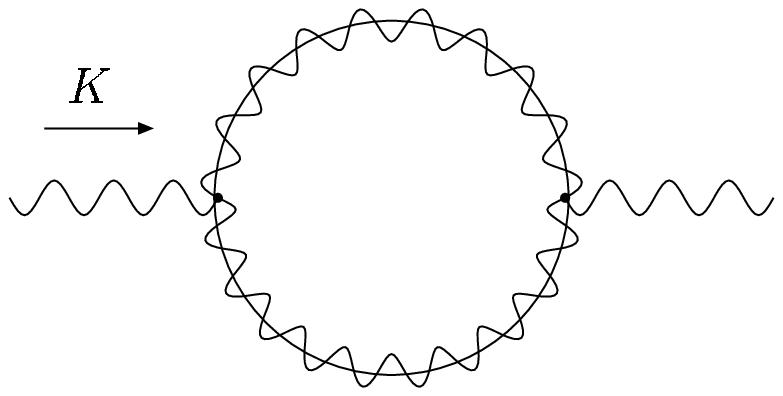}
    \hspace{20pt}%
  } \subfloat[Fermion contributions]{
    \label{fig:HTL-fermion-contr}
     \includegraphics[width=0.4\textwidth]{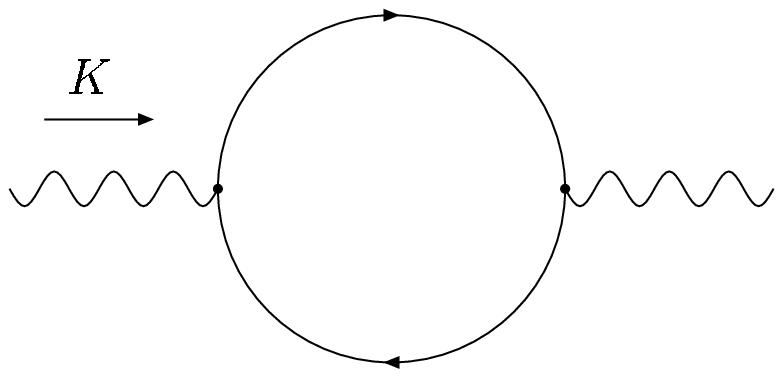}%
   }\\ 
   \subfloat[Scalar contributions]{
    \label{fig:HTL-scalar-contr}
     \includegraphics[width=0.4\textwidth]{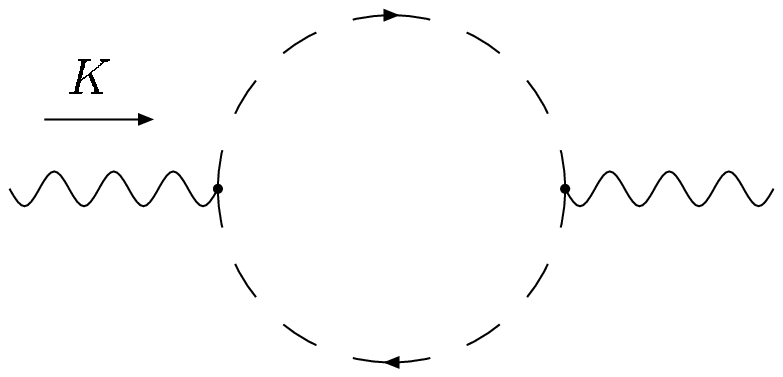}  \hspace{20pt}%
     \includegraphics[width=0.4\textwidth]{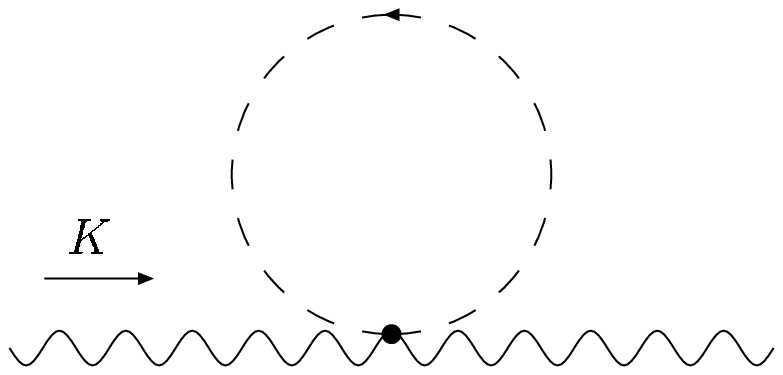}%
   } 
   \caption[Hard Thermal Loop self-energies]{HTL self-energy contributions to the
     gauge boson polarization tensor~$\P^{(\a)}_{\m\n}$. The internal
     loop momenta are \textit{hard},~i.e., of order $T$. In the
     resummation these contributions become relevant for \textit{soft} gauge boson
     momenta $K\sim gT$.}
\label{fig:HTL-energies}
\end{figure}

The HTL self-energies which contribute to the gauge boson polarization
tensor~$\P^{(\a)}_{\m\n}$ are given by one-loop diagrams such as the
ones shown in Fig.~\ref{fig:HTL-energies}. They carry \textit{soft}
external momenta $K=(k_0, \mathbf{k})$ with $k_0$ and $k=|\mathbf{k}|$
of order~$gT$ and \textit{hard} internal loop momenta of order~$T$.
In the HTL approximation of hard loop momenta, only a small part of
the integration region in the one-loop diagrams contributes. Hard
thermal loops are gauge invariant and the corresponding self-energy
satisfies the Ward identity \mbox{$K^\m\P^{(\a)}_{\m\n}(K) =0$}. They
are exclusively due to thermal fluctuations and are therefore
ultraviolet finite \cite{Braaten:1989mz}.

At finite temperatures, the thermal bath constitutes a privileged
rest frame. The polarization tensor $\P^{(\a)}_{\m\n}$ can be
decomposed into two independent propagating modes which are chosen to be the
longitudinal and the transverse parts of the self-energy of the gauge
bosons $A^{(\a)}_\m$. Both are physical and read respectively
\begin{subequations}
  \begin{align}
    \label{eq:long-and-trans-self-energies-DEF}
    \P_\mathrm{L}^{(\a)}(K) &= \P^{(\a)}_{00}(K)\; , \\
    \P_\mathrm{T}^{(\a)}(K) &= \frac{1}{2} \left( \d_{ij} - \frac{k_i
        k_j}{k^2}\right) \P^{(\a)}_{ij}(K)\; ,
  \end{align}
\end{subequations}

In the high temperature limit, $ k_0, k \ll T $, to
leading order in the gauge couplings $g_{\a}$, these components of the
self-energy are given by~\cite{Weldon:1982aq}
\begin{subequations}
    \label{eq:long-and-trans-self-energies}
  \begin{align}
    \label{eq:long-self-energies}
    \P_\mathrm{L}^{(\a)}(k_0, k) & = -3 \, m_{\a}^2\, \left( 1 - \frac{k_0}{2k}
      \ln{\frac{k_0 + k  }{k_0 -k }} \right) \; , \\
    \label{eq:trans-self-energies}
    \P_\mathrm{T}^{(\a)}(k_0, k) & = \frac{3}{2} \,m_{\a}^2 \,\frac{k_0^2}{k^2}
    \left[ 1- \left(1-\frac{k^2}{k_0^2}\right) \frac{k_0}{2 k}
      \ln{\frac{k_0 + k  }{k_0 -k  }} \right] \; ,
  \end{align}
\end{subequations}
where $m_\a$ denotes the thermal masses of the corresponding gauge
boson~$A^{(\a)}_\m$. Note that the self-energy
components~(\ref{eq:long-and-trans-self-energies}) depend on the gauge
boson momenta in a non-trivial way and that they have an imaginary
part for~\mbox{$k_0^2 < k^2 $}.

In the static limit, $k_0\rightarrow 0$, the longitudinal part of the
self-energy reduces to $ \P_\mathrm{L}^{(\a)}(0, k) = - 3 m_{\a}^2$ which is
analogous to the Debye screening of static electric fields in a QED
plasma with inverse screening length $\l_\mathrm{D}^{-1} = \sqrt{3} m_{\a} $.
There is no static magnetic screening since $\P_\mathrm{T}^{(\a)}(0,
k)=0 $. Nevertheless, the transverse self-energy $\P_\mathrm{T}$ approaches zero
sufficiently slow so that quantities in which the magnetic
divergence is only logarithmic are regularized~\cite{Braaten:1991dd}.

In SUSY extensions of the Standard Model, the hard thermal loops shown
in Fig.~\ref{fig:HTL-energies} contribute to the self-energy tensor
and thereby to the thermal gauge boson mass $m_\a$,
  \begin{align}
    m^2_\a & = m^2_{\a,\,\text{gauge bosons}} + m^2_{\a,\,\text{gauginos}} +
    m^2_{\a,\,\text{fermions}} + m^2_{\a,\,\text{scalars}}\; .
  \end{align}
  For the non-abelian gauge group $\text{SU}(N_\a)$, the contributions
  are given by~\cite{Bolz:2000xi,Braaten:1989mz}
\begin{subequations}
  \label{eq:thermal-masses-single-contr-nonabelian}
  \begin{align}
    m^2_{\a,\,\text{gauge bosons}} & = \frac{g_\a^2 T^2}{9} \sum_{b,\,c} |f^{(\a)\,abc}|^2 =
    N_\a \,\frac{g_\a^2 T^2}{9}\; , \\
    m^2_{\a,\,\text{gauginos}} & = \frac{g_\a^2 T^2}{18} \sum_{b,\,c} |f^{(\a)\,abc}|^2 =
    N_\a\,\frac{g_\a^2 T^2}{18} \; , \\
    m^2_{\a,\,\text{fermions}} & = n_\a \,\frac{ g_\a^2 T^2}{18} \sum_{i,\,j} |\gen
    aij^{(\a)}|^2 = n_\a  \,\frac{ g_\a^2 T^2}{36} \; ,\\
    m^2_{\a,\,\text{scalars}} & = n_\a \,\frac{ g_\a^2 T^2}{9} \sum_{i,\,j} |\gen
    aij^{(\a)}|^2 = n_\a \,\frac{ g_\a^2 T^2}{18}\; .
  \end{align}
\end{subequations}
For $\a= 2$ and $\a = 3$ with $n_\a$ defined in~(\ref{eq:nsu2 and
  nsu3}), this determines the thermal masses of the wino and the
gluino, respectively. They read
\begin{align}
\label{eq:thermal-masses-nonabelian}
  m^2_\a & = \frac{ g_\a^2 T^2}{6} \left ( N_\a + \frac{n_\a}{2}
  \right).
\end{align}
Analogously, one finds for $\uone$:
\begin{align}
  \label{eq:thermal-mass-U1}
  m_1^2 & = \frac{g_1^2 n_1 T^2}{6} 
\end{align}
which determines the thermal bino mass. In the $\uone$ case,
$m^2_{1,\,\text{gauge bosons}} = 0$ and
$m^2_{1,\,\text{gauginos}} = 0$ since the diagrams of
Fig.~\ref{fig:HTL-gauge-bososn-contr} are absent.

In covariant gauge, the HTL-resummed gauge
boson propagator has the form~\cite{Pisarski:1989cs,Kalashnikov:1979cy}
\begin{align}
   \label{eq:HTL-gaugeboson-prop}
  i\D^{(\a)}_{\mu\nu}(K) = i\left( 
  A_{\mu\nu} \D^{(\a)}_{\text{T}} + B_{\mu\nu} \D^{(\a)}_{\text{L}} + C_{\mu\nu} \xi  \right)\;,
\end{align}
with the tensorial quantities
\begin{subequations}
     \label{eq:tensors}
  \begin{align}
    A_{\mu\nu} & = -g_{\mu\nu} -\frac{1}{k^2}\left[ K^2 v_\mu v_\nu -
      K\cdot v (K_\mu v_\nu + K_\nu v_\mu) + K_\mu K_\nu
    \right]\;,  \\
    B_{\mu\nu} & = v_\mu v_\nu - \frac{K\cdot v}{K^2}(K_\mu v_\nu +
    K_\nu v_\mu) + \left(
      \frac{K\cdot v}{K^2}\right)^2 K_\mu K_\nu\;,  \\
    C_{\mu\nu} & = \frac{K_\mu K_\nu}{\left(K^2\right)^2}\;. \label{eq:gauge-dep-part-C}
  \end{align}
\end{subequations}
The velocity of the thermal bath is denoted by $v$ and the gauge
fixing parameter by $\xi$. 
The transverse and longitudinal propagators are
\begin{subequations}
    \label{eq:transv-and-long-propagators}
  \begin{align}
    \D^{(\a)}_{\text{T}}(k_0,k) & =  \frac{1}{k_0^2 - k^2 - \Pi^{(\a)}_{\text{T}}(k_0,k)}\;, \\
    \D^{(\a)}_{\text{L}}(k_0,k) & = \frac{1}{k^2 - \Pi^{(\a)}_{\text{L}}(k_0,k)}\;,
  \end{align}
\end{subequations}
which have spectral representations~\cite{Pisarski:1989cs}
\begin{align}
  \label{eq:spectral-repr}
    \D^{(\a)}_{\text{T/L}}(k_0,k)  =  \int_{-\infty}^\infty d\o
   {1\over k_0-\o}  \rho^{(\a)}_{\text{L/T}}(\o,k) \; .
\end{align}
For $|\o|<k$, the spectral densities $\rho^{(\a)}_{L/T}$  are given by
\begin{subequations}
    \label{eq:gauge-boson-spectra}
  \begin{align}
    \rho^{(\a)}_{\text{T}}(\o,k) & =
    \frac{3}{4m_\a^{2}}\frac{x}{(1-x^2)\left[A_{\text{T}}(x)^2+(z+B_{\text{T}}(x))^2\right]}\;,
\\
    \rho^{(\a)}_{\text{L}}(\o,k) & = \frac{3}{4m_\a^{2}} \frac{2
      x}{A_{\text{L}}(x)^2+(z+B_{\text{L}}(x))^2}\; ,
  \end{align}
\end{subequations}
with $x = \o/k$ and $z = k^2/m_\a^2$ and
\begin{subequations}
  \begin{align}
    A_{\text{T}}(x) &= \frac{3}{4}\pi x\;, & B_{\text{T}}(x) &= \frac{3}{4}\left(
      2\frac{x^2}{1-x^2} + x
      \ln{\frac{1+x}{1-x}}\right)\;,\\
    A_{\text{L}}(x) &= \frac{3}{2}\pi x\;, & B_{\text{L}}(x) &= \frac{3}{2}\left( 2 - x
      \ln{\frac{1+x}{1-x}}\right)\;.
  \end{align}
\end{subequations}

Let us now turn to the calculation of the soft part of the gravitino
production rate.  In the previous section we have seen that the
squared matrix elements for the $2\rightarrow 2 $ scatterings given in
Table~\ref{tab:squared-matrix-elements} contain the factor
\begin{align}
  \label{eq:generic-prefactor}
  |\mathcal{M}_i|^2 \propto \left( 1 +
    \frac{M_\a^2}{3 \mgrav^2} \right)\; .
\end{align}
The first term results from the helicity $\pm 3/2$ states of the
gravitino, the second term from the helicity $\pm 1/2$ states of the
gravitino which represent the goldstino components.  For the gravitino
self-energy~$\S_{\gr}$, it has been shown up to two loop order in the
gauge couplings that one obtains the same
factor~(\ref{eq:generic-prefactor}); cf.~\cite{Bolz:2000fu}. Indeed,
we employ the effective theory for light gravitinos given
in~(\ref{eq:goldstino-lagrangian}) to calculate the production rate
for the helicity $\pm 1/2$ components of the gravitino in terms of the
imaginary part of the goldstino self-energy~$ \S^{(\a)}_\psi(P) $,
namely,
\begin{align}
  \label{eq:soft-production-rate}
  E \, \frac{d\G^{(\a)}_{\psi}}{d^3 p} \left.\vphantom{
      \frac{d\G_{\psi}}{d^3 p} }\right|_{\text{soft}} & = - \frac{1
  }{(2 \p)^3} \ffermi{E}\,
  \text{Im}\,\S^{(\a)}_\psi(E+i\ve,\mathbf{p}) |_{k<k^*}\; .
\end{align}
This leads to the full rate by replacing $M_\a^2/3\mgrav^2$ with
the prefactor (\ref{eq:generic-prefactor}). Note that we have
introduced the cutoff $k^*$  since we
consider here only soft three-momentum transfers.

\begin{figure}[tb]
  \centering
   \includegraphics[width=0.6\textwidth]{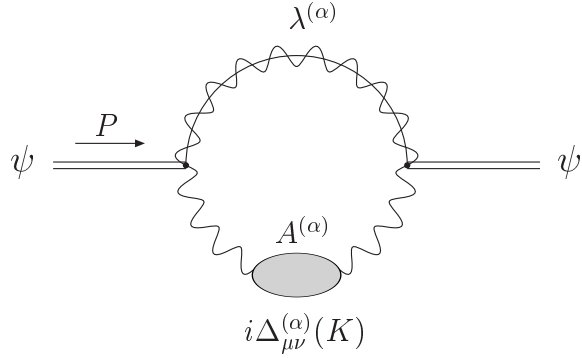}
 \caption[Goldstino self-energy]{The leading contribution to the imaginary
   part of the goldstino self-energy in the effective theory of light
   gravitinos. The blob indicates the HTL-resummed gauge boson
   propagator~(\ref{eq:HTL-gaugeboson-prop}).}
\label{fig:gravitino-selfenergy}
\end{figure}

The leading contribution to the self-energy is given by the gauge
boson--gaugino loop shown in Fig.~\ref{fig:gravitino-selfenergy};
see the discussion below~(\ref{eq:goldstino-lagrangian}). Here the
HTL-resummed gauge boson propagator~(\ref{eq:HTL-gaugeboson-prop}) is
indicated by the blob. Accordingly, one finds for the leading
contribution to the self-energy\footnote{We commit a certain abuse of
  terminology since a self-energy contribution is usually referred to
  as the expression which one obtains from the amputated diagram of
  Fig.~\ref{fig:gravitino-selfenergy}. The inclusion of the spinors
  $u(P)$ and $\overline{u}(P)$ is necessary to
  interpret~(\ref{eq:soft-production-rate}) in terms of a probability
  (see \cite{Weldon:1983jn}).}
\begin{align}
\label{eq:goldstino-self-energy-v1}
 -i\,\S^{(\a)}_\psi(P) & =  \frac{ X^{(\a)} M_\a^2}{24 \mgrav^2 \mplanck^2}
  \sum_{s=\pm 1/2} \int \frac{d^4 K}{(2\p)^4}\, \mathrm{tr}\, \left\{
    \overline{u}^s(P) \left[ \slashed K , \g^\n \right] \frac{i
      \slashed Q}{Q^2} i\, \D^{(\a)}_{\m\n} \left[ \g^\m , \slashed K \right]
    u^s(P) \right\}\; .
\end{align}
Working in the high-energy limit, we have neglected the masses in the
gaugino propagators. The polarization sum for the goldstino spinors
$u^s(P)$ and $\overline{u}^s(P)$ is included. The gauge bosons carry
soft momenta $K$ and the gaugino momenta are given by $Q = P-K$. The
multiplicity factors $X^{(\a)}$ count the number of gauge
bosons/gauginos in the loop for the corresponding gauge group
$\mathcal{G}_\a$, namely,
\begin{subequations}
  \label{eq:multiplicities-soft-rate}
  \begin{alignat}{2}
    X^{(\a)} & = N_\a^2 -1  &&\quad  \text{for} \quad \text{SU}(N_\a)    \; , \\ 
    X^{(1)} & = 1 &&\quad \text{for} \quad \uone \;.
  \end{alignat}
\end{subequations}
One computes~(\ref{eq:goldstino-self-energy-v1}) in the imaginary time
formalism where the energies of the particles are given by their
Matsubara frequencies. For the gauge bosons, the frequencies are even,
$k_0 = 2\p i n T$, and the integral over $k_0$ turns into a sum over
discrete energies,~i.e.,
\begin{align}
  \int \frac{d k_0 }{2\p} &\rightarrow i\,T\,\sum_{n=-\infty}^\infty\; .
\end{align}
After performing the polarization sum,
the self-energy reads
\begin{align}
  \label{eq:goldstino-self-energy-v2}
  \S^{(\a)}_\psi(P) & = \frac{4}{3}\frac{X^{(\a)} m_\a^2 T}{ \mgrav^2
    \mplanck^2} \sum_{k_0} \int \frac{d^3k}{(2\pi)^3}  \frac{1}{Q^2} \left(D_{\text{L}} \D_{\text{L}} +
    D_{\text{T}} \D_{\text{T}}\right)\;,
\end{align}
with the Dirac traces:
\begin{subequations}
  \begin{align}
    \label{eq:traces}
    D_{\text{T}} & = \frac{1}{32}\mbox{tr}\left\{\slashed P \, [\slashed
      K,\g^\nu]\
      \slashed Q \, [\slashed K,\g^\mu] A_{\mu\nu}\right\}\;,\\
    D_{\text{L}} & = \frac{1}{32}\mbox{tr}\left\{\slashed P \, [\slashed
      K,\g^\nu]\ \slashed Q \, [\slashed K,\g^\mu]
      B_{\mu\nu}\right\}\;.
  \end{align}
\end{subequations}
Note that the dependence on the gauge fixing parameter $\xi$~ drops
out since~(\ref{eq:gauge-dep-part-C}) contracted with the gauge boson
momentum $K$ vanishes in the Dirac trace.

Using the spectral representations of the
propagators~(\ref{eq:spectral-repr}), the summation over the Matsubara
frequencies can be  performed conveniently with the Saclay
method~\cite{Pisarski:1987wc}. The result reads after analytic
continuation from discrete energy values $p_0$ to continuous real
goldstino energies~$E$~\cite{Bolz:2000fu}
\begin{align}
   \frac{d\G^{(\a)}_{\psi}}{d^3 p} \left.\vphantom{
      \frac{d\G_{\psi}}{d^3 p} }\right|_{\text{soft}} 
&= \frac{ X^{(\a)} m_{\a}^2\,T}{48\pi^4 \mplanck^2\mgrav^2} \,\ffermi{E}\int_0^{k^*}\! dk k^3 
\int_{-k}^k {d\o\over \o} \nonumber\\&\times
\left[\rho_{\text{L}}(\o,k)\left(1-{\o^2\over k^2}\right) 
+ \rho_{\text{T}}(\o,k)\left(1-{\o^2\over k^2}\right)^2 \right].
\end{align}
The structure of the integrand is identical to the one obtained for
the axion production rate~\cite{Braaten:1991dd} where the analytic
dependence on the cutoff $k^*$ is extracted. Thus, after performing
the integrations, one finds
\begin{align}
 \frac{d\G^{(\a)}_{\psi}}{d^3 p} \left.\vphantom{
      \frac{d\G_{\psi}}{d^3 p} }\right|_{\text{soft}}
  & =\ffermi{E} \,\frac{  X^{(\a)}  m_\a^2 M_\a^2\, T}{32\pi^4 \mplanck^2 \mgrav^2}
  \left[\ln\left(\frac{k^{*2}}{m_\a^2}\right)-1.379\right].
\end{align}
The replacement of $M_\a^2/3\mgrav^2$ with the
factor~(\ref{eq:generic-prefactor}) for the full theory yields the
final result for the soft part of the gravitino production
rate~(\ref{eq:soft-production-rate-gravitino})
\begin{align}
  \label{eq:soft-production-rate-FINAL}
 \gammasoft
  & =  \ffermi{E} \sum_{\a=1}^3   \left( 1 +
    \frac{M_\a^2}{3 \mgrav^2} \right)\,\frac{3  X^{(\a)}  m_\a^2\, T}{32\pi^4 \mplanck^2 }
  \left[\ln\left(\frac{k^{*2}}{m_\a^2}\right)-1.379\right]
\end{align}
with the multiplicites $ X^{(\a)}$ given
in~(\ref{eq:multiplicities-soft-rate}) and the thermal
 masses $m_\a$ given in~(\ref{eq:thermal-masses-nonabelian})
and~(\ref{eq:thermal-mass-U1}).

Adding the results for the hard and the soft part of the gravitino
production rate, one finds that the logarithmic dependence on $k^*$
cancels out. We will do so by calculating the Boltzmann collision term
which is the crucial quantity for all further calculations.

\section{The Boltzmann Collision Term}
\label{sec:boltzmann-coll-term}

The quantity we are interested in is the gravitino number
density\footnote{The factor of four accounts for
  the internal degrees of freedom of the massive gravitino.}
\begin{align}
  n_{\gr}= 4 \int \frac{d^3p}{(2\p)^3}\,f_{\gr}(E,t)\;. 
\end{align}
Its evolution with cosmic time $t$ is governed by the
Boltzmann equation
\begin{align}
\label{eq:boltzmann-equation}
  \frac{dn_{\gr}}{dt} + 3 H n_{\gr} &= C_{\gr}\; .
\end{align}
The second term on the left-hand side accounts for the dilution of
gravitinos due to the expansion of the Universe, which is described by
the Hubble parameter~$H$.  For negligible gravitino disappearance
processes, the collision term $C_{\gr}$ on the right-hand side of the
Boltzmann equation describes gravitino production processes. It is
obtained by integrating the thermal gravitino production rate
\begin{align}
  C_{\gr} & = \int \frac{d^3p}{E} \;\left[ E \frac{d\G_{\gr}}{d^3 p}\right] =  \int{d^3
    p} \left[  \gammahard + \gammasoft   \right] .
\end{align}

Consider first the soft
part~(\ref{eq:soft-production-rate-FINAL}). The corresponding 
collision term reads
\begin{align}
  C^{(\alpha)}_{\gr ,\,\text{soft}} &= \left( 1 + \frac{M_\a^2}{3
      \mgrav^2} \right)\,\frac{3 X^{(\a)} m_\a^2\, T}{8\pi^3
    \mplanck^2 }
  \left[\ln\left(\frac{k^{*2}}{m_\a^2}\right)-1.379\right] 
  \int_{0}^\infty dE\,\frac{E^2}{e^{E/T}+1} \; ,
\end{align}
The final integration over the Fermi-Dirac distribution function is
easily performed. The result reads
\begin{align}
  \label{eq:collision-soft-FINAL}
  C_{\gr ,\,\text{soft}} &= \sum_{\a = 1}^3 \left( 1 + \frac{M_\a^2}{3
      \mgrav^2} \right)\,\frac{9 X^{(\a)} \zeta(3)  m_\a^2\, T^4}{16 \pi^3
    \mplanck^2 }
  \left[\ln\left(\frac{k^{*2}}{m_\a^2}\right)-1.379\right] \; .
\end{align}

Now we turn to the hard part~(\ref{eq:prodrate-BFB-FFF}). The
production rates for the BFB, FFF, and BBF processes are given
in~(\ref{eq:hard-prod-rate-BFB-FFF-BBF-with-coefficients}). The
numerical integrations are performed using
\texttt{VEGAS}~\cite{Hahn:2004fe}. We get:
\begin{align}
  C^{\BFB}_{\gr ,\,\text{hard}} &= 
  \sum_{\alpha = 1}^3
  \frac{g_\a^2\,T^6}{\mplanck^2} \left( 1+\frac{M^2_{ \alpha
      }}{3\mgrav^2}\right)
  \nonumber \\
  &\times \Bigg\{\, c_{\BFB,\,1 }^{(\alpha)} \left[
    \frac{\zeta(3)}{32 \p^3} \left( \ln{\left( \frac{2 T }{ k^* }
        \right)} + 0.9930 \right) - 11.1362\times 10^{-4} \right]
  \nonumber \\
  & \quad + c_{\BFB,\,2 }^{(\alpha)}
  \left[\vphantom{\frac{\zeta(3)T^6}{32 \p^3} } - 1.3284\times 10^{-4}
  \,\right]\Bigg\}
\end{align}
\begin{align}
  C^{\FFF}_{\gr ,\,\text{hard}} &= 
  \sum_{\alpha = 1}^3
  \frac{g_\a^2\,T^6}{\mplanck^2} \left( 1+\frac{M^2_{ \alpha
      }}{3\mgrav^2}\right)
  \nonumber \\
  &\times \Bigg\{\, c_{\FFF,\,1 }^{(\alpha)} \left[
    \frac{\zeta(3)}{64 \p^3} \left( \ln{\left( \frac{2 T }{ k^* }
        \right)} + 1.6862 \right) - 6.9992\times 10^{-4} \right]
  \nonumber \\
  & \quad + c_{\FFF,\,3 }^{(\alpha)}
  \left[\vphantom{\frac{\zeta(3)T^6}{32 \p^3} } \, 0.5039\times 10^{-4}
  \,\right]\Bigg\}
\end{align}
\begin{align}
  C^{\BBF}_{\gr ,\,\text{hard}} &= 
  \sum_{\alpha = 1}^3
  \frac{g_\a^2\,T^6}{\mplanck^2} \left( 1+\frac{M^2_{ \alpha
      }}{3\mgrav^2}\right)
  \nonumber \\ &\times
   \Bigg\{\, 
    c_{\BBF,\,2 }^{(\alpha)} \left[\vphantom{\frac{\zeta(3)T^6}{32
          \p^3} } - 1.2975\times 10^{-4}\,\right] +  
    c_{\BBF,\,3 }^{(\alpha)} \left[\vphantom{\frac{\zeta(3)T^6}{32
          \p^3} }\,  0.8647\times 10^{-4}\,\right]  \Bigg\}
\end{align}

In the sum of the soft and hard parts, the logarithmic dependence on
$k^*$ cancels out as anticipated. Plugging in the squared thermal
masses~(\ref{eq:thermal-masses-nonabelian})
and~(\ref{eq:thermal-mass-U1}), the final result can be brought into
the following form
\begin{align}
  \label{eq:COLLISION-TERM-FINAL}
  C_{\gr} &= \sum_{\alpha = 1}^3 \left( 1+\frac{M^2_{ \alpha
      }}{3\mgrav^2}\right) \frac{3\,\zeta(3) T^6 }{16 \pi^3
    \mplanck^2}\, c_\a \, g_\a^2  \ln{\left( \frac{ k_\a}{g_\a}\right) }
\end{align}
This is one of the main results of this thesis. The coefficients are
given by $c_\a= (\,11\,,\,27\,,72\,)$ and the scales in the logarithms by
$k_\a = (\,1.266\,,\, 1.312\,,\, 1.271\,)$. These numbers are
associated with the gauge groups $\uone$, $\sutwo$, and $\suthree$,
respectively. The temperature $T$ provides the scale for the
evaluation of the gaugino mass parameters $M_\a = (\,M_1\,,\, M_2\,,\,
M_3\,)$ and  the gauge couplings $g_\a = (\,g'\,,\, g\,,\,
g_{\text{s}}\,)$.

The error in the hard production rate of Ref.~\cite{Bolz:2000fu}
manifests itself in a larger coefficient $k_3=1.271$ for the
$\suthree$ contribution than $1.163$ obtained from~\cite{Bolz:2000fu}.

Recall that the Braaten--Yuan prescription~\cite{Braaten:1991dd} relies
on the weak coupling limit~$g\ll 1$.  Because of the large value of
the strong coupling constant,~e.g., $g_{\text{s}}(100\,\GEV) \simeq
2.5$, the results for the production rates and the corresponding
collision term require high temperatures~$T\gg 10^6\,\GEV$, where, for
example, \mbox{$g_{\text{s}}(10^6\,\GEV) \simeq 0.99$} and
\mbox{$g_{\text{s}}(10^9\,\GEV) \simeq 0.88$}.

Since we neglect gravitino disappearance processes, the collision
term~(\ref{eq:COLLISION-TERM-FINAL}) acts as pure source term in the
Boltzmann equation~(\ref{eq:boltzmann-equation}).  For small
temperatures, the logarithm turns negative for the $\suthree$ part.
Thus, the result becomes unphysical and shall not be trusted when
extrapolating to small temperatures.


\cleardoublepage

\chapter{Gravitino Cosmology}
\label{cha:gravitino-cosmology}

The new result for the Boltzmann collision
term~(\ref{eq:COLLISION-TERM-FINAL}) has important implications for
gravitino cosmology. In this chapter, we compute the gravitino yield
which describes the primordial gravitino abundance. This quantity is
crucial for phenomenological considerations of stable and unstable
gravitinos. For gravitino dark matter scenarios, we calculate the
relic gravitino density from thermal production. This consequently
yields an upper bound on the reheating temperature of the Universe.

\section{Gravitino Yield from Thermal Production}
\label{sec:the early universe}

The observed isotropy and homogeneity of the Universe on large scales
allows us to express the overall geometry of the Universe in terms of
the Robertson-Walker metric with the line element
\begin{align}
   ds^2=dt^2-R(t)^2\left[{dr^2\over1-k\,r^2}+r^2\left(d\q^2
        +\sin^2\q d\f^2\right)\right]\;,
\end{align}
where $(t,r,\q,\f)$ are comoving coordinates. By a rescaling of $r$,
the curvature parameter $k$ can be assigned the discrete values $k=
1,\,-1,\, \text{or } 0 $, corresponding to spatially closed, open, or
flat geometries.  The evolution of the scale factor $R$ is described
by the Friedmann equation
\begin{align}
\label{eq:friedmann-eqn}
  H^2 \equiv \left(\frac{\dot{R}}{R}\right)^2 = {8\p G_{\text{N}}
    \over 3}\, \r - \frac{k}{R^2} \; ,
\end{align}
which defines the  Hubble parameter $H$. The total energy density
of the Universe is denoted by $\rho$. The derivative of the scale
factor~$R$ with respect to cosmic time~$t$ is written as~$\dot{R}$.

In the radiation dominated epoch of the Universe, $\rho$ is given in
good approximation by
\begin{align}
\label{eq:rho-rad-dom}
  \r& =g_*{\p^2\over30}T^4\;,
\end{align}
where $T$ is the photon temperature and $g_*$ denotes the effectively
massless degrees of freedom, i.e., those species with mass $m_i \ll T_i$,
\begin{align}
  g_*& =\sum_{i=\text{bosons}} g_{i} \left(\frac{T_i}{T} \right)^4
  +{7\over8}\sum_{i=\text{fermions}} g_{\text{i}}\left(\frac{T_i}{T}
  \right)^4\; .
\end{align}

The Hubble rate in the radiation dominated epoch is given by ($k=0$)
\begin{align}
\label{eq:Hubble-rate in rad-dom}
  H(T)&= \sqrt{\frac{g_* \p^2}{90}} \frac{T^2}{\mplanck}\;,
\end{align}
where $\mplanck$ is the reduced Planck mass~(\ref{eq:reduced-planck-mass}). During
this epoch, time and temperature are related via  $H(T)=1/(2t)$.

The entropy density of the Universe, defined as $s\equiv (\r + p)
/T $, is dominated by the contribution of relativistic
particles for which $p=\r/3$ holds. Hence, one finds
\begin{align}
  \label{eq:entropy density in rad-dom}
  s &= g_{*S} \frac{4 \p^2}{90} T^3\; ,
\end{align}
where 
\begin{align}
  \label{eq:g-star-s}
   g_{*S} = \sum_{i=\text{bosons}} g_{i} \left(\frac{T_i}{T} \right)^3
  +{7\over8}\sum_{i=\text{fermions}} g_i\left(\frac{T_i}{T}
  \right)^3 \; .
\end{align}

In the previous section, the Boltzmann
equation~(\ref{eq:boltzmann-equation}) has been written in terms of the
gravitino number density~$n_{\gr}$. 
It is useful to scale out
the expansion by dividing the gravitino number density $n_{\gr}$ by the
entropy density $s$. This defines the yield variable:
\begin{align}
  \label{eq:def-yield}
  \yield \equiv \frac{n_{\gr}}{s}\; .
\end{align}
With the conservation of entropy per comoving volume, $s R^3=
\text{const.}$, the Boltzmann equation~(\ref{eq:boltzmann-equation})
can be rewritten as
\begin{align}
  \label{eq:boltzmann eqn in terms of the yield}
  \frac{d \yield}{dt} &= \frac{\coll}{s}\; .
\end{align}
Using $ dt= - dT\big/[H(T)T]$, the
gravitino yield is obtained by integrating
\begin{align}
\label{eq:differential gravitino yield}
  d\yield = - \frac{\coll (T) dT }{s(T) H(T) T} \; .
\end{align}

We consider thermal gravitino production beginning after completion of
the reheating phase where the temperature of the primordial plasma is
the reheating temperature~$\TR$.  We assume that any initial gravitino
population has been diluted away by inflation,~i.e., $\yield (\TR)=0$.
Hence, the gravitino yield at the temperature $T'$ is
\begin{align}
  \label{eq:yield before integration}
 \yield (T') = - \int_{T_R}^{T'} dT
  \frac{\coll (T)}{s(T) H(T) T}\; .
\end{align}

Unstable gravitinos have typically long lifetimes because their
interactions are suppressed by $\mplanck$. In particular, when the
gravitino is lighter than $\apprle 20\, \text{TeV}$, its lifetime
becomes longer than $\apprge 1\, \text{s}$~\cite{Kohri:2005wn}. Hence,
unstable gravitinos may decay during and/or after big-bang
nucleosynthesis where $t_{\text{\tiny BBN }}\simeq 1\,\text{s}$
and $T_{\text{\tiny BBN }}\simeq 1\,\text{MeV}$. Thus, the yield of
gravitinos from thermal production prior to their decay is
obtained for $T'=T_{\text{\tiny BBN }}$ ($\mgrav \apprle
20\,\text{TeV}$). 
Note that the $T^6$ dependence of the collision
term~(\ref{eq:COLLISION-TERM-FINAL}) cancels out in the integrand
of~(\ref{eq:yield before integration}). Furthermore, recall that we
consider scenarios in which $\TR \apprge 10^6\,\GEV \gg T_{\text{\tiny BBN }} $ (see
Section~\ref{sec:boltzmann-coll-term}). Thus, with the collision
term~(\ref{eq:COLLISION-TERM-FINAL}), we can solve the Boltzmann
equation to good approximation analytically.  We find
\begin{align}
  \label{eq:yield-FINAL}
 \yield( T_{\text{\tiny BBN }} )  \simeq \sum_{\a=1}^3
  \left( 1+\frac{M^2_{ \alpha }(\TR)}{3\mgrav^2}\right) y_\a\, g_{\a}\,(\TR)^2 
 \ln{\left( \frac{ k_\a}{g_\a(\TR)}\right) }\left(
   \frac{\TR}{10^{10}\,\GEV} \right)
\end{align}
with $y_\a = (\, 0.653 \, , \, 1.604 \, ,\, 4.276\, )\times 10^{-12}$
for $\uone$, $\sutwo$, and $\suthree$, respectively. The scales $k_\a$
in the logarithms are given at the end of
Sec.~\ref{sec:boltzmann-coll-term}. Here, we have used that after
reheating, at temperature $\TR$, all particles of the MSSM are in
thermal equilibrium and relativistic, for which $g_*(\TR) =
g_{*S}(\TR) = 915/4 .$

The yield~(\ref{eq:yield-FINAL}) is the starting point for studies of
cosmological constraints in scenarios with unstable gravitinos. In the
remainder of this chapter we will consider the case of a stable
gravitino.

\section{Gravitino Dark Matter}
\label{sec:grav-dark-matt}

In the following, we focus on scenarios in which the gravitino is the
LSP and stable due to R-parity conservation.
Gravitinos, once produced, will thus contribute to the present value of
the energy density $\r$ since they do not decay. Because of the large redshift, the
gravitino energy density at the present time $t_0$ is  \mbox{$\r_{\gr}(t_0)
  = \mgrav\, n_{\gr}(t_0)$}. 
We have seen that the thermal production of
gravitinos is efficient only during the very early hot radiation
dominated epoch so that
\begin{align}
  \label{eq:relic-yield}
  \yield (T_0) \simeq \yield (T_{\text{\tiny BBN}})  
\end{align}
for stable gravitinos. Here, 
\begin{align}
   T_0 &= 2.725\, K = 2.348 \times 10^{-13}\, \GEV 
\end{align}
is the present temperature of the Universe~\cite{Yao:2006px}.

Thus, the present day density parameter of thermally
produced gravitinos is given by:
\begin{align}
  \label{eq:relic-density-PRE-final}
  \ogravtp &= \frac{\rho_{\gr}(t_0)}{\rhocrit (t_0)}h^2 =
  \frac{\mgrav\, \yield(T_0)\, s(T_0)\, h^2 }{\rhocrit(T_0)}\; .
\end{align}

Here, the dimensionless quantity $h$ is used to parameterize the
Hubble constant $H_0 = H(T_0)= 100\, h\,
\text{km}\,\text{s}^{-1}\,\text{Mpc}^{-1}$ and the present value of
the critical density reads\footnote{$h = 0.73^{+0.04}_{-0.03}$
  \cite{Yao:2006px}}~\cite{Yao:2006px}
\begin{subequations}
  \begin{align}
    \rhocrit(t_0) /h^2 &= 8.096 \times 10^{-47}\, \GEV^4\; .
  \end{align}
\end{subequations}
The entropy density $s(T_0)$ is obtained from (\ref{eq:entropy density
  in rad-dom}) with $g_{*S} (T_0) = 43/11$.

Thus, from~~(\ref{eq:yield-FINAL}), (\ref{eq:relic-yield}), and
(\ref{eq:relic-density-PRE-final}) we find the result for the relic
density of thermally produced gravitinos to leading order in the
Standard Model gauge couplings:
\begin{align}
  \label{eq:relic gravitino abundance TP}
       \ogravtp
        &=
        \sum_{\a=1}^{3}
        \left(1+\frac{M_\a(\TR)^2}{3\mgrav^2}\right)\omega_\a\, g_\a(\TR)^2 
        \ln\left(\frac{k_\a}{g_\a(\TR)}\right)
        \left(\frac{\mgrav}{100\,\GEV}\right)
        \left(\frac{\TR}{10^{10}\,\GEV}\right)\; 
\end{align}
with $\omega_\a=(\,0.018\,,\,0.044\,,\,0.117\,)$ for $\uone$,
$\sutwo$, and $\suthree$, respectively.

The relic gravitino density $\ogravtp$ is essentially linear in the
reheating temperature $\TR$. Recall that the production of the
helicity $\pm 3/2$ states of gravitinos is described by the first term
in the generic factor $(1 + M_\a^2 /3\mgrav^2 )$.  Thus, the relative
weights for the helicity $\pm 3/2$ production of gravitinos for the
different factors of the Standard Model gauge group are basically
given by $\o_\a\,g_\a(T_R)^2$---which is model-independent.  The
production of the helicity $\pm 1/2$ states depends on the ratio of
 squared gaugino masses $M_\a^2$ to gravitino mass $\mgrav$.  Hence, the
thermal production becomes more efficient for light gravitinos.

\begin{figure}[bt]
\centering
\subfloat[]{
    \label{fig:gauge running}
     \includegraphics[width=0.5\textwidth]{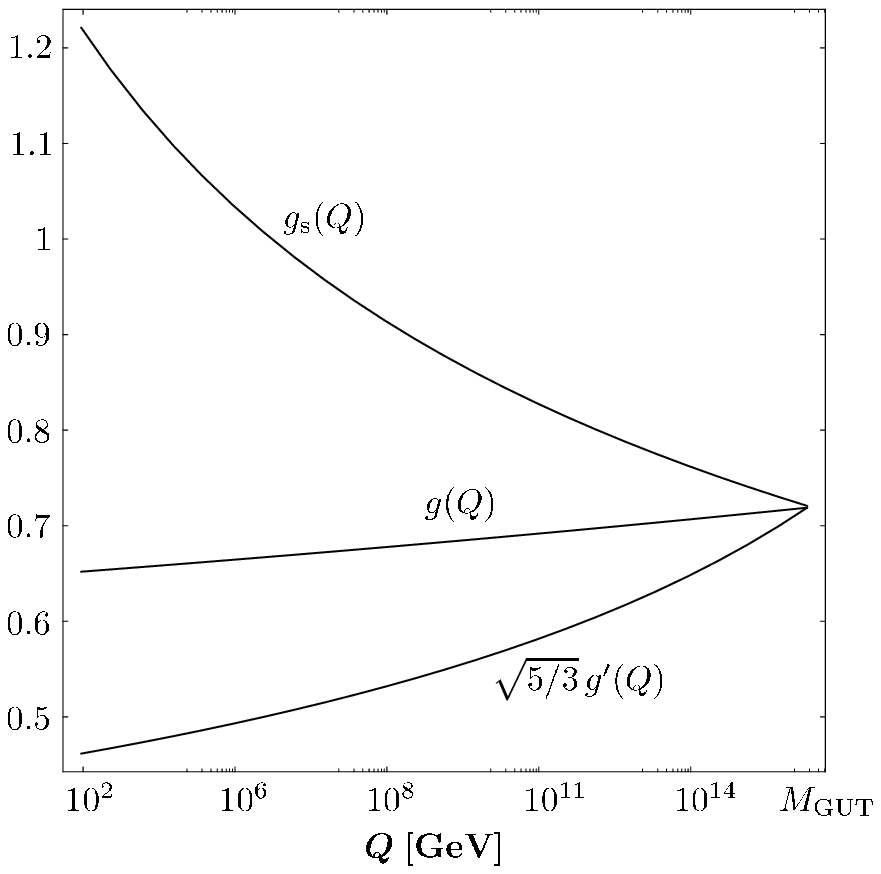}}%
   \subfloat[]{
    \label{fig:gaugino running}
    \includegraphics[width=0.5\textwidth]{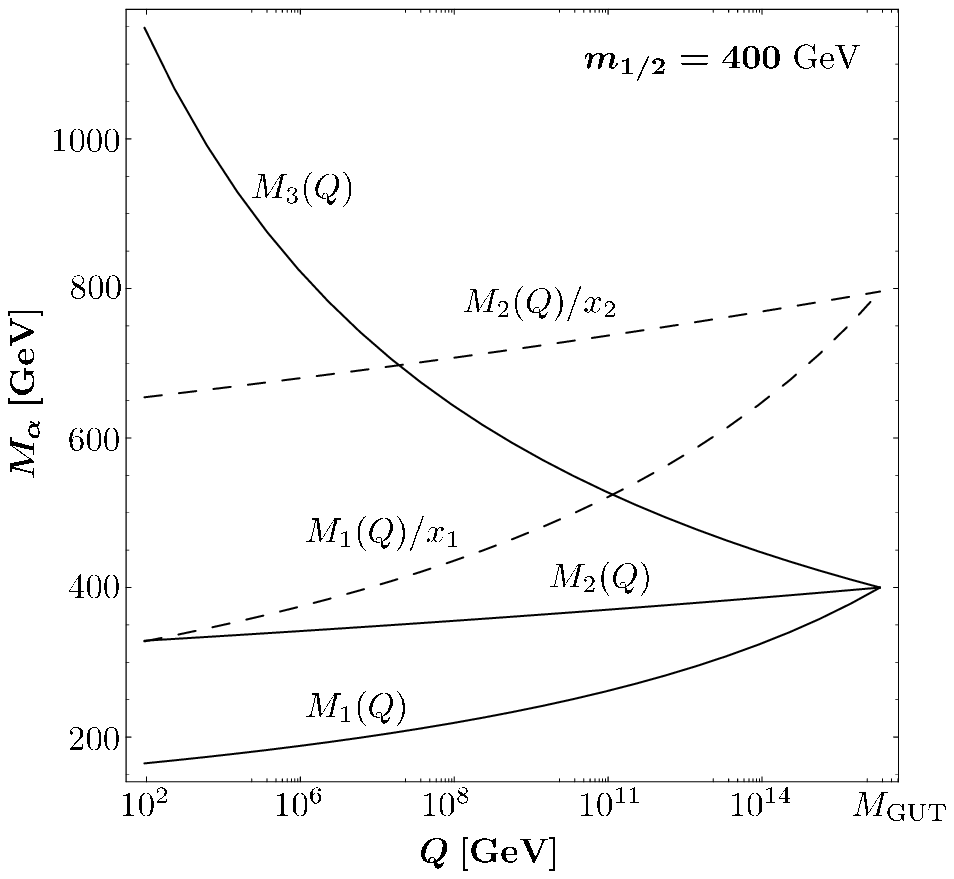}}%
  \caption[Renormalization Group running in the MSSM]{One-loop renormalization group
    running of the gauge couplings (a) and the gaugino mass
    parameters (b) in the MSSM. The point of gauge coupling
    unification is denoted as $\mgut$. The runnings of the gaugino
    masses are inferred from (\ref{eq:GUTrelation for gaugino masses})
    for universal $M_{1,2,3}=\monetwo$ at $\mgut$ with a value of
    $\monetwo = 400\,\GEV$. They are given in (b) as solid
    lines. Dashed lines in (b) show a non-universal scenario with $0.5\,
    M_{1,2}=M_{3}=\monetwo$ at $\mgut$,~i.e., $x_{1,2}=2$. }
\label{fig:RG-runnings}
\end{figure}

In order to calculate numbers from~(\ref{eq:relic gravitino abundance
  TP}), we need to evaluate the gauge couplings $g_\a$ and the gaugino
mass parameters $M_\a$ at the scale provided by the reheating
temperature $\TR$. At the one-loop level, the renormalization group
(RG) equations for the gauge couplings read
\begin{align}
  \label{eq:RGE gauge couplings}
  \frac{d\,g_{\a}(Q)}{d\ln{Q/Q_0}}  &= \frac{\beta_\a^{(1)}}{16\p^2}\, g_{\a}(Q)\,
\end{align}
where $Q$ is the scale of evaluation and $Q_0$ is some input
scale. One can solve (\ref{eq:RGE gauge couplings}) analytically:
\begin{align}
  g_{\a}(\TR)& = \left[ g_\a(m_{\text{Z}})^{-2} - \frac{\beta_\a^{(1)}
    }{ 8\p^2} \ln{ \left(\frac{\TR}{m_{\text{Z}}} \right)
    }\right]^{-1/2}
\end{align}
with $Q_0 = m_{\text{Z}} \simeq 91.19\,\GEV$ \cite{Yao:2006px} and
$Q=\TR$.  The beta-function coefficients $\beta_\a^{(1)} = (\,11\,,\,
1\,,\,-3\,)$ correspond to $g_{\a} =(\,g'\,,\, g\,,\,g_{\text{s}}\,)$
in the MSSM, respectively. In Fig.~\ref{fig:gauge running} the point
of gauge coupling unification is referred to as the grand unification
(GUT) scale \mbox{$\mgut \simeq 2\times 10^{16}\,\GEV$}. It is defined
as the point where the GUT-normalized hypercharge coupling,
\begin{align}
  \label{eq:GUT normalized uone coupling}
  \hat{g}_1 = \sqrt{5/3}\, g'\;,
\end{align}
 the weak coupling, $g$, and the strong coupling, $g_{\text{s}}$, meet. 

The one-loop RG equations for the gaugino masses $M_\a$ are given by
the expression analogous to~(\ref{eq:RGE gauge couplings}) and the
same coefficients $\beta_\a^{(1)}$ as above. Often, universal boundary
conditions are considered for the RG equations in which the gaugino
masses unify at $\mgut$. Since our result for the relic gravitino
abundance depends on $M_\a$, we will also consider scenarios in which
the gaugino masses do not unify at $\mgut$. We can parameterize this
by writing
\begin{align}
  \label{eq:GUTrelation for gaugino masses}
  \frac{1}{x_1}\, \frac{M_1(Q)}{ \hat{g}_1(Q)^2 }& = \frac{1}{x_2}\,
  \frac{M_2(Q)}{ g_2(Q)^2 } = \frac{M_3(Q)}{ g_3(Q)^2 }
\end{align}
which holds at any scale $Q$.\footnote{Up to small two loop effects
  and possible (unknown) threshold effects close to $\mgut$
  \cite{Martin:1997ns}.} For the gaugino masses,
we choose the input scale $Q_0$ to be $\mgut$. This defines the gaugino mass
parameter $\monetwo$, namely, 
\begin{align}
  \label{eq:definition of m12}
  \monetwo \equiv M_3(\mgut) = M_{2}(\mgut)/x_{2} =M_{1}(\mgut)/x_{1} \; ,
\end{align}
so that for a unifying scenario,~i.e., for $x_{1,2}=1$, we have
$\monetwo = M_1(\mgut) = M_2(\mgut) = M_3(\mgut) $. As illustrated for
$\monetwo=400\,\GEV$, fixing
$\monetwo$ determines all gaugino mass parameters $M_\a$ .

\begin{figure}[bt]
\centering
\subfloat[]{
    \label{fig:M12 is 400}
     \includegraphics[width=0.8\textwidth]{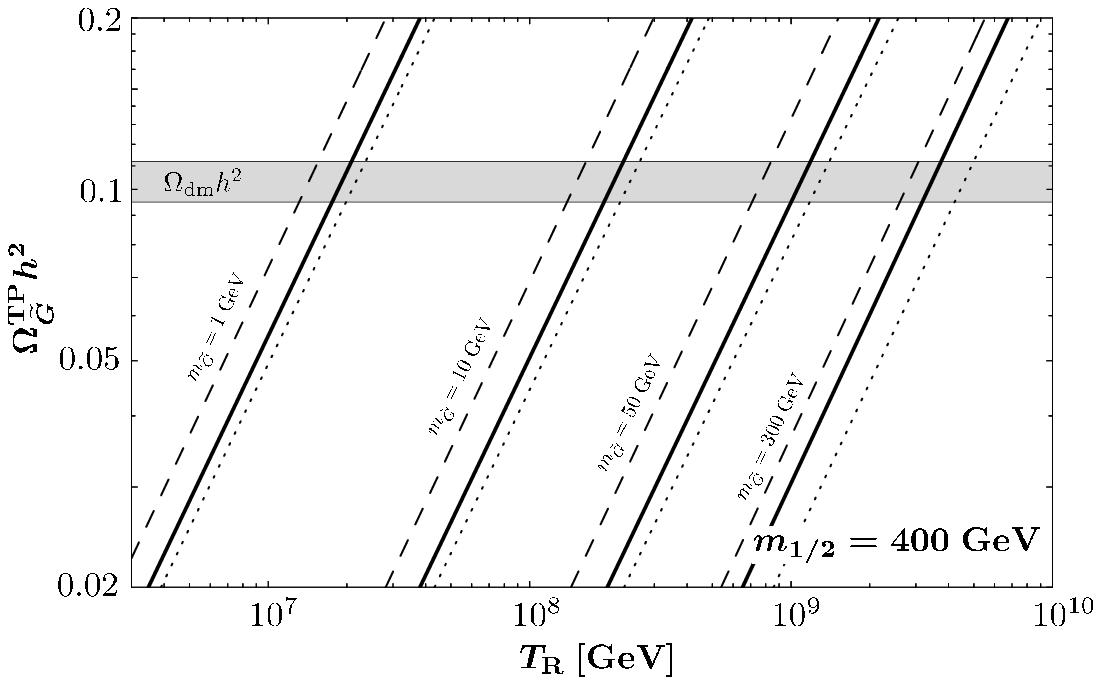}}
\\[20pt]
   \subfloat[]{
    \label{fig:M12 is 1500}
    \includegraphics[width=0.8\textwidth]{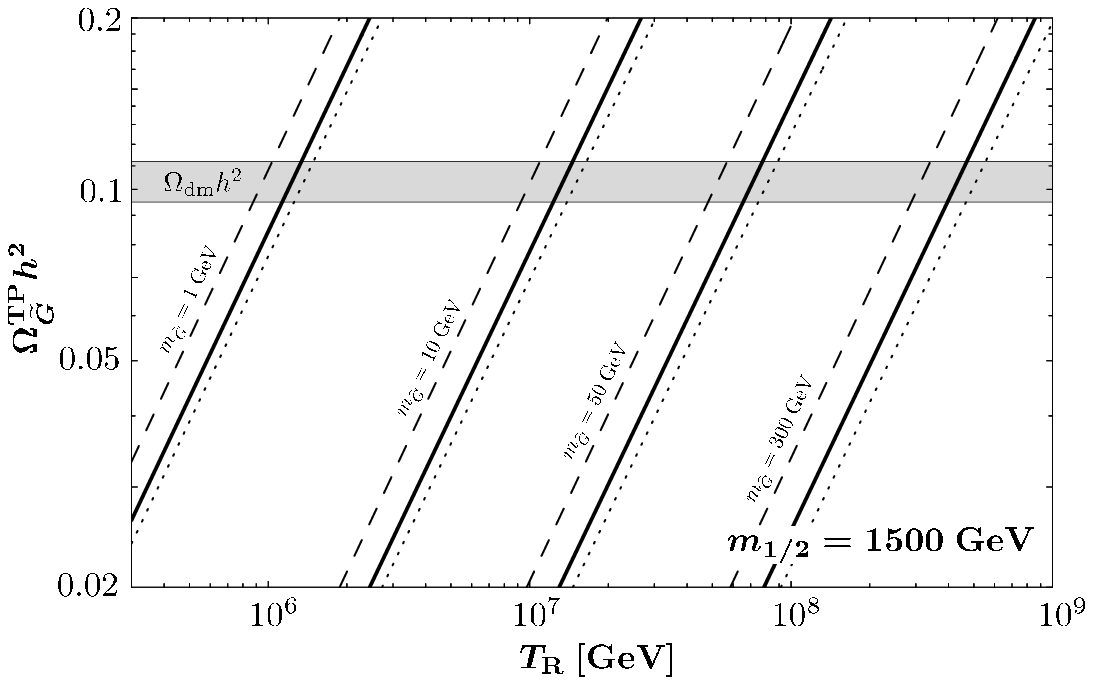}}%
\\[10pt]
  \caption[Relic gravitino abundance]{The relic gravitino density from
    thermal production, $\ogravtp$, as a
    function of $\TR$.  The solid and dashed curves show the
    $\smgroup$ results for universal ($M_{1,2,3}=m_{1/2}$) and
    non-universal ($0.5\, M_{1,2}=M_3=m_{1/2}$) gaugino masses at
    $\mgut$, respectively.  The dotted curves show the new result of
    the $\suthree$ contribution for $M_3=m_{1/2}$ at
    $\mgut$. The grey band indicates the dark matter density~$\omegadm$.}
\label{fig:Relic-Abundance-TP}
\end{figure}

Taking the RG evolution into account, we compute the relic gravitino
density. Figure~\ref{fig:Relic-Abundance-TP} shows $\ogravtp$ as a
function of the reheating temperature $\TR$ for gravitino masses
$\mgrav = 1,\,10,\,50,\, \text{and}\, 300\,\GEV$. We consider two
representative values of the gaugino mass parameter $\monetwo=400\,
\GEV\,\text{and}\,1500\, \GEV$ in Figs.~\ref{fig:M12 is 400}\, and
\ref{fig:M12 is 1500}, respectively. The solid curves represent the
gravitino density for universal gaugino masses $M_{1,2,3}=m_{1/2}$ at
$\mgut$ while the dashed curves show a non-universal scenario where
$0.5\, M_{1,2}=M_3=m_{1/2}$ at $\mgut$, i.e., $x_{1,2}=2$.  The
corrected result for the $\suthree$ contribution is given by the
dotted lines. The grey band indicates the dark matter
density~\cite{Yao:2006px}
\begin{align}
        \omegadm  & =0.105^{+0.007}_{-0.010}
        \ .
\label{Eq:OmegaCDMobs}
\end{align}
and thus shows the parameter region in which the thermally produced
gravitinos provide the observed dark matter density.

We find that electroweak processes enhance $\O_{\gr}^{\text{TP}}$ by
about $20\%$ for universal gaugino masses at $\mgut$. In non-universal
cases, $M_{1,2}>M_3$ at $\mgut$, the electro\-weak contributions are
more important. For $0.5\, M_{1,2}=M_3$ at $\mgut$, they provide about
$40\%$ of $\ogravtp$. Moreover, with our new $k_3$ value---see
Eqs.~(\ref{eq:COLLISION-TERM-FINAL}) and (\ref{eq:relic gravitino
  abundance TP})---we find an enhancement of about $30\%$ of the
$\suthree$ contribution to relic density in comparison to the result
given in~\cite{Bolz:2000fu}.

\section{Upper Bounds on the Reheating Temperature}
\label{sec:upper-bound-TR}

\begin{figure}[bt]
\centering

\subfloat[]{
    \label{fig:M12 is 400 upperbound TR}
     \includegraphics[width=0.8\textwidth]{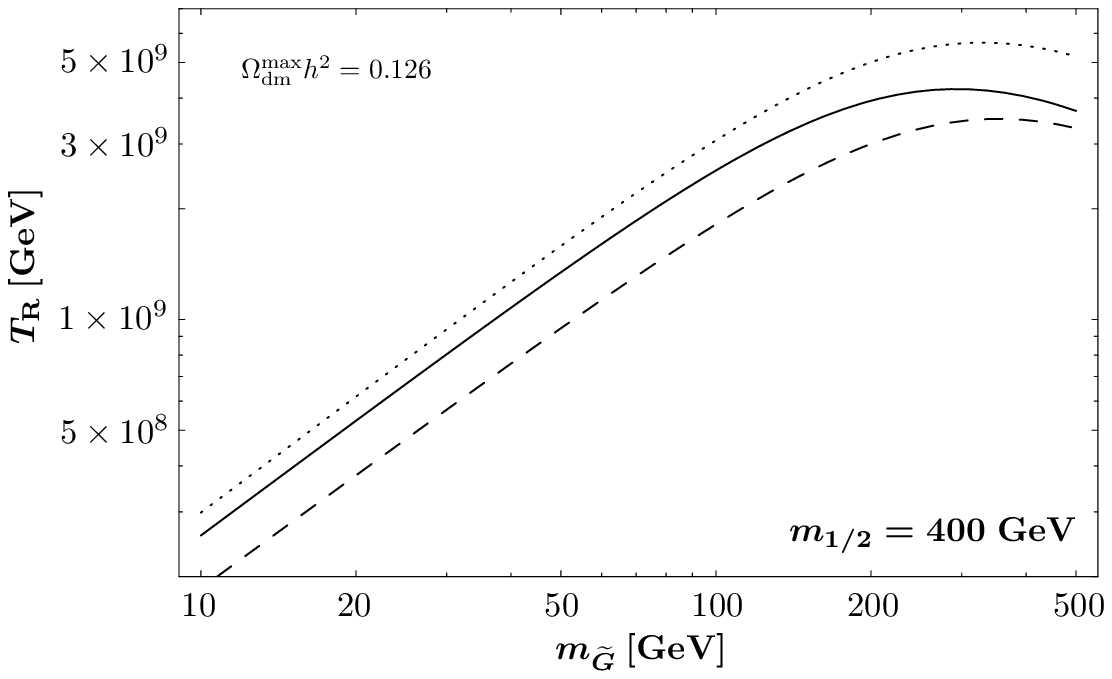}}
\\[20pt]
   \subfloat[]{
    \label{fig:M12 is 1500 upperbound TR}
    \includegraphics[width=0.8\textwidth]{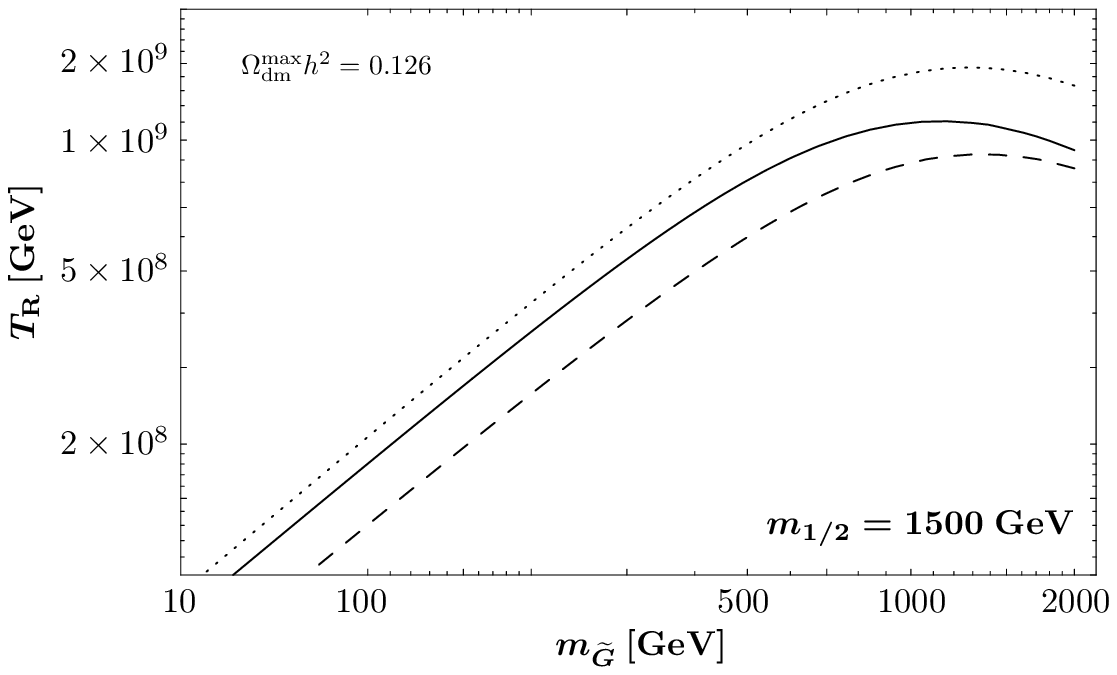}}%
\\[10pt]
\caption[Upper bounds on the reheating temperature]{Upper bounds on
  $\TR$ from $\O_{\gr}^{\text{TP}} \leq \omegaDMmax$. The solid and
  dashed curves show the limits inferred from the full $\smgroup$
  result~(\ref{eq:relic gravitino abundance TP}) for $M_{1,2,3}=m_{1/2}$ and
 $0.5\, M_{1,2}=M_3=m_{1/2}$ at $\mgut$, respectively. The dotted
 curves show the limit on $\TR$ for our new result of the $\suthree$
 contribution. }
\label{fig:Upper-bound-TR}
\end{figure}

In the gravitino dark matter scenario we can derive an upper bound on
the reheating temperature $\TR$ from $\O_{\gr}^{\text{TP}} \leq
\O_{\text{dm}}$, once  $m_{1/2}$ is specified.
Figure~\ref{fig:Upper-bound-TR} shows the upper limits on $\TR$ for
$m_{1/2}=400\,\GEV$ and $m_{1/2}=1500\,\GEV$, respectively. The solid
and dashed curves give the upper bounds on $\TR$ inferred from our
$\smgroup$ result of $\ogravtp$ for universal ($M_{1,2,3}=m_{1/2})$ and
non-universal ($0.5\, M_{1,2}=M_3=m_{1/2}$) gaugino masses at $\mgut$, respectively.
The dotted curves show the $\suthree$ limits for $M_3=m_{1/2}$ at
$\mgut$. We have adopted
\begin{align}
\label{eq:omegaDMmax}
  \omegaDMmax h^2 & = 0.126
\end{align}
as a nominal $3\sigma$ upper limit on $\omegadm$. 

Note that for higher values of $m_{1/2}$ the bounds on $\TR$ are more
stringent; cf.~(\ref{eq:relic gravitino abundance TP}).  For small
$\mgrav$, the thermal gravitino production is very efficient.  Then
$\ogravtp \sim M_\a^2/3\mgrav$ since the production of helicity $\pm
1/2$ states dominates. For large values of $\mgrav$, the upper limit
on $\TR$ becomes more severe again. Then $\ogravtp \sim \mgrav $ since
the production of the helicity $\pm 3/2$ states constitutes the
dominant part. The bounds on $\TR$ will become more severe when one
includes non-thermal production processes such as late-decays of the
next-to-lightest supersymmetric particle (NLSP) into the gravitino.
This, however, depends on the details of the realized SUSY model while
the bounds in Fig.~\ref{fig:Upper-bound-TR} are rather model
independent.


\cleardoublepage

\chapter{Testing Leptogenesis at Colliders}
\label{sec:coll-tests-lept}

The smallness of the neutrino masses can be understood naturally in
terms of the see-saw mechanism~\cite{Minkowski:1977sc,Yanagida:1980xy} once the
Standard Model is extended with right-handed neutrinos which have
heavy Majorana masses and only Yukawa couplings. For a reheating
temperature after inflation, $\TR$, which is larger or not
much smaller than the masses of the heavy neutrinos, these particles
are produced in thermal reactions in the early Universe. The
CP-violating out-of-equilibrium decays of the heavy neutrinos generate
a lepton asymmetry that is converted into a baryon asymmetry by
sphaleron processes~\cite{Fukugita:1986hr}. This mechanism, known as
thermal leptogenesis, can explain the cosmic baryon asymmetry for
$\TR \apprge 3\times 10^9\,\GEV$~\cite{Buchmuller:2004nz}.

One will face severe cosmological constraints on $\TR$ if
supersymmetry is discovered.  We have seen that gravitinos are
produced efficiently in the hot primordial plasma.  Because of their
extremely weak interactions, unstable gravitinos with $\mgrav \apprle
5~\text{TeV}$ have long lifetimes, $\tau_{\gr} \apprge 100~\text{s}$,
and decay after BBN.  The associated
decay products affect the abundances of the primordial light elements.
Demanding that the successful BBN predictions are preserved, bounds on
the abundance of gravitinos before their decay can be derived which
imply $\TR \apprle 10^8\,\GEV$ for $\mgrav \apprle
5~\text{TeV}$~\cite{Kohri:2005wn}. Thus, the temperatures needed for
thermal leptogenesis are excluded.

Let us therefore consider SUSY scenarios in which a gravitino with
$\mgrav \apprge 10~\GEV$ is the LSP and stable due to R-parity
conservation. These scenarios are particularly attractive for two
reasons: (i) the gravitino LSP can be dark matter and (ii) thermal
leptogenesis can still be a viable explanation of the baryon
asymmetry~\cite{Bolz:1998ek}.

\section{Collider Predictions of Leptogenesis}

Thermal leptogenesis requires $\TR \apprge 3\times
10^9\,\GEV$~\cite{Buchmuller:2004nz}.  This condition together with
the constraint $\O_{\gr}^{\text{TP}} \leq \omegaDMmax $ [see
Eqn.~(\ref{eq:omegaDMmax})] leads to upper limits on the gaugino
masses.  The $\suthree$ result for $\O_{\gr}^{\text{TP}}$ implies
limits on the gluino mass~\cite{Bolz:2000fu,Fujii:2003nr}.  With our
$\smgroup$ result, the limits on the gluino mass $M_3$ become more
stringent because of the new $k_3$
value~(\ref{eq:COLLISION-TERM-FINAL}) and the additional electroweak
contributions.  Moreover, as a prediction of thermal leptogenesis, we
obtain upper limits on the electro\-weak gaugino mass parameters
$M_{1,2}$.  At the Large Hadron Collider (LHC) and the International
Linear Collider (ILC), these limits will be probed in measurements of
the masses of the neutralinos and charginos, which are typically
lighter than the gluino. If the superparticle spectrum does not
respect these bounds, one will be able to exclude standard thermal
leptogenesis.

\begin{figure}[tb]
\begin{center}
\includegraphics[width=3.25in]{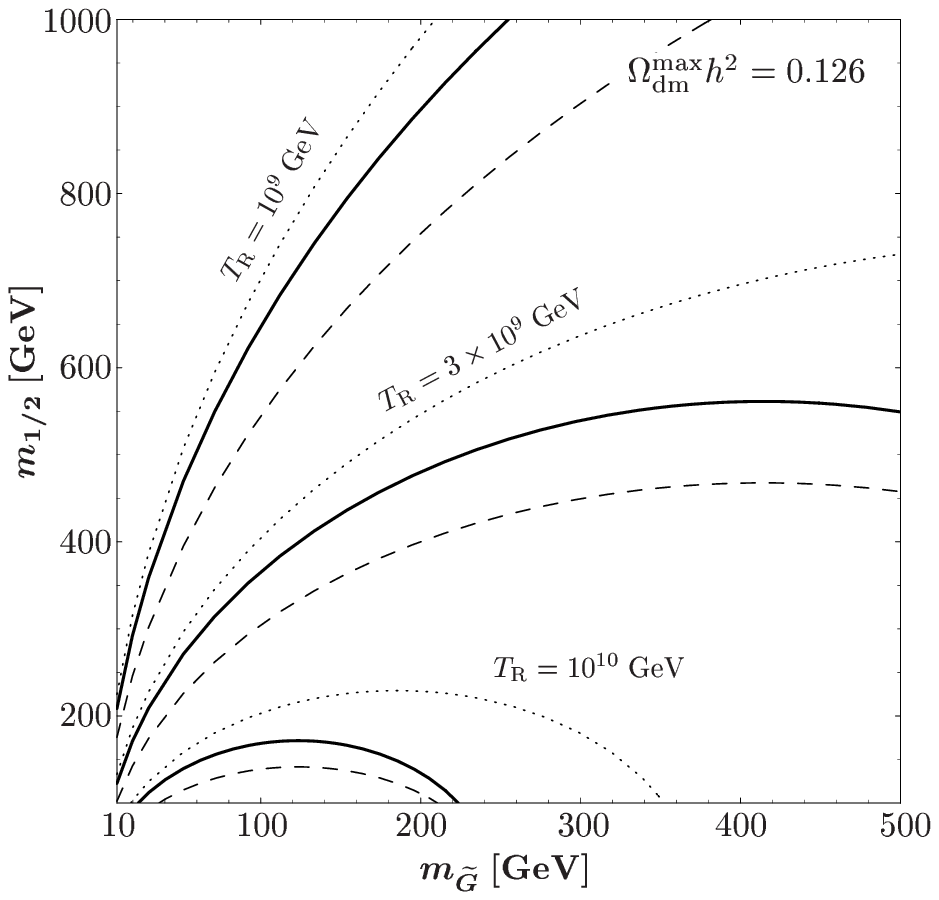} 
\caption[Upper limits on the gaugino masses]{\small Upper limits on the gaugino mass parameter $m_{1/2}$
  from $\Omega_{\widetilde{G}}^{\mathrm{TP}} \leq \omegaDMmax$ for the
  indicated values of $\TR$. The solid and dashed curves
  show our $\smgroup$
  results for universal ($M_{1,2,3}=m_{1/2}$) and non-universal
  ($0.5\, M_{1,2}=M_3=m_{1/2}$) gaugino masses at $\mgut$,
  respectively.  The dotted curves show the $\suthree$ limits for
  $M_3=m_{1/2}$ at $\mgut$.}
\label{Fig:UpperLimitM12}
\end{center}
\end{figure}
Figure~\ref{Fig:UpperLimitM12} shows the gaugino mass bounds for $\TR=10^9$, $3\times
10^9$, and $10^{10}~\GEV$ evolved to $\mgut$, i.e., in terms of
limits on the gaugino mass parameter $m_{1/2}$.
With the observed superparticle spectrum, one will be able to evaluate
the gaugino mass parameters $M_{1,2,3}$ at $\mgut$ using the SUSY
renormalization group
equations~\cite{Baer:2000gf,Blair:2002pg,Lafaye:2004cn,Bechtle:2004pc}.
While the determination of $M_{1,2}$ at low energies depends on
details of the SUSY model that will be probed at
colliders~\cite{Martin:1997ns}, the bounds shown in
Fig.~\ref{Fig:UpperLimitM12} depend mainly on the $M_i$ relation at
$\mgut$. This is illustrated by the solid and dashed curves obtained
with $M_{1,2,3}=m_{1/2}$ and $0.5\, M_{1,2}=M_3=m_{1/2}$,
respectively. The dotted curves represent the $\suthree$ limits for
$M_3=m_{1/2}$ at $\mgut$ and emphasize the importance of the
electroweak contributions.

\section{Decays of the Next-to-Lightest Supersymmetric Particle}

With a gravitino LSP of $\mgrav \apprge 10~\GEV$, the next-to-lightest
SUSY particle (NLSP) has a long lifetime of $\tau_{\text{NLSP}}\apprge
10^6\,\text{s}$~\cite{Feng:2004mt,Steffen:2006hw}.  After decoupling
from the primordial plasma, each NLSP decays into one gravitino LSP
and Standard Model particles. The resulting relic density of these
non-thermally produced gravitinos is given by
\begin{equation}
        \Omega_{\widetilde{G}}^{\text{NTP}}h^2
        = \frac{\mgrav}{m_{\text{NLSP}}}\,\Omega_{\text{NLSP}}h^2
        \ ,
\label{Eq:GravitinoDensityNTP}
\end{equation}
where $m_{\text{NLSP}}$ is the mass of the NLSP and $\Omega_{\text{NLSP}}h^2$ is
the relic density that the NLSP would have today, if it had not
decayed. As shown below, more severe limits on $m_{1/2}$ are obtained
with $\Omega_{\widetilde{G}}^{\text{NTP}}h^2$ taken into account.  Moreover,
since the NLSP decays take place after BBN, the emitted Standard Model
particles can affect the abundance of the primordial light elements.
Successful BBN predictions thus imply bounds on $\mgrav$ and
$m_{\text{NLSP}}$~\cite{Feng:2004mt,Steffen:2006hw}. From these cosmological
constraints it has been found that thermal leptogenesis remains viable
only in the cases of a charged slepton NLSP or a sneutrino
NLSP~\cite{Fujii:2003nr,Cerdeno:2005eu}.

\section{Collider Tests of Leptogenesis}

Thermal leptogenesis will predict a lower bound on the gravitino mass
$\mgrav$ once the masses of the Standard Model superpartners are known.
With a charged slepton as the lightest Standard Model superpartner, it
could even be possible to identify the gravitino as the LSP and to
measure its mass $\mgrav$ at future
colliders~\cite{Buchmuller:2004rq,Brandenburg:2005he,Martyn:2006as,Steffen:2005cn}.
Confronting the measured $\mgrav$ with the predicted lower bound will
then allow us to decide about the viability of thermal leptogenesis.

To be specific, let us assume that the analysis of the observed
spectrum~\cite{Lafaye:2004cn,Bechtle:2004pc} will point to the universality of the
soft SUSY breaking parameters at $\mgut$ and, in particular, to the
minimal supergravity (mSUGRA) scenario with the gaugino mass parameter
$m_{1/2}=400~\GEV$, the scalar mass parameter $m_0=150~\GEV$, the
trilinear coupling $A_0=-150$, a positive higgsino mass parameter,
$\mu>0$, and the mixing angle $\tan\beta=30$ in the Higgs sector.
A striking feature of the spectrum will then be the appearance of the
lighter stau $\stau$ with $\mst=143.4~\GEV$ as the lightest Standard
Model superpartner~\cite{Djouadi:2002ze}.
In the considered gravitino LSP case, $ 10~\GEV \apprle \mgrav < \mst $, this
stau is the NLSP and decays with a lifetime of
$\tau_{\stau}\apprge 10^6\,\text{s}$
into the gravitino.
For the identified mSUGRA scenario and the considered reheating
temperatures, the cosmological abundance of the $\stau$ NLSP prior to
decay can be computed from 
$\Omega_{\text{NLSP}}h^2=\Omega_{\stau}h^2\simeq 3.83\times 10^{-3}$,
which is provided by the computer program
\texttt{micrOMEGAs}~\cite{Belanger:2001fz}.
For given $\mgrav$, this abundance determines
$\Omega_{\gr}^{\text{NTP}}h^2$ and the release of electromagnetic
(EM) and hadronic energy in $\stau$ NLSP decays governing the
cosmological constraints~\cite{Feng:2004mt,Steffen:2006hw}.

\begin{figure}[t]
\begin{center}
\includegraphics[width=3.25in]{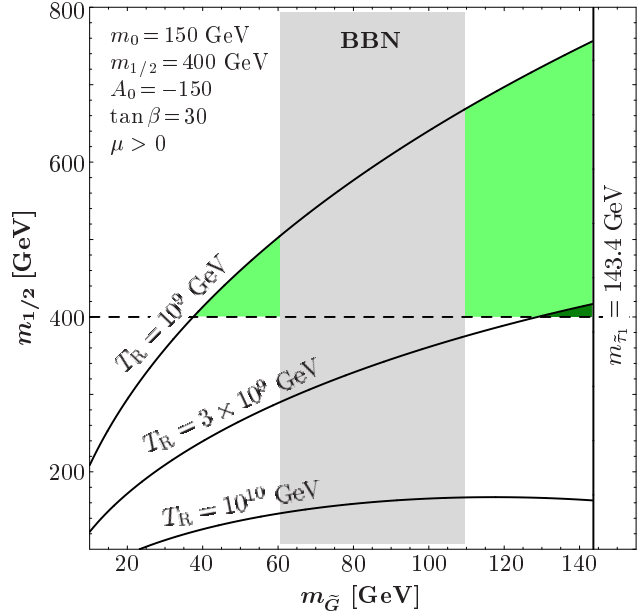} 
\caption[Probing the viability of leptogenesis]{\small Probing the viability of thermal leptogenesis. The
  solid curves show the limits on the gaugino mass parameter $m_{1/2}$
  from
  $\Omega_{\widetilde{G}}^{\mathrm{TP}}+\Omega_{\widetilde{G}}^{\text{NTP}}
  \leq \omegaDMmax$
  for $\TR=10^9$, $3\times 10^9$, and $10^{10}~\GEV$.  The
  dashed line indicates the $m_{1/2}$ value of the considered
  scenario. The vertical solid line is given by the $\stau$ NLSP mass
  which limits the gravitino LSP mass from above:
  $\mgrav<\mst=143.4~\GEV$. The $\mgrav$ values at which temperatures
  above $3\times 10^9~\GEV$ and $10^9~\GEV$ remain allowed are
  indicated by the dark-shaded (dark-green) and medium-shaded
  (light-green) regions, respectively.  The $\mgrav$ values within the
  light-shaded (grey) region are excluded by BBN constraints.}
\label{Fig:ProbingTLGViability}
\end{center}
\end{figure}

Figure~\ref{Fig:ProbingTLGViability}
allows us to probe the viability of thermal leptogenesis in the
considered mSUGRA scenario.\footnote{Thermal leptogenesis requires
  right-handed neutrinos and thus an extended mSUGRA scenario. This
  could ma\-ni\-fest itself in the masses of the third generation
  sleptons~\cite{Baer:2000gf}. Since the effects are typically small,
  we leave a systematic investigation of extended scenarios for future
  work.} 
From the constraint
$\Omega_{\widetilde{G}}^{\mathrm{TP}}+\Omega_{\widetilde{G}}^{\text{NTP}}
\leq
\omegaDMmax$,
we obtain the solid curves which provide the upper limits on $m_{1/2}$
for $\TR=10^9$, $3\times 10^9$, and $10^{10}~\GEV$.
The dashed line indicates the $m_{1/2}$ value of the considered
scenario. The vertical solid line is given by $\mst=143.4~\GEV$ which
limits $\mgrav$ from above.
In the considered scenario, the $m_{1/2}$ value exceeds the $m_{1/2}$
limits for $\TR\apprge 10^{10}~\GEV$. Thus, temperatures
above $10^{10}~\GEV$ can be excluded.
Temperatures above $3\times 10^9~\GEV$ and $10^9~\GEV$ remain allowed
for $\mgrav$ values indicated by the dark-shaded (dark-green) and
medium-shaded (light-green) regions, respectively.
The $\mgrav$ values indicated by the light-shaded (grey) region are
excluded by BBN constraints for late $\stau$ NLSP decays.\footnote{We
  use the conservative BBN bounds considered in~\cite{Steffen:2006hw}.
  The average EM energy release in one $\stau$ NLSP decay is assumed
  to be $E_{\tau}/2$, where $E_{\tau}$ is the energy of the tau
  emitted in the dominant 2-body decay $\stau\to\gr\tau$ (cf.\
  Fig.~16 of Ref.~\cite{Steffen:2006hw}). With an EM energy release
  below $E_{\tau}/2$, the grey band can become smaller. For less
  conservative BBN constraints and/or enhanced EM energy release, the
  excluded $\mgrav$ region becomes larger.}

Here thermal leptogenesis, $\TR\apprge 3\times 10^9\,\GEV$, predicts
$\mgrav\apprge 130~\GEV$ and thus a $\stau$ lifetime of
$\tau_{\stau}>10^{11}~\text{s}$~\cite{Feng:2004mt,Steffen:2006hw}.  If
decays of long-lived $\stau$'s can be analyzed at colliders giving
evidence for the gravitino
LSP~\cite{Buchmuller:2004rq,Brandenburg:2005he,Martyn:2006as,Steffen:2005cn},
there will be the possibility to determine $\mgrav$ in the laboratory:
From a measurement of the lifetime $\tau_{\stau}$ governed by the
decay $\stau\to\gr\tau$, $\mgrav$ can be extracted using the
supergravity prediction for the associated partial width,
\begin{equation}
        \tau_{\stau} 
        \simeq \Gamma^{-1}(\stau\to\gr\tau)
        = \frac{48 \pi \mgrav^2 \mplanck^2}{m_{\stau}^5} 
        \left(1-\frac{\mgrav^2}{m_{\stau}^2}\right)^{-4}
\label{Eq:StauLifetime}
\end{equation}
Moreover, for $\mgrav\apprge 0.1\,\mst$, $\mgrav$ can be infered
kinematically from the energy of the tau, $E_{\tau}$, emitted in the
2-body decay
$\stau\to\gr\tau$~\cite{Buchmuller:2004rq,Martyn:2006as}:
\begin{equation}
        \mgrav = \sqrt{m_{\stau}^2-m_{\tau}^2-2 m_{\stau} E_{\tau}}
        \ . 
\label{Eq:GravitinoMassKinematically}
\end{equation}
While $\mgrav$ within the dark-shaded (dark-green) region will favor
thermal leptogenesis, any $\mgrav$ outside of the medium-shaded
(light-green) region will require either non-standard mechanisms
lowering the $\TR$ value needed for thermal leptogenesis or
an alternative explanation of the cosmic baryon asymmetry.


\cleardoublepage

\chapter{Conclusions}
\label{cha:conclusions}

The starting point of this thesis has been a discussion of the general
supergravity Lagrangian in four spacetime dimensions. We have shown
how supersymmetry can be broken spontaneously. For a simple
gravity-mediation scenario, we have explicitly carried out the
transition to a softly-broken supersymmetric theory as the
low-energy limit of supergravity. Relating the obtained effective
theory with the MSSM in the high-energy limit of unbroken electroweak
symmetry and identifying the relevant gravitino interactions, we have
derived all Feynman rules necessary for the calculation of the
regeneration of gravitinos after inflation.

We have used the Braaten--Yuan~\cite{Braaten:1991dd} prescription
together with the hard thermal loop resummation
technique~\cite{Braaten:1989mz} to calculate the thermal gravitino
production rate.  We have considered the regeneration of gravitinos
that starts after completion of reheating.  As one of the main results
of this thesis, we present the Boltzmann collision
term~(\ref{eq:COLLISION-TERM-FINAL}) to leading order in the
$\smgroup$ couplings. It acts as the source term in the Boltzmann
equation which governs the time evolution of the gravitino number
density in the thermal bath. The collision term has been obtained in a
gauge invariant way within the framework of thermal field theory and
does not depend on arbitrary cutoffs.  Our result includes for the
first time the $\sutwo\times\uone$ sector.  Moreover, we correct an
error in the previously known $\suthree$ result which was obtained in
Ref.~\cite{Bolz:2000fu}. 

With direct implications for gravitino cosmology, we obtain the
gravitino yield~(\ref{eq:yield-FINAL}) from thermal production by
solving the Boltzmann equation. The yield parametrizes the primordial
abundance of gravitinos in the early Universe. Hence, it is a crucial
quantity for both scenarios with stable gravitinos and scenarios with
unstable gravitinos.  Focusing on gravitino dark matter scenarios, we
have obtained the relic abundance of thermally produced gravitinos,
$\O_{\gr}^{\text{TP}}$.  It depends on the reheating temperature
$\TR$, the gravitino mass $\mgrav$, and the gaugino masses $M_\a$.  We
find that electroweak processes enhance $\O_{\gr}^{\text{TP}}$ by
about $20\%$ for universal gaugino masses at the grand unification
scale $\mgut$. For non-universal scenarios, the electroweak
contributions become more important. Furthermore, with our new
$\suthree$ result, we find an enhancement of the $\suthree$
contribution to  $\O_{\gr}^{\text{TP}}$ by about
$30\%$ as compared to the relic density obtained
in~\cite{Bolz:2000fu}. We also give an update for the upper bounds on
the reheating temperature depending on the specified gaugino mass
parameter $\monetwo$.

With the result for the relic density of thermally produced gravitino
LSPs, new gravitino and gaugino mass bounds emerge as a prediction of
thermal leptogenesis which requires $\TR \apprge 3\times 10^9\,\GEV$.
If supersymmetry is realized in nature, these bounds will be
accessible at the LHC and the ILC. For certain SUSY parameter regions,
a charged slepton will be the NLSP. We have studied a
 scenario in which the lighter stau is the NLSP. Here we have
also taken into account the non-thermal gravitino production from the
stau decays into the gravitino LSP. There exists then the exciting
possibility to identify the gravitino as the LSP and to measure its
mass. Confronting the measured gravitino mass with the predicted
bounds will then allow for a unique test of the viability of thermal
leptogenesis in the laboratory.


\cleardoublepage
\begin{appendix}
\chapter{Conventions and Spinor Notation}
\label{ch:appendix-spinor-alg}

The flat-space Lorentz metric is given by\footnote{Though we make a
  distinction between Einstein and Lorentz indices in the beginning
  of Chapter~\ref{cha:supergravity}, greek indices stand for flat
  spacetime indices in the other chapters of this thesis. }
 \begin{align}
    \label{metric}
    \h_{\m\n}=\h^{\m\n}\equiv\mathrm{diag}(+1,-1,-1,-1).
 \end{align}
We fix the sign of the completely antisymmetric tensor
$\ve_{\m\n\rho\s}$ by choosing
 \begin{align}
    \ve_{0123}\equiv-1.
 \end{align}
Greek indices $\m,\n,\dots =0,\dots,3$ denote space-time indices.

\section{Weyl Spinors}
A two-component complex undotted Weyl spinor (left-handed Weyl spinor)
$\x_\a$ transforms in the
$\left(\mathbf{\frac{1}{2}},\mathbf{0}\right)$ matrix representation
of the Lorentz group $\text{SO}(3,1)$, i.e.\ under
$\text{SL}(2,\mathbbm{C})$, while the dotted Weyl spinor (right-handed
Weyl spinor) $\overline{\x}_{\dot{\a}}$ is in the conjugate
representation $\left(\mathbf{0},\mathbf{\frac{1}{2}}\right)$. Both
spinors are related by hermitian conjugation, i.e.\ $\left
  (\x_\a\right )^\dagger=\overline{\x}_{\dot{\a}}$ and
$\left(\overline{\x}_{\dot{\a}}\right)^\dagger=\x_\a\,$.  Explicitly,
for a Lorentz transformation $M\in\mbox{SL}(2,\mathbbm{C})$:
\begin{subequations}
  \begin{alignat}{2}
    \x_{\a}' &= {M_{\a }}^{\b } \x_{\b}, & \qquad \overline{\x}'_{\dot{\a}}&=
    \tensor{M}{^*_{\dot{\a }}^{\dot{\b }}}\,\overline{\x}_{\dot{\b }}\;,\\
     {\x'}^{\,\a }&=M\indices{^{-1}_\b^\a} \x^{\b},&
     {\overline{\x}'}^{\,\dot{\a }}&=\tensor{(M^*)}{^{-1}_{\dot{\b }}^{\dot{\a}}}\,
     \overline{\x}^{\,\dot{\b}}\;.
  \end{alignat}
\end{subequations}
Spinor indices are pulled by the Lorentz invariant $\ve$-tensors
\begin{subequations}
  \label{eq:App-epsilons}
  \begin{alignat}{2}
    \ve_{\a\b} &\equiv\begin{pmatrix}0&-1\\1&0\\
    \end{pmatrix}\;,\quad\quad&
    \ve^{\a\b} &\equiv\begin{pmatrix}0&1\\-1&0\\ \end{pmatrix}\;,
    \label{eq:App-epsilons-undotted}  \\
    \ve_{\dot{\a} \dot{\b}} &\equiv\begin{pmatrix}0&-1\\1&0\\
    \end{pmatrix}\;,&
    \ve^{\dot{\a}\dot{\b}}&\equiv\begin{pmatrix}0&1\\-1&0\\
    \end{pmatrix}\;,
  \end{alignat}
\end{subequations}
namely,
  \begin{eqnarray}
    \x_{\a }=\ve_{\a\b}\x^{\b},&\quad&\x^{\a}=\ve^{\a\b}\x_{\b}\;,\\
    \overline{\x}_{\,\dot{\a}}=\ve_{\dot{\a}\dot{\b}}
    \overline{\x}^{\,\dot{\b}},&\quad&\overline{\x}^{\,\dot{\a}}=
    \ve^{\dot{\a}\dot{\b}}\overline{\x}_{\,\dot{\b}}\;.
  \end{eqnarray}
  Furthermore, we define the Pauli sigma matrices (index 1,2,3) with
  lower Lorentz indices:
\begin{subequations}
\label{sigmas}
  \begin{alignat}{2} 
    \s_0&\equiv\begin{pmatrix}1&0\\0&1\\ \end{pmatrix}\;,
    \quad\quad& \s_1&\equiv\begin{pmatrix}0&1\\1&0\\
    \end{pmatrix}\;, \\
    \s_2&\equiv\begin{pmatrix}0&-i\\i&0\\ \end{pmatrix}\;,
    \quad\quad& \s_3&\equiv\begin{pmatrix}1&0\\0&-1\\
    \end{pmatrix}\;.
  \end{alignat}
\end{subequations}
The standard convention for the contraction of anticommuting
Weyl spinors is
\begin{subequations}
\begin{align}
    \x\h\equiv\x^\a\h_\a=\ve^{\a\b}\x_{\b}\h_{\a}=
    -\ve^{\a\b}\h_{\a}\x_{\b}=\h\x \; ,\\
    \overline{\x}\overline{\h}\equiv\overline{\x}_{\,\dot{\a}}\overline{\h}^{\,\dot{\a}}=
    \ve^{\dot{\a}\dot{\b}}\overline{\x}_{\,\dot{\b}}\overline{\h}_{\,\dot{\a}}=
    -\ve^{\dot{\a}\dot{\b}}\overline{\x}_{\,\dot{\b}}\overline{\h}_{\,\dot{\a}}=
    \overline{\h}\overline{\x}\; .
\end{align}
\end{subequations} 
Note the spinor index structure of the sigma matrices
$\tensor{\s}{^{\,\m}_{\a\dot\a}}$\;. One defines 
\begin{align}
  \tensor{\overline{\s}}{^{\,\m\,}^{\dot\a\a}}\equiv
  \ve^{\dot{\a}\dot{\b}}\ve^{\a\b} \tensor{\s}{^{\,\m}_{\b\dot\b}}\; ,
\end{align}
as well as
\begin{subequations}
\begin{align}
  \tensor{\s}{^{\,\m\n}_\a^\b}&\equiv\frac{1}{4}\left (
    \tensor{\s}{^{\,\m}_{\a\dot\a}}
    \tensor{\overline{\s}}{^{\,\n\,}^{\dot\a\b}}-
    \tensor{\s}{^{\,\n}_{\a\dot\a}}
    \tensor{\overline{\s}}{^{\,\m\,}^{\dot\a\b}}
  \right )\; ,  \\
  \tensor{\overline{\s}}{^{\,\m\n\,\dot\a}_{\dot\b}}&\equiv\frac{1}{4}\left
    ( \tensor{\overline{\s}}{^{\,\m\,}_{\dot\a\a}}
    \tensor{\s}{^{\,\n}_{\a\dot\b}}-
    \tensor{\overline{\s}}{^{\,\n\,}_{\dot\a\a}}
    \tensor{\s}{^{\,\m}_{\a\dot\b}} \right )\; .
\end{align} 
\end{subequations}

\section{Four-component Spinors}
In the Weyl basis, the Dirac $\gamma$ matrices read
\begin{align}
    \gamma_\m=\begin{pmatrix}0&\s_\m\\\overline{\s}_\m&0\\
    \end{pmatrix}.
\end{align}
They satisfy the Clifford algebra
\begin{align}
  \{\gamma_\m,\gamma_\n\}=2\h_{\m\n}
\end{align}
and anticommute with $\gamma_5\equiv
i\gamma^0\gamma^1\gamma^2\gamma^3$, i.e.\ $\{\gamma_\m,\gamma_5\}=0$.
In this representation\footnote{The sign-convention for $\gamma_5$
  results from the choice (\ref{sigmas}) which allowed for a clean transition in
  Lorentz signatures.}: 
\begin{align}
  \gamma_5=\begin{pmatrix}\mathbbm{1}&0\\0&-\mathbbm{1}\\
    \end{pmatrix}
\end{align}
We can write a Dirac spinor in terms of a left-handed and a right-handed
Weyl spinor
\begin{align}
\label{eq:def-dirac-spinor}
    \j_{\text{\tiny (D)}}=\begin{pmatrix}\x_\a\\\overline\h^{\,\dot\a}\end{pmatrix},
\end{align}
and its adjoint spinor as
\begin{align}
    \overline{\j}_{\text{\tiny (D)}}\equiv{\j_{\text{\tiny (D)}}}^\dagger\gamma^0=
    \begin{pmatrix}\h^\a&\overline\x_{\dot\a}\end{pmatrix}
\end{align}
With the chiral projectors $P_L=\frac{1}{2}(\mathbbm{1}+\gamma_5)$ and
$P_R=\frac{1}{2}(\mathbbm{1}-\gamma_5)$  left-handed and right-handed
four-spinors are given as
\begin{subequations}
\begin{align} \j_L\equiv P_L\j_{\text{\tiny (D)}}&=
    \begin{pmatrix}
    \mathbbm{1}&0\\0&0
    \end{pmatrix}
    \begin{pmatrix}\x_\a\\\overline\h^{\,\dot\a}\end{pmatrix}=
    \begin{pmatrix}\x_\a\\0\end{pmatrix}\intertext{and}
    \j_R\equiv P_R\j_{\text{\tiny (D)}}&=
    \begin{pmatrix}
    0&0\\0&\mathbbm{1}
    \end{pmatrix}
    \begin{pmatrix}\x_\a\\\overline\h^{\,\dot\a}\end{pmatrix}=
    \begin{pmatrix}0\\\overline\h^{\dot\a}\end{pmatrix}\; ,
\end{align}
\end{subequations}
respectively. For the adjoints of the chiral spinors one finds
$\overline{\j}_L=\overline{\j}_{\text{\tiny (D)}}P_R$ and
\mbox{$\overline{\j}_R=\overline{\j}_{\text{\tiny (D)}}P_L$}.

By virtue of the charge conjugation matrix $C$ an equivalent
realization of the Clifford algebra is given by the transposed
$\g$ matrices:
\begin{align}
    C^{-1}\gamma_\m C&=-\gamma_\m^T, \intertext{with}
    C^\dagger=C^T=C^{-1}&=-C\quad\textrm{and}\quad
    C^2=-\mathbbm{1}.
\end{align}
The matrix $C$ can be written as
 \begin{align} C=
i\gamma^2\gamma^0=\begin{pmatrix}\ve_{\a\b}&0\\0&\ve^{\dot\a\dot\b}\end{pmatrix},
\end{align}
so that the charge-conjugated Dirac spinor of~(\ref{eq:def-dirac-spinor})
then reads
\begin{align}
\j_{\text{\tiny (D)}}^{\,c}&\equiv
C{\overline{\j}_{\text{\tiny (D)}}^{\;T}}=\begin{pmatrix}\h_\a\\\overline\x^{\,\dot\a}\end{pmatrix}.
\end{align}
A Majorana spinor is equal to its own charge-conjugate, i.e.,
$\j_{\text{\tiny (M)}}=\j_{\text{\tiny (M)}}^{\,c}$, so that it can be written as
 \begin{align}
   \label{eq:definition-majorana-field}
\j_{\text{\tiny (M)}}=\begin{pmatrix} \x_\a\\  \overline\x^{\,\dot\a}\end{pmatrix}.
\end{align}


\cleardoublepage
\chapter{Feynman rules}
\label{cha:feynman-rules}

Here we provide the complete set of the Feynman rules which is
necessary for the calculations performed in Chapter~\ref{cha:electr-therm-grav}.
The method of a continuous fermion flow is addressed in
Section~\ref{sec:feynman-rules}. For details, see~\cite{Denner:1992vz}.

The momentum $P$ always flows from the left to the
right for the external lines and propagators shown below. Furthermore, momenta
are assumed to flow towards the vertices.

\smallskip \noindent\textbf{External Lines}\smallskip

\begin{itemize}
\item Gauginos $\l^{(\a)}$ and matter fermions $\chi_L$  \vspace*{0.3cm}\\
\hspace*{1cm}\includegraphics[scale=0.7]{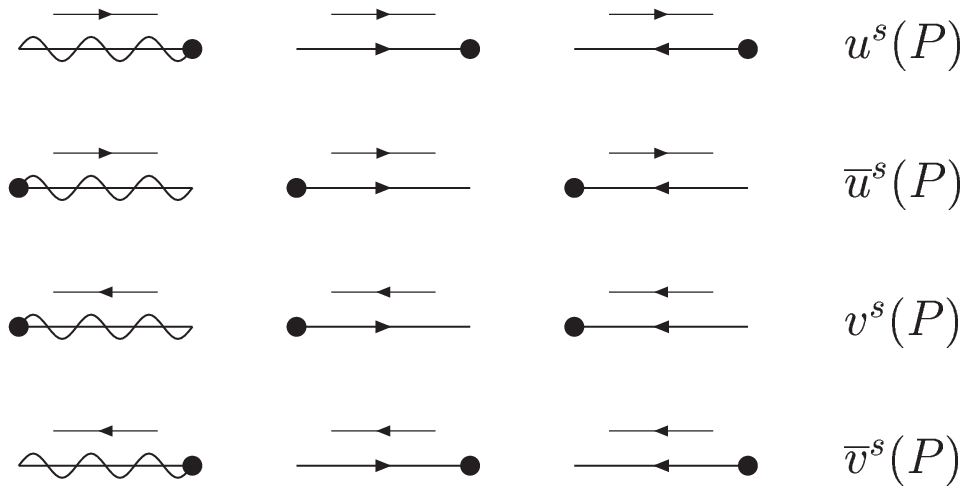}%
\item Gauge bosons  $A^{(\a)}$\vspace*{0.3cm} \\
\hspace*{1cm}\includegraphics[scale=0.7]{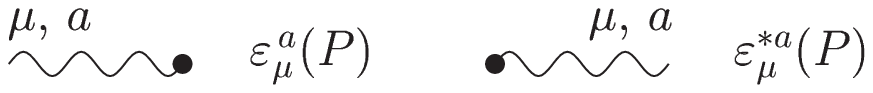}%
\item Gravitinos $\psi_\m$\vspace*{0.3cm} \\
\hspace*{1cm}\includegraphics[scale=0.7]{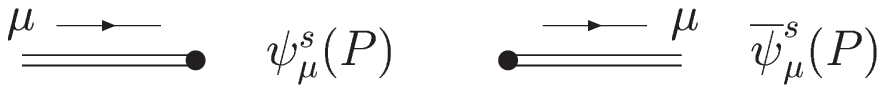}%
\end{itemize}

\newpage

\smallskip \noindent\textbf{Propagators}\smallskip

\begin{itemize}
\item Gauginos $\l^{(\a)}$  \vspace*{-0.1cm}\\
\parbox[t][1cm][c]{0.35\textwidth}{
\hspace*{1cm}\includegraphics[scale=0.7]{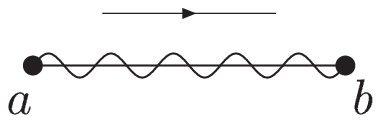}
}%
\parbox[t][1cm][c]{0.2\linewidth}{
 $$ \frac{i(\slashed{P}+M_\a)}{P^2 - M_\a^2}\,\d^{ab}$$}
\item Matter fields $\chi_L$\footnote{Note that for unbroken
    electroweak symmetry, the only non-vanishing masses for the chiral
    matter fermions arise from the higgsino mass parameter $\mu$;
    see~(\ref{eq:yukakwas and higgsino mass term final}) and (\ref{eq:superpotential-MSSM}).} \vspace*{-0.1cm}\\
\parbox[t][1cm][c]{0.35\textwidth}{
\hspace*{1cm}\includegraphics[scale=0.7]{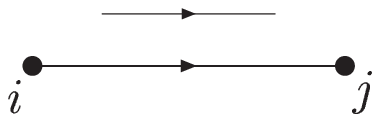}
}%
\parbox[t][1cm][c]{0.2\linewidth}{
 $$ \frac{i(\slashed{P}+m_{\chi})}{P^2 - m_{\chi}^2}\,\d^{ij}$$} \vspace*{0.2cm} \\
\parbox[t][1cm][c]{0.35\textwidth}{
\hspace*{1cm}\includegraphics[scale=0.7]{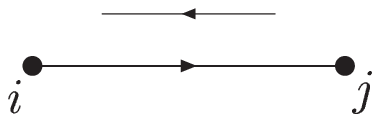}
}%
\parbox[t][1cm][c]{0.2\linewidth}{
 $$ \frac{i(- \slashed{P}+m_{\chi})}{P^2 - m_{\chi}^2}\,\d^{ij}$$}
\item Scalars $\f $ \vspace*{-0.1cm}\\
\parbox[t][1cm][c]{0.35\textwidth}{
\hspace*{1cm}\includegraphics[scale=0.7]{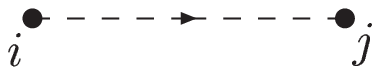}
}%
\parbox[t][1cm][c]{0.2\linewidth}{
 $$ \frac{i}{P^2 - m_\f^2}\,\d^{ij}$$}
\item Gauge bosons $A^{(\a)}$ \vspace*{-0.1cm}  \\
\parbox[t][1cm][c]{0.35\textwidth}{
\hspace*{1cm}\includegraphics[scale=0.7]{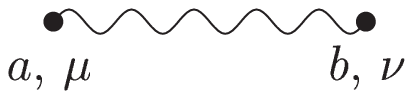}
}%
\parbox[t][1cm][c]{0.2\linewidth}{
 $$ i \left[ -\frac{g_{\m\n}}{P^2} + (1 - \xi) \frac{P_\m P_\n}{(P^2)^2}  \right] \d^{ab}  $$}
\end{itemize}

\smallskip \noindent\textbf{Relevant Gauge Vertices from
  Eq.~(\ref{eq:Lsusy-gauge})}\smallskip

\hfill\includegraphics[width=0.45\textwidth]{./feynmangraphs/V1.eps}%
\includegraphics[width=0.45\textwidth]{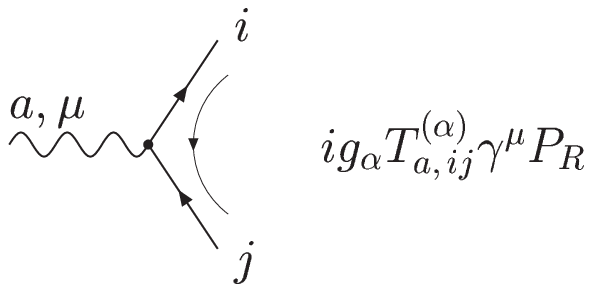}\hspace*{\fill}

\hfill\includegraphics[width=0.45\textwidth]{./feynmangraphs/V3.eps}%
\includegraphics[width=0.45\textwidth]{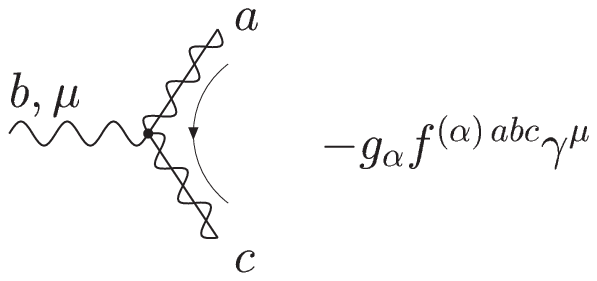}\hspace*{\fill}

\hfill\includegraphics[width=0.45\textwidth]{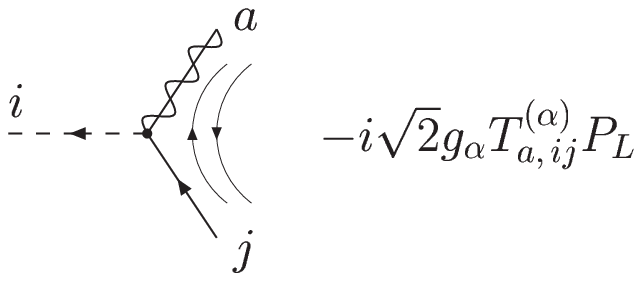}%
\includegraphics[width=0.45\textwidth]{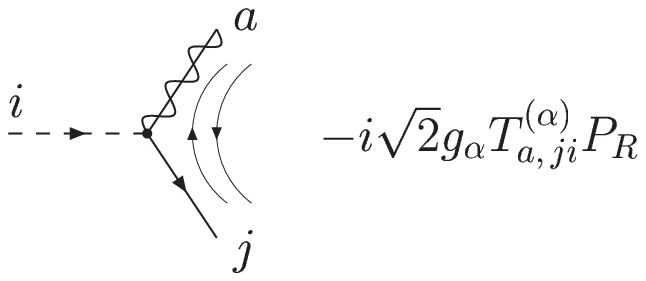}\hspace*{\fill}

\hfill\includegraphics[width=0.45\textwidth]{./feynmangraphs/V2.eps}%
\includegraphics[width=0.45\textwidth]{./feynmangraphs/V5.eps}\hspace*{\fill}

\smallskip \noindent\textbf{Gravitino Vertices from Eq.~(\ref{eq:gravitino
    interaction lagrangian})}\smallskip

\hfill\includegraphics[width=0.45\textwidth]{./feynmangraphs/G1.eps}%
\includegraphics[width=0.45\textwidth]{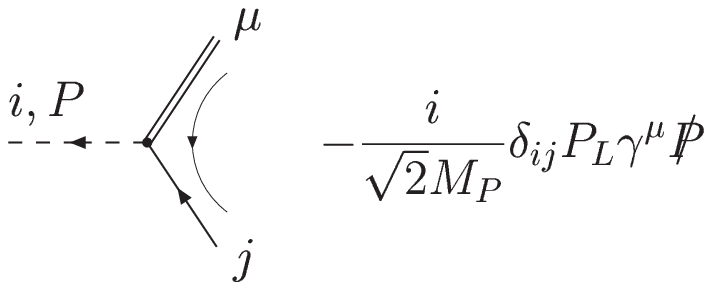}\hspace*{\fill}

\hfill\includegraphics[width=0.45\textwidth]{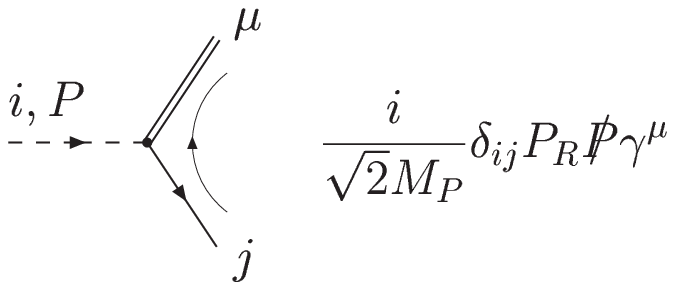}%
\includegraphics[width=0.45\textwidth]{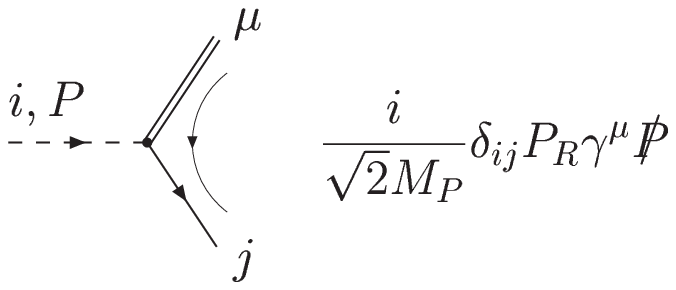}\hspace*{\fill}

\hfill\includegraphics[width=0.45\textwidth]{./feynmangraphs/G2.eps}%
\includegraphics[width=0.45\textwidth]{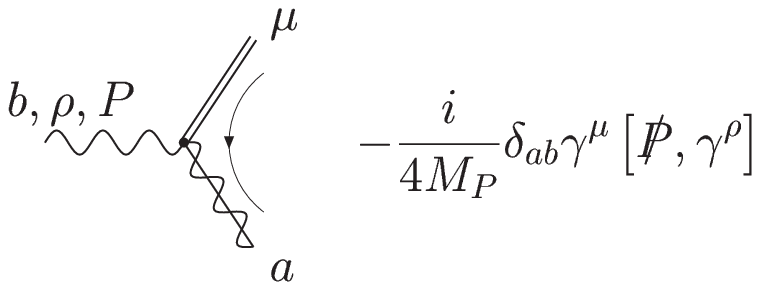}\hspace*{\fill}

\hfill\includegraphics[width=0.45\textwidth]{./feynmangraphs/G3.eps}%
\includegraphics[width=0.45\textwidth]{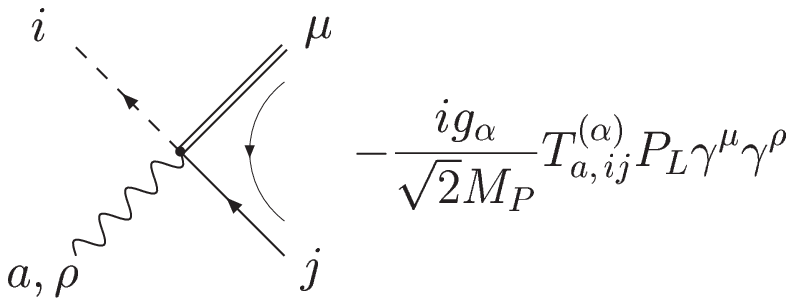}\hspace*{\fill}

\hfill\includegraphics[width=0.45\textwidth]{./feynmangraphs/G5.eps}%
\includegraphics[width=0.45\textwidth]{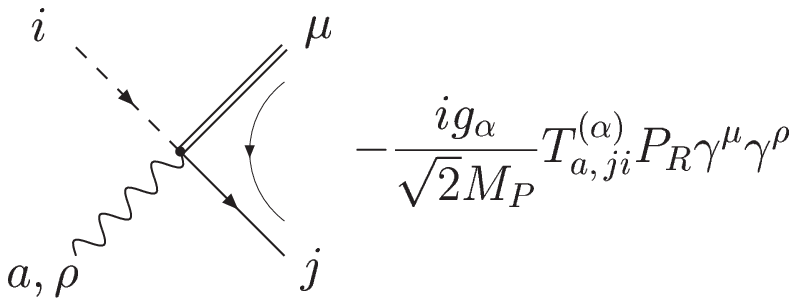}\hspace*{\fill}

\hfill\includegraphics[width=0.45\textwidth]{./feynmangraphs/G4.eps}%
\includegraphics[width=0.45\textwidth]{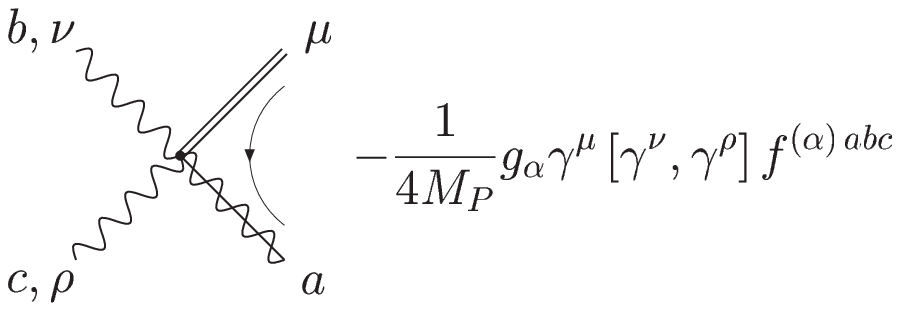}\hspace*{\fill}

\newpage

\smallskip \noindent\textbf{Gravitino Vertices from the Effective
  Theory~(\ref{eq:goldstino-lagrangian})}\smallskip

Note that in the effective theory for light gravitinos, the external
gravitinos are treated as Majorana fermions.  

\hfill\includegraphics[width=0.45\textwidth]{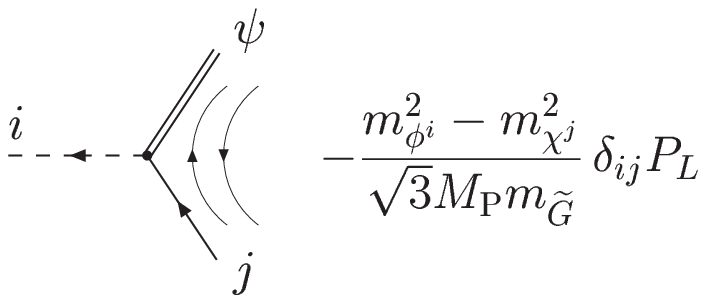}%
\includegraphics[width=0.45\textwidth]{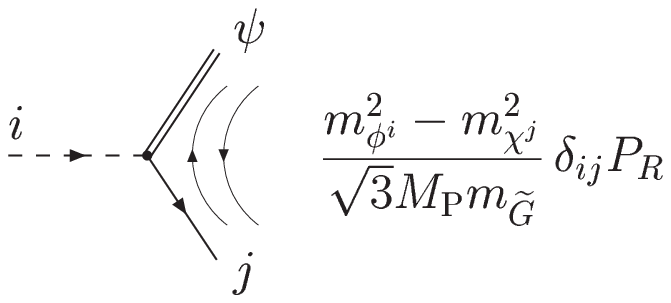}\hspace*{\fill}

\hfill\includegraphics[width=0.45\textwidth]{./feynmangraphs/G3light.eps}%
\includegraphics[width=0.45\textwidth]{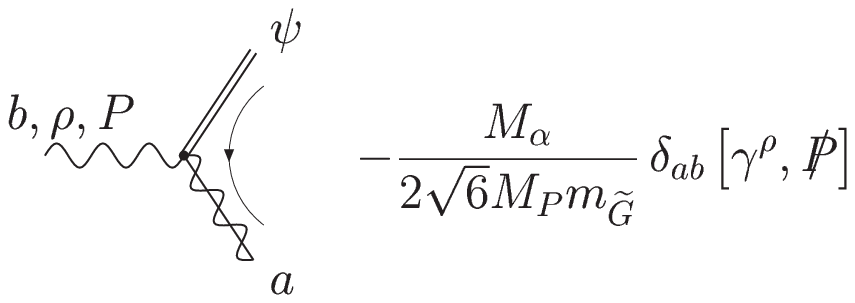}\hspace*{\fill}

\hfill\includegraphics[width=0.45\textwidth]{./feynmangraphs/G4light.eps}%
\includegraphics[width=0.45\textwidth]{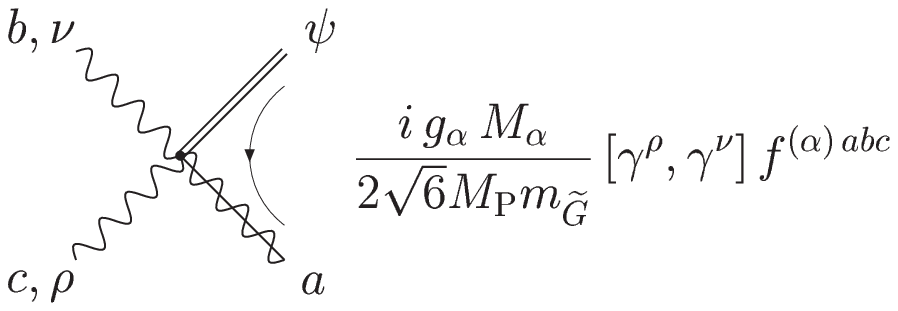}\hspace*{\fill}

\hfill\includegraphics[width=0.45\textwidth]{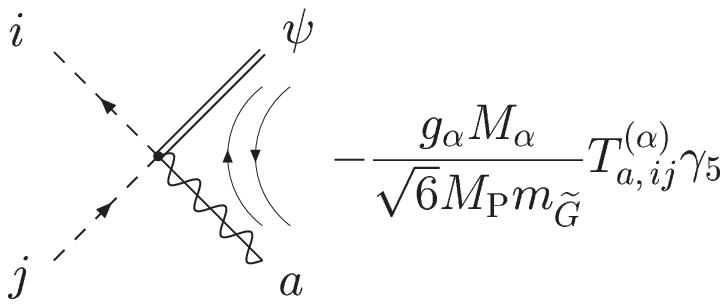}\hspace*{0.5\textwidth}


\cleardoublepage

\chapter{Hard Production Rate}
\label{cha:Appendix-hard-production-rate}

In this Appendix, we derive the hard part of the gravitino production
rate for the full SM gauge group $\smgroup$. We will follow  the
approach of \cite{Bolz:2000fu} but show in detail how to perform the
integrations and point out where differences between our $\suthree$
result and the one obtained in \cite{Bolz:2000fu} emerge.

Recall the definition for the
production rate~(\ref{eq:hard-production-rate})
\begin{align}
\label{eq:hard-production-rate-appendix}
\gammahard & =  {
  1\over (2\pi)^3 2E}  \int {d\Omega_p \over 4\pi} \int
\left[\prod_{i=1}^3 \frac{ d^3 p_i }{ (2\pi)^3 2E_i} \right] (2\pi)^4 \delta^4( P_1+ P_2- P_3- P)\nonumber \\
&\times  \sum_{\alpha = 1}^3 \left( f_{\BFB} |M^{ ( \alpha ) }_{\BFB}|^2 +  f_{\BBF} |M^{ ( \alpha ) }_{\BBF}|^2 
  + f_{\FFF} |M^{ ( \alpha ) }_{\FFF}|^2\right)\Theta(|{\mathbf p}_1-{\mathbf p}_3|-k^*)\ ,
\end{align}
where the shorthand notation for the products of the quantum
statistical distribution functions is given in~(\ref{eq:distributions-product}).
In the following, we show in detail how to obtain
\begin{align}
  \label{eq:prodrate-BFB-FFF-app}
  \gammahard & =   \gammahard^{(\BFB)} +  \gammahard^{(\FFF)}+  \gammahard^{(\BBF)} .
\end{align}

\section{Production Rate for BFB Processes}
\label{sec:production-rate-bfb}

To keep better track of the various contributions, we use the notation
 \begin{align}
  \label{eq:hard-prod-rate-intermsofI}
  \gammahard^{(\BFB)} & =   \sum_{\alpha = 1}^3 
  \frac{g_\a^2}{\mplanck^2} \left( 1+\frac{M^2_{ \alpha
      }}{3\mgrav^2}\right)   \sum_{i=1}^3
  \left[ c_{ \BFB,\, i }^{(\alpha)}\, I_{ \BFB}^{|M_i|^2}
    \right]\;,
\end{align}
which was introduced in
Eq.~(\ref{eq:hard-prod-rate-BFB-FFF-BBF-with-coefficients}). The
integrals $I_{\BFB}^{|M_i|^2}$ will be calculated below. The thermal
bath provides a distinguished frame of reference. It is the frame, in
which the quantum statistical distribution functions $f_{\text{B}}$
and $f_{\text{F}}$ have their simple
form~(\ref{eq:bose-fermi-distributions}). Hence, we perform all
integrations in the rest frame of the  plasma.

\subsection[Contribution from \texorpdfstring{$| M_1
  |^2$}{M1}]{Contribution from \texorpdfstring{\boldmath$| M_1
    |^2$}{M1}}
\label{sec:BFB-M1}

We have to compute the integral
 \begin{align}
  \label{eq:I-BFB-1}
  I_{\BFB}^{|M_1|^2} &=  {
    1\over (2\pi)^3 2E}  \int {d\Omega_p \over 4\pi} \int
  \left[\prod_{i=1}^3 \frac{ d^3 p_i }{ (2\pi)^3 2E_i}
  \right] (2\pi)^4 \delta^4( P_1+ P_2- P_3- P)\nonumber \\
  & \times  f_{\BFB} |M_1|^2  \Theta( |\mathbf{p}_1 - \mathbf{p}_3 | -
  k^*  ) \; .
\end{align}
The squared matrix element
\begin{align}
  |M_1|^2 =  -t-2s-\frac{2s^2}{t}
\end{align}
becomes singular for $t = \mbox{$(P_1 - P_3)^2$} \rightarrow 0$. It proves
useful to relate the phase space integrations to the reference three
momentum $\mathbf{k} \equiv \mathbf{p}_1 - \mathbf{p}_3$. The Lorentz
invariant measures can then be written as
 \begin{align}
  \label{eq:delta1}
  \frac{ d^3 p_1}{2E_1} &= \delta(P_1^2) \Theta(E_1)dE_1d^3
  p_1 \nonumber \\
  &= \int d^3 k\, \delta^3({\mathbf k} +{\mathbf p}_3-{\mathbf p}_1)
  \delta(P_1^2) \Theta(E_1)dE_1d^3 p_1 \nonumber \\
  &= \delta(E_1^2-| {\mathbf{ k}}+{\mathbf p}_3
  |^2)\Theta(E_1)dE_1d^3 k
\end{align}
and
 \begin{align}
  \label{eq:delta2}
  \frac{d^3 p_2}{2E_2} \delta^4 (P_1+P_2-P-P_3) &=
  \frac{\delta(E_2 - | \mathbf{p}_2 |)}{2 | \mathbf{p}_2 | } 
  \Theta(E_2) dE_2  d^3 p_2 \nonumber \\
  &\times 
  \delta (E_1+E_2-E-E_3)  \delta^3 (\mathbf{ p}_1 + \mathbf{p}_2 -
  \mathbf{p} -\mathbf{p}_3) \nonumber \\
  &= \frac{\delta(E+E_3-E_1 - | \mathbf{p} +\mathbf{p}_3 -
    \mathbf{p}_1  |)}{2 | \mathbf{p} +\mathbf{p}_3 -
    \mathbf{p}_1  |}  
  \Theta(E+E_3-E_1) \nonumber \\
  &= \delta((E+E_3-E_1)^2- |{\mathbf p}-{\mathbf k}|^2)\Theta(E+E_3-E_1)\;.
\end{align}
In order to perform the angular integrations, we are free to choose
\begin{align}
\label{eq:k-frame}
{\mathbf k} &= k\ (0,0,1)^T,\nonumber \\ 
{\mathbf p} &= E\ (0,\sin\tilde{\theta},\cos\tilde{\theta})^T,\nonumber \\
{\mathbf p}_3 &= E_3\ (\cos\phi\sin\theta,\sin\phi\sin\theta,\cos\theta)^T\;,
\end{align} 
where we have made the approximation that the typical energies in the
thermal bath are much higher than the rest masses of the particles
involved in the scattering.
This yields
\begin{subequations}
  \begin{align}
    s&=
    (P_1+P_2)^2=(P+P_3)^2=2EE_3(1-\sin\theta\sin\phi\sin\tilde{\theta}
    -\cos\theta\cos\tilde{\theta})\;, \\
    t&= (P_1-P_3)^2=(E_1-E_3)^2-k^2\;,
  \end{align}
\end{subequations}
and allows us to write $|M_1(t,s)|^2 $ as $ |
M_1(E,E_1,E_2,k,\theta,\tilde\theta ,\phi) |^2$. For the
$\delta$-functions in (\ref{eq:delta1}) and (\ref{eq:delta2}), we find
 \begin{align}
 \delta((E+E_3-E_1)^2-|{\mathbf p}-{\mathbf k}|^2)&=
\frac{1}{2kE}
\delta\left(\cos\tilde\theta-\frac{E^2+k^2-(E+E_3-E_1)^2}
{2kE}\right)\;,\nonumber \\
\delta(E_1^2-|{\mathbf k}+{\mathbf p}_3|^2)&=
\frac{1}{2kE_3}
\delta\left(\cos\theta-\frac{E_1^2-E_3^2-k^2}{2kE_3}\right)\;.
\end{align}
Equation (\ref{eq:I-BFB-1}) then reads
 \begin{align}
  \label{eq::I-BFB-1-version2}
  I_{\BFB}^{|M_1|^2} &= \frac{1}{2^{14}\pi^9 E^2} \int
  (-d\cos{\tilde\theta}) d\tilde\phi  \int dE_1 \int dE_3
  (-d\cos{\theta}) d\phi
  \int dk d\Omega_k \nonumber \\ 
  & \times  \delta\left(\cos\theta - 
    \frac{E_1^2-E_3^2-k^2}{2kE_3}
  \right)
  \delta \left(\cos\tilde\theta -
    \frac{E^2+k^2-(E+E_3-E_1)^2}{2kE}
  \right)  \nonumber \\
 & \times  f_{\BFB} \left| M_1(E,E_1,E_2,k,\theta,\tilde\theta
   ,\phi)\right |^2  \Theta(k-k^*) \Theta(E_1) \Theta(E_3) \Theta(E+E_3-E_1)\;.
\end{align}
The integrations over $\cos{ \theta }$ and $\cos{ \widetilde \theta}$
yield restrictions on the integration range. Using also the
$\Theta$-functions in (\ref{eq::I-BFB-1-version2}) and positivity of
$k$ we find
  \begin{align}
\label{eq:restricitionsM1}
  \cos\theta < 1 &  \Leftrightarrow  k > E_1 - E_3\;,  \nonumber  \\
  \cos\theta > -1 & \Leftrightarrow   E_3 - E_1 <  k < E_1 + E_3\;,  \nonumber \\
  \cos{ \tilde {\theta}} < 1  & \Leftrightarrow   E_1 - E_3 < k < 2E + E_3
  -E_1\;, \nonumber  \\
  \cos{ \tilde{ \theta} } > -1 & \Leftrightarrow  k >  E_3 - E_1\;.
\end{align}  
The relations (\ref{eq:restricitionsM1}) are all fulfilled with the
inclusion of \mbox{$\Theta(k-|E_1-E_3|)$} and \mbox{$ \Theta(2E+E_3-E_1-
  k)$}. 

Furthermore, it is easy to integrate out all angles and we find
  \begin{align}
  \label{eq:BFB-g-form-forM1}
   I_{\BFB}^{|M_1|^2} &=  \frac{3}{2^{12}\pi^6 E^2} \int  dE_1  dE_3
 dk  f_{\BFB}\,  g_{\BFB}^{|M_1|^2} \, \Omega
\end{align}  
where
  \begin{align}
  \label{eq:def-g-M1}
  g_{\BFB}^{|M_1|^2} &\equiv \left((E_1-E_3)^2-k^2\right)
  \left[
    -1+\frac{2}{3}\frac{E_1^2+E_3^2+2EE_2}{k^2}
  -\frac{(E_3+E_1)^2(E+E_2)^2}{k^4}
  \right]\; ,
\end{align}  
and
  \begin{align}
\label{eq:omegathetas}
\Omega &= \Theta(k-k^*)\Theta(k-|E_1-E_3|)\nonumber \\
&\times \Theta(E_1+E_3- k)\Theta(2E+E_3-E_1- k)\nonumber\\
&\times\Theta(E_1)\Theta(E_3)\Theta(E+E_3-E_1)\;.
\end{align}  
Here, $E_2 = E+ E_3 - E_1 $ is  understood.
Using
\begin{align*}
 \Theta(E_1+E_3-k)& =  1 - \Theta(k-E_1-E_3) 
\end{align*}
and
\begin{align*}
\Theta(k-E_1-E_3) & =    \Theta(k-E_1-E_3)\Theta(k-|E_1-E_3|)\;,
\end{align*}
we can rewrite (\ref{eq:omegathetas}) as
  \begin{align}
\label{eq:omegethetas2}
\Omega &= \Big[\Theta(k-k^*)
\Theta(k-|E_1-E_3|) \Theta(2E+E_3-E_1-k)\nonumber \\
&-\Theta(k-k^*) 
\Theta(k-E_1-E_3) \Theta(2E+E_3-E_1-k)
\Big] \nonumber \\ &\times
\Theta(E_1) \Theta(E_3)\Theta(E+E_3-E_1)\;.
\end{align}  
Now we insert $ 1= \Theta(k^*- E_1-E_3)+\Theta(E_1+E_3-k^*)$ in the
second term in the brackets of (\ref{eq:omegethetas2}) and use 
\begin{align*}
&\Theta(k-k^*) 
\Theta(k-E_1-E_3)\Theta(k^*-E_1-E_3)=
\Theta(k-k^*)\Theta(k^*-E_1-E_3)\;,
\end{align*}
and
\begin{align*}
&\Theta(k-k^*) 
\Theta(k-E_1-E_3)\Theta(E_1+E_3-k^*) =
\Theta(k-E_1-E_3)\Theta(E_1+E_3-k^*)\;.
\end{align*}
This leaves us with a sum of three integrals which we have to compute:
  \begin{align}
  \label{eq:g1plusg2plusg3}
  I_{\BFB}^{|M_1|^2} &= g_{\BFB,\,1}^{|M_1|^2} +
  g_{\BFB,\,2}^{|M_1|^2} + g_{\BFB,\,3}^{|M_1|^2}\;,
\end{align}  
namely,
  \begin{align}
  \label{eq:BFB-g1}
  g_{\BFB,\,1}^{|M_1|^2} & = \frac{3}{2^{12}\pi^6 E^2}  \int_0^\infty  dE_1 \int_0^\infty
  dE_3\, f_{\BFB} 
  \Theta(E+E_3-E_1) \nonumber \\ &\times  \int  dk \, 
  \Theta(k-k^*)  \Theta(k-|E_1-E_3|)  \Theta(2E+E_3-E_1-k)  
  g_{\BFB}^{|M_1|^2}\;,
\end{align}  
  \begin{align}
  \label{eq:BFB-g2}
  g_{\BFB,\,2}^{|M_1|^2} & = -\frac{3}{2^{12}\pi^6 E^2}  \int_0^\infty  dE_1 \int_0^\infty
  dE_3\, f_{\BFB} 
  \Theta(E+E_3-E_1) \Theta( E_1 + E_3 - k^*)  \nonumber \\ &\times  \int  dk \, 
  \Theta(k - E_1 -E_3)  \Theta(2E+E_3-E_1-k)  g_{\BFB}^{|M_1|^2}\;,
\end{align}  
  \begin{align}
  \label{eq:BFB-g3}
  g_{\BFB,\,3}^{|M_1|^2} & = -\frac{3}{2^{12}\pi^6 E^2}  \int_0^\infty  dE_1 \int_0^\infty
  dE_3\, f_{\BFB} 
  \Theta(E+E_3-E_1) \Theta( k^* -E_1 - E_3 )  \nonumber \\ &\times  \int  dk \, 
   \Theta(k-k^*) \Theta(2E+E_3-E_1-k)  g_{\BFB}^{|M_1|^2}\;,
\end{align}  

\smallskip \noindent\textbf{Calculation of \boldmath$ g_{\BFB,\,1}^{|M_1|^2} $}\smallskip

\noindent The $\Theta$-functions for in the integral $ g_{\BFB,\,1}^{|M_1|^2} $
can be further manipulated. We multiply (\ref{eq:BFB-g1}) by 
\begin{align*}
1 &=\Theta(k^*-|E_1-E_3|) + \Theta(|E_1-E_3|-k^*)\;,
\end{align*}
which allows us to split the integral into two
parts
\begin{align}
  g_{\BFB,\,1}^{|M_1|^2}  &\equiv  g_{\BFB,\,11}^{|M_1|^2} +  g_{\BFB,\,12}^{|M_1|^2}\;.
\end{align}
Note that
\begin{align*}
& \Theta(k-k^*)\Theta(k-|E_1-E_3|)\Theta(k^*-|E_1-E_3|) \\
&\quad =\Theta(k-k^*)
\Theta(k^*-|E_1-E_3|)\;,
\end{align*} 
and
\begin{align*}
& \Theta(k-k^*)\Theta(k-|E_1-E_3|)
\Theta(|E_1-E_3|-k^*) \\
& \quad  = \Theta(k-|E_1-E_3|)
\Theta(|E_1-E_3|-k^*)\;,
\end{align*} 
so that we find
  \begin{align}
  \label{eq:BFB-g11}
  g_{\BFB,\,11}^{|M_1|^2} &= \frac{3}{2^{12} \pi^6 E^2}
  \int_0^{\infty} dE_3\int_0^{\infty}dE_1 
  f_{FBF}\Theta(E+E_3-E_1)
  \Theta(k^*-|E_1-E_3|) \nonumber \\ &\times \int dk\,
  \Theta(k-k^*)\Theta(2E+E_3-E_1-k)
  g_{\BFB}^{|M_1|^2}\; ,
\end{align}
and  
\begin{align}
    \label{eq:BFB-g12}
    g_{\BFB,\,12}^{|M_1|^2} &=
    \frac{3}{2^{12}\pi^6}\int_0^{\infty}dE_3\int_0^{\infty}dE_1
    f_{FBF}\Theta(E+E_3-E_1) \Theta(|E_1-E_3|-k^*) \nonumber \\
    &\times \int dk \, \Theta(k-|E_1-E_3|)\Theta(2E+E_3-E_1-k)
    g_{\BFB}^{|M_1|^2}\;.
\end{align}  
The integration over $dk$ in $ g_{\BFB,\,11}^{|M_1|^2}$ yields a lengthy
expression. From  \mbox{$\Theta(k^*-|E_1-E_3|)$} in (\ref{eq:BFB-g11})
we see that the $E_1$ integration only contributes for
\begin{align*}
  E_3-k^*<E_1<E_3+k^*.
\end{align*}
Hence, in the limit $ k^* \rightarrow 0$, we can set $E_1=E_3$ in
$f_{FBF}$ and find
  \begin{align}
 \label{eq:BFB-g11-v2}
 g_{\BFB,\,11}^{|M_1|^2} &=  \frac{1}{48 \pi^6}\ffermi{E}
\int_0^{\infty} dE_3 E_3^2\fbose{E_3} (1 +\fbose{E_3}).
\end{align}  
This integral can be evaluated analytically by choosing the proper
series expansion for the exponentials which reduces the integral to a
Laplace transformation on the summands. The transformed series can
then be resummed \cite{Wheelon:1953}.  For (\ref{eq:BFB-g11-v2}) we
find
  \begin{align}
  \label{eq:laplace1}
  g_{\BFB,\,11}^{|M_1|^2} &=  \frac{T^3 \ffermi{E} }{48 \pi^6} 
  \int_0^{\infty} dx \frac{ x^2 e^x }{ (e^x - 1)^2 } \nonumber\\
  &= \frac{T^3 \ffermi{E} }{48 \pi^6}  \sum_{n=1}^\infty n
  \int_0^\infty dx x^2 e^{-nx} \nonumber\\
  &= \frac{T^3 \ffermi{E} }{24 \pi^6} 
  \sum_{n=1}^\infty \frac{1}{n^2} =  \frac{T^3 \ffermi{E} \zeta(2)
  }{24 \pi^6}\; ,
\end{align}  
where $\zeta(x) $ is the Riemann Zeta function.

Now consider $g_{\BFB,\,12}^{|M_1|^2}$. In order to remove the absolute
value in the \mbox{$\Theta$-function} of (\ref{eq:BFB-g12}) we insert $ 1 =
\Theta(E_1-E_3)+\Theta(E_3-E_1) $ and thereby split
$g_{\BFB,\,12}^{|M_1|^2}$ into two parts,
  \begin{align}
  g_{\BFB,\,12}^{|M_1|^2} =  g_{\BFB,\,121}^{|M_1|^2} +
  g_{\BFB,\,122}^{|M_1|^2}\; ,
\end{align}  
where
  \begin{align} 
  g_{\BFB,\,121}^{|M_1|^2} &= 
  \frac{3}{2^{12}\pi^6 E^2}\int_0^{\infty}dE_3\int_0^{\infty}dE_1 f_{\BFB}\Theta(E+E_3-E_1)
  \Theta(E_1-E_3-k^*) \nonumber \\ &\times \int dk\, 
  \Theta(k-E_1+E_3)\Theta(2E+E_3-E_1-k) g_{\BFB}^{|M_1|^2} \; , 
  \label{eq:BFB-121-M1}  \\
  g_{\BFB,\,122}^{|M_1|^2} &= 
  \frac{3}{2^{12}\pi^6 E^2}\int_0^{\infty}dE_3\int_0^{\infty}dE_1
  f_{\BFB}
  \Theta(E_3-E_1-k^*) \nonumber \\ &\times \int dk \,
  \Theta(k-E_3+E_1)\Theta(2E+E_3-E_1-k) g_{\BFB}^{|M_1|^2} \label{eq:BFB-122-M1}\; .
\end{align}  
Performing the $k$ integration yields 
  \begin{align}
  \label{eq:BFB-121-M1-version2}
  g_{\BFB,\,121}^{|M_1|^2} &= \frac{1}{2^8 \pi^6 E^2}
\int_0^{\infty}dE_3\int_{E_3+k^*}^{E_3+E}dE_1
 f_{\BFB}
\frac{(E_1^2+E_3^2)E_2^2}{E_1-E_3}\;, \\
  \label{eq:BFB-122-M1-version2}
  g_{\BFB,\,122}^{|M_1|^2}  &= -\frac{1}{2^8 \pi^6 E^2}\int_0^{\infty} 
dE_3\int_0^{E_3-k^*} dE_1 f_{\BFB}
\frac{E^2(E_1^2+E_3^2)}{E_1-E_3}\;.
\end{align}  
Since
  \begin{align}
  \label{eq:logdep}
  \frac{ d  }{ dE_1  } \ln{\left( \frac{ | E_1 - E_3  |}{ E_3  } \right) } = {1 \over
  E_1 -E_3}\;,
\end{align}  
an integration by parts will extract the logarithmic dependence on
$k^*$ in (\ref{eq:BFB-121-M1-version2}) and
(\ref{eq:BFB-122-M1-version2}).  Let us consider first equation
(\ref{eq:BFB-121-M1-version2}) which is then given by the sum of the surface
term and the remaining integral, namely,
  \begin{align}
  g_{\BFB,\,121}^{|M_1|^2} =  g_{\BFB,\,121}^{|M_1|^2\,\text{surface}} +  g_{
    \BFB, \,121}^{|M_1|^2\,\text{partial}}\; .
\end{align}  
The surface term is
  \begin{align}
  \label{eq:BFB-121-surface}
  g_{\BFB,\,121}^{|M_1|^2\,\text{surface}} &= \frac{1}{2^8 \pi^6 E^2}
  \int_0^{\infty}dE_3 \left[  \ln{\left( \frac{ | E_1 - E_3  |}{ E_3
        } \right)}  f_{\BFB} (E_1^2+E_3^2) E_2^2  \right]_{E_1 = E_3 +
    k^*}^{E_1 = E_3 + E} \nonumber \\
  &= -  \frac{\ffermi{E}  }{2^7 \pi^6 } \int_0^{\infty}dE_3 
   \ln{\left( \frac{ k^* }{ E_3 } \right)} \frac{ E_3^2\, e^{ E_3/T } }{
  (e^{ E_3/T } -1)^2 }\;,
\end{align}  
where we used $E_2 = E +E_3 -E_1$ and set $k^* \rightarrow 0$ in the
distribution functions after evaluating the borders.  For the analytic
integration of (\ref{eq:BFB-121-surface}) we write the logarithm as
$\ln(k^* / E_3) = \ln k^* - \ln E_3 $. The integration of the first
term is then the same as in (\ref{eq:laplace1}). 

For the second part we have to evaluate an integral of the form
\begin{align}
  \label{eq:laplace-g112-surface-BFB}
 & \int_0^{\infty}dx\,  x^2  \ln x\, \frac{  e^{ x } }{
    (e^{ x } -1)^2 } \nonumber \\
& \qquad  = \sum_{n=1}^{\infty} n \int_0^{\infty}dx\, x^2
  \ln x\;  e^{ - n x } \nonumber \\
&  \qquad=  \sum_{n=1}^{\infty} \left( \frac{3}{n^2} - \frac{2 \gamma }{ n^2
    }  -  \frac{ 2 \ln n }{n^2}  \right) =   { \pi^2 \over 2} -
    {\gamma \pi^2 \over 3} -{ 2 \zeta'(2)} \; .
\end{align}
Thus, for (\ref{eq:BFB-121-surface}) we find
  \begin{align}
  \label{eq:BFB-g121-surface-final}
   g_{\BFB,\,121}^{|M_1|^2\,\text{surface}} &= \frac{f_F(E) T^3}{
     3\cdot 2^7
     \pi^4}
    \left[  \ln{\left( \frac{ T }{ k^* } \right)} 
     + {3 \over 2} - \gamma +
         {\zeta '(2) \over \zeta(2)}  \right]\;,
\end{align}  
where we used $ \zeta(2) = \pi^2 / 6 $. In the remaining integral, we can
set $k^* \rightarrow 0$ and obtain
  \begin{align}
  g_{ \BFB, \,121}^{|M_1|^2\,\text{partial}}
  &= -  \frac{1}{2^8 \pi^6 E^2}    \int_0^{\infty}dE_3
  \int_0^{ E +E_3  }dE_1\,
  \ln\left(\frac{|E_1-E_3|}{E_3}\right)\nonumber \\ &\times
  \Theta(E_1-E_3)
  \frac{d}{dE_1}\left[ f_{\BFB}E_2^2(E_1^2+E_3^2)\right]\;.
\end{align}  
Now we turn to the second integral
(\ref{eq:BFB-122-M1-version2}). Note that for
the surface term 
  \begin{align}
  \label{eq:BFB-122-surface}
  g_{\BFB,\,122}^{|M_1|^2\,\text{surface}} &= \frac{1}{2^8 \pi^6}
  \int_0^{\infty}dE_3 \left[  \ln{\left( \frac{ | E_1 - E_3  |}{ E_3
        } \right)}  f_{\BFB} (E_1^2+E_3^2)
  \right]_{E_1 = 0}^{E_1 = E_3 - k^*} 
\end{align}  
we get an additional contribution from the lower integral border,
  \begin{align}
  \label{eq:limit-g122-borders}
  \lim_{E_1\rightarrow 0} \left[ \ln{\left( \frac{ | E_1 - E_3  |}{ E_3
      } \right)} \fbose{E_1} (E_1^2+E_3^2) \right]& = - E_3 \,T\;.
\end{align}  
 In fact, the upper border yields the same contribution as in
(\ref{eq:BFB-g121-surface-final}), so that we can write
  \begin{align}
  \label{eq:g122surface-plus-rest}
  g_{\BFB,\,122}^{|M_1|^2\,\text{surface}} &=
  g_{\BFB,\,121}^{|M_1|^2\,\text{surface}} +  g_{\BFB,\,122}^{|M_1|^2\,\text{rest}} 
\end{align}  
with (in the limit $ k^* \rightarrow 0$)
  \begin{align}
  \label{eq:BFB-g122-rest}
  g_{\BFB,\,122}^{|M_1|^2\,\text{rest}} &= - \frac{T}{2^8 \pi^6 }
  \int_0^{\infty}dE_3\, E_3
  f_F(E + E_3) [ 1 +\fbose{E_3} ] \nonumber \\
  &= - \frac{T\ffermi{E}}{2^8 \pi^6 } \int_0^{\infty}dE_3\, E_3 \left[
    \fbose{E_3} + \ffermi{E + E_3} \right] \; .
\end{align} 
This can be written as with $E_3= x\,T$ and $a \equiv E/T$ as 
 \begin{align}
  \label{eq:BFB-g122-rest-v2}
  g_{\BFB,\,122}^{|M_1|^2\,\text{rest}} &= 
  - \frac{T^3 \ffermi{E} }{2^8 \pi^6 }
  \sum_{n=1}^\infty  \int_0^{\infty}dx\, x \left[ e^{-nx} - (-1)^{n}
  e^{-nx} e^{-na} \right] \nonumber \\
 &=  - \frac{T^3 \ffermi{E} }{2^8 \pi^6 }  \sum_{n=1}^\infty 
 \left[ \frac{1}{n^2} -    (-1)^{n}  \frac{e^{-na} }{n^2} \right]\;,
\end{align}
and be resummed to
  \begin{align}
  \label{eq:g122-rest-final}
   g_{\BFB,\,122}^{|M_1|^2\,\text{rest}} &=   \frac{T^3\ffermi{E} }{2^8
     \pi^6 } \left[ \text{Li}_2\left(-e^{-{E/ T}}\right)  - \zeta(2)
   \right]\; .
\end{align}  
The dilogarithm $ \text{Li}_2(x)$ is defined as
\begin{align}
  \label{eq:dilog}
   \text{Li}_2(x ) & = - \int_{ 0 }^x dt {\ln{(1-t)}\over t}\; .
\end{align}
The authors of \cite{Bolz:2000fu} find a contribution like
(\ref{eq:g122-rest-final}) for the \mbox{$\BFB$-part} of the gravitino
production rate, but do not specify from where it emerges. In
contrast, we find below in~(\ref{eq:g2-surface-final}) that
$ g_{\BFB,\,122}^{|M_1|^2\,\text{rest}} $ exactly cancels out and thus we do not get
(\ref{eq:g122-rest-final}) as part of the production rate.

For the remaining integral, we find for \mbox{$k^* \rightarrow 0$ }
after an integration by parts
  \begin{align}
  g_{ \BFB, \,122}^{|M_1|^2\,\text{partial}}
  &=  \frac{1}{2^8 \pi^6 E^2} \int_0^{\infty} dE_3
  \int_0^{ E +E_3  }dE_1\,
  \ln\left(\frac{|E_1-E_3|}{E_3}\right)\nonumber \\ &\times
  \Theta(E_3-E_1)
  \frac{d}{dE_1}\left[ f_{\BFB} E^2(E_1^2+E_3^2)\right]\;.
\end{align}  

\smallskip \noindent\textbf{Calculation of \boldmath$ g_{\BFB,\,2}^{|M_1|^2} $}\smallskip

\noindent In equation (\ref{eq:BFB-g2}), the $\Theta$-functions allow for a contribution
of the integrand only if
  \begin{align*}
E_1+E_3 < 2E+E_3-E_1 \Leftrightarrow E_1 < E\;.
\end{align*}  
We can take the limit $k^* \rightarrow 0$ and perform the $k$
integration which gives
  \begin{align}
 g_{\BFB,\,2}^{|M_1|^2} &=  \frac{1}{2^{8}\pi^6 E^2}  \int_0^\infty dE_3 \int_0^{E}dE_1 f_{\BFB} (E_1-E)\left[(E+E_1)E_3+E_1(E-E_1)\right]\;.
\end{align}  
This can be rewritten with $E_2=E+E_3-E_1$ as
  \begin{align}
  g_{\BFB,\,2}^{|M_1|^2}& = - \frac{1}{2^{8}\pi^6 E^2} \int_0^\infty
  dE_3 \int_0^{E}dE_1 f_{\BFB}  {1\over E_1 - E_3} \left( E_1^2E_2^2
    -E^2 E_3^2   \right)\; .
\end{align}  
An integration by parts yields a  surface term and we write
  \begin{align}
  \label{eq:BFB-g2-sum}
  g_{\BFB,\,2}^{|M_1|^2}& =  g_{\BFB,\,2}^{|M_1|^2\,\text{surface}}  +  
  g_{\BFB,\,2}^{|M_1|^2\,\text{partial}} .
\end{align}  
For the lower border $E_1= 0$ of the surface term
  \begin{align}
  \label{eq:g2-surface-borders}
  g_{\BFB,\,2}^{|M_1|^2\,\text{surface}} & = - \frac{1}{2^{8}\pi^6 E^2} \int_0^\infty
  dE_3 \left[  f_{\BFB} \ln{\left(\frac{|E_1-E_3|}{E_3}\right)} 
  \right. \nonumber \\ &\times \left. \vphantom{ \ln{\left(\frac{|E_1-E_3|}{E_3}\right)}}
    ( E_1^2 (E+E_3-E_1)^2 - E^2 E_3^2 ) \right]_{E_1=0}^{E_1=E}\; ,
\end{align}  
the Bose-distribution $f_B(E1)$ becomes singular. In the limit $E_1 \rightarrow
0$, we obtain
  \begin{align}
  \label{eq:g2-surface-final}
  g_{\BFB,\,2}^{|M_1|^2\,\text{surface}} & =  \frac{T}{2^{8}\pi^6 E^2} \int_0^\infty
  dE_3 E_3 f_F(E + E_3) [ 1 +\fbose{E_3} ]  \nonumber \\ &= - g_{\BFB,\,121}^{|M_1|^2\,\text{rest}}\;.
\end{align}  
We see that the arising dilogarithm (and $\zeta$-function) from the
integral (\ref{eq:BFB-g122-rest}) cancels with
(\ref{eq:g2-surface-final}).

In the partial term,  we include $\Theta(E - E_1)$ in order to change
the upper integral border and find
  \begin{align}
  \label{eq:BFB-M1-g2-partial}
    g_{\BFB,\,2}^{|M_1|^2\,\text{partial}}& =  \frac{1}{2^{8}\pi^6 E^2} \int_0^\infty
  dE_3  \int_0^{E+E_3}
  dE_1 \ln{\left(\frac{|E_1-E_3|}{E_3}\right)} 
  \nonumber \\ &\times \Theta(E - E_1)  {d\over dE_1} \left[
     f_{\BFB} (   E_1^2 E_2^2 - E^2 E_3^2 ) \right].
\end{align}

\smallskip \noindent\textbf{Calculation of \boldmath$ g_{\BFB,\,3}^{|M_1|^2} $}\smallskip

The $k$ integration in (\ref{eq:BFB-g3}) can be carried out directly
and leads a lenghty result. From the $\Theta$-function $\Theta( k^*
-E_1 - E_3 )$ in (\ref{eq:BFB-g3}) it follows that $E_1 < k^*$ and
$E_2 < k^*$. Thus, one finds that the integration over $k$ yields an
expression of order $k^*$ and hence
\begin{align}
  \label{eq:BFB-g3-final}
   g_{\BFB,\,3}^{|M_1|^2} & = 0 \quad \text{for} \quad k^*\rightarrow 0\; .
\end{align}

\smallskip \noindent\textbf{Result for \boldmath$I_{\BFB}^{|M_1|^2} $}\smallskip

Collecting the results, we find for the contributions from $|M_1|^2$
  \begin{align}
 \label{eq:I-BFB-M1-final}
 I_{\BFB}^{|M_1|^2} &= g_{\BFB,\,11}^{|M_1|^2} + 2\cdot
 g_{\BFB,\,121}^{|M_1|^2\,\text{surface}} + g_{ \BFB,
   \,121}^{|M_1|^2\,\text{partial}}+ g_{ \BFB,
   \,122}^{|M_1|^2\,\text{partial}} +
 g_{\BFB,\,2}^{|M_1|^2\,\text{partial}} \nonumber \\
 & =  \frac{T^3 \ffermi{E}}{192 \p^4} \left[ \ln{\left( \frac{ T }{ k^* }
     \right)} + {17 \over 6} - \gamma + {\zeta '(2) \over \zeta(2)} \right] \nonumber\\
  & \quad + \frac{1}{256\pi^6 } \int_0^\infty
  dE_3  \int_0^{E+E_3}
  dE_1  \ln{\left(\frac{|E_1-E_3|}{E_3}\right)} \nonumber \\
  & \quad \times \Bigg\{ - \Theta(E_1-E_3)
  \frac{d}{dE_1}\left[ f_{\BFB}\frac{E_2^2}{E^2} (E_1^2+E_3^2)\right] \nonumber \\
  & \; \hphantom{ \quad\times \Bigg\{ } 
    +  \Theta(E_3-E_1)
  \frac{d}{dE_1}\left[ f_{\BFB} (E_1^2+E_3^2)\right] \nonumber \\
  &  \; \hphantom{ \quad\times \Bigg\{ }   +  \Theta(E - E_1)  {d\over dE_1} \left[
     f_{\BFB} (  \frac{ E_1^2 E_2^2}{E^2} -  E_3^2 ) \right] \Bigg\}\; .
\end{align}

This concludes our analysis of $|M_1|^2$ and we
can turn to the integrals containing $|M_2|^2$.

\subsection[Contribution from \texorpdfstring{$| M_2 |^2$}{M2}]{Contribution from \texorpdfstring{\boldmath$| M_2 |^2$}{M2}}
\label{sec:BFB-M2}

The expression to evaluate is
  \begin{align}
  \label{eq:I-BFB-2}
  I_{\BFB}^{|M_2|^2} &=  {
    1\over (2\pi)^3 2E}  \int {d\Omega_p \over 4\pi} \int
  \left[\prod_{i=1}^3 \frac{ d^3 p_i }{ (2\pi)^3 2E_i}
  \right] (2\pi)^4 \delta^4( P_1+ P_2- P- P_3)\nonumber \\
  & \times  f_{\BFB} |M_2|^2 \Theta( |\mathbf{p}_1 - \mathbf{p}_3 | - k^*  )\; .
\end{align}  
Since there is no singular behavior of $ |M_2|^2 = t$ for $t
\rightarrow 0$, we  set $k^* \rightarrow 0$ from the very beginning.
We use the center of mass momentum $\mathbf{q}\equiv \mathbf{p} +
\mathbf{p_3}$ as a reference momentum which yields in an analogous
manner to (\ref{eq:delta1}) and (\ref{eq:delta2}) the integration
measures
  \begin{align}
  \label{eq:M3delta1}
  &\frac{d^3 p_1}{2E_1} \delta^4 (P_1+P_2-P-P_3) \nonumber \\ & \quad=
  \delta((E+E_3-E_2)^2-|{\mathbf q}-{\mathbf p_2}|^2) \Theta(E + E_3 -E_2)\;,
\end{align}  
and
  \begin{align}
  \label{eq:M3delta2}
  \frac{ d^3 p_3}{2E_3} &= 
  \delta(E_3^2-| {\mathbf{ q }} -{\mathbf p}|^2)\Theta(E_3) dE_3 d^3q\;.
\end{align}  
Rotational invariance allows us to choose a frame where
  \begin{align}
  \label{eq:q-frame}
{\mathbf q} &= q\ (0,0,1)^T\;,\nonumber \\ 
{\mathbf p} &= E\ (0,\sin\tilde{\theta},\cos\tilde{\theta})^T\;,\nonumber \\
{\mathbf p}_2 &= E_2\ (\cos\phi \sin\theta,\sin\phi \sin\theta,\cos\theta)^T\;.
\end{align}  
Again, we have assumed that the typical energies in the plasma are
much higher as compared to the rest masses of the external particles
which are involved in the scattering. The squared matrix element $|M_2|^2
= t$ in system (\ref{eq:q-frame}) reads
\begin{align}
  \label{eq:t-inq-system}
  |M_2|^2 &= -2 E E_2 \left( 1- \cos\theta \cos\tilde\theta -
    \sin\theta \sin\tilde\theta \sin\phi \right)\; .
\end{align}
For the $\delta$-functions we find
  \begin{align}
 \delta((E+E_3-E_2)^2-|{\mathbf q}-{\mathbf p_2}|^2) &=
\frac{1}{2q E_2}
\delta\left(\cos\theta-\frac{E^2+q^2-(E+E_3-E_2)^2}
{2qE_2}\right)\;,\nonumber \\
 \delta(E_3^2-| {\mathbf{ q }} -{\mathbf p}|^2)&=
\frac{1}{2qE}
\delta\left(\cos\tilde\theta-\frac{E^2-E_3^2+q^2}{2qE}\right)\;.
\end{align}  
Equation (\ref{eq:I-BFB-2}) thus reads 
  \begin{align}
  \label{eq::I-BFB-2-version2}
  I_{\BFB}^{|M_2|^2} &= \frac{1}{2^{14}\pi^9 E^2} \int
  (-d\cos{\tilde\theta}) d\tilde\phi 
  (-d\cos{\theta}) d\phi dE_2 dE_3
   dq d\Omega_q \nonumber \\ 
  & \times  \delta\left(\cos\theta-\frac{E^2+q^2-(E+E_3-E_2)^2}{2qE_2} \right) 
  \delta\left(\cos\tilde\theta-\frac{E^2-E_3^2+q^2}{2qE}\right)  \nonumber \\
 & \times  f_{\BFB} | M_2 |^2  \Theta(E_2) \Theta(E_3) \Theta(E+E_3-E_2)\;.
\end{align}  
From the integration over the $\delta$-functions, we find the
phase-space restrictions 
  \begin{align}
\label{eq:restricitionsM2}
  \cos\theta < 1  & \Leftrightarrow  2E_2 -E_3 -E  <  q  <  E + E_3\;, \nonumber \\
  \cos\theta> -1 & \Leftrightarrow   q >  E + E_3 -2 E_2\; , \nonumber \\
  \cos{ \tilde {\theta}} < 1  & \Leftrightarrow   E - E_3 < q < E + E_3\;, \nonumber \\
  \cos{ \tilde{ \theta} }> -1 & \Leftrightarrow  q >  E_3 - E\;,
\end{align}  
which yield the $\Theta$-functions \mbox{$\Theta(q-|E-E_3|)$}, \mbox{$
  \Theta(E+E_3- q)$} and \mbox{$ \Theta(q- |2E_2 - E_3
  -E|)$}. We can perform the remaining angular integrations and
find for Eq.~(\ref{eq::I-BFB-2-version2})
  \begin{align}
  \label{eq:I-BFB-M2-version3}
   I_{\BFB}^{|M_2|^2} &=  \frac{1}{2^{11}\pi^6 E^2} \int  dE_2  dE_3
 dq  f_{\BFB}\,  g_{\BFB}^{|M_2|^2} \, \Omega
\end{align}  
with
  \begin{align}
 g_{\BFB}^{|M_2|^2} &= \left((E_3+E)^2-q^2\right) \left[-1 + \frac{E_3^2-2 E_2
   E_3-E^2+2 E_2 E}{q^2}\right]   
\end{align}  
and 
  \begin{align}
  \label{eq:omegathetasM2}
  \Omega &= \Theta(q- |2E_2 - E_3  -E|)  \Theta(q-|E-E_3|) \nonumber \\
  &\times \Theta(E+E_3- q) \Theta(E+E_3-E_2)   \Theta(E_2) \Theta(E_3)  \;.
\end{align}  
Now we use
\begin{align*}
  \Theta(q- |2E_2 - E_3  -E|) &= 1 -  \Theta(|2E_2 - E_3  -E| -q)
\end{align*}
and
\begin{align*}
 \Theta(|2E_2 - E_3  -E| -q) \Theta(E+E_3- q) &=  \Theta(|2E_2 - E_3  -E| -q)
\end{align*}
to split (\ref{eq:I-BFB-M2-version3}) into two parts, namely,
  \begin{align}
  \label{eq:M2-g1plusg2}
   I_{\BFB}^{|M_2|^2} &=  g_{\BFB,\,1}^{|M_2|^2} +  g_{\BFB,\,2}^{|M_2|^2}\;,
\end{align}  
with
  \begin{align}\label{eq:BFB-M2-g1}
  g_{\BFB,\,1}^{|M_2|^2} &=  \frac{1}{2^{11}\pi^6 E^2}
  \int_{0}^{\infty} dE_2 \int_{0}^{\infty}  dE_3 \Theta(E+E_3-E_2)\nonumber \\
  &\times
  \int dq  \Theta(q-|E-E_3|) \Theta(E+E_3- q) f_{\BFB} g_{\BFB}^{|M_2|^2},
\end{align}  
  \begin{align}\label{eq:BFB-M2-g2}
  g_{\BFB,\,2}^{|M_2|^2} &= - \frac{1}{2^{11}\pi^6 E^2}
  \int_{0}^{\infty} dE_2 \int_{0}^{\infty}  dE_3 \Theta(E+E_3-E_2)\nonumber \\
  &\times
  \int dq  \Theta(q-|E-E_3|) \Theta(|2E_2 - E_3  -E| -q) f_{\BFB} g_{\BFB}^{|M_2|^2}.
\end{align}  
We can take away the absolute value in $ \Theta(q-|E-E_3|)$ if we
insert
\begin{align}
\label{eq:theta-insertion-for-M2-BFB}
  1 &=  \Theta(E - E_3)  +  \Theta(E_3 - E)\;,
\end{align}
which splits  again the integrals (\ref{eq:BFB-M2-g1}) and (\ref{eq:BFB-M2-g2})  into
  \begin{align}
   g_{\BFB,\,1}^{|M_2|^2} &= g_{\BFB,\,11}^{|M_2|^2} +
   g_{\BFB,\,12}^{|M_2|^2}\,,  \label{eq:BFB-M2-g1-v2} \\
    g_{\BFB,\,2}^{|M_2|^2} &= g_{\BFB,\,21}^{|M_2|^2} +
    g_{\BFB,\,22}^{|M_2|^2}\,. \label{eq:BFB-M2-g2-v2}
\end{align}  
Let us consider first (\ref{eq:BFB-M2-g1-v2}) which then reads
  \begin{align}
  g_{\BFB,\,11}^{|M_2|^2} &= \frac{1}{2^{11}\pi^6 E^2}
  \int_{0}^{\infty} dE_3 \int_{0}^{E+E_3}  dE_2
  f_{\BFB}  
  \int_{E-E_3}^{E+E_3} dq\, \Theta(E - E_3)  g_{\BFB}^{|M_2|^2}\;,  \label{eq:M2-g11} \\
 g_{\BFB,\,12}^{|M_2|^2} &=  \frac{1}{2^{11}\pi^6 E^2}
  \int_{0}^{\infty} dE_3 \int_{0}^{E+E_3}  dE_2 
   f_{\BFB}  
  \int_{E_3-E}^{E+E_3} dq\, \Theta(E_3 - E)  g_{\BFB}^{|M_2|^2}\;.\label{eq:M2-g12}
\end{align}  
Integration over $q$ gives
  \begin{align}
  g_{\BFB,\,11}^{|M_2|^2} &=
   \frac{1}{2^{8}\pi^6 E^2} \int_{0}^{\infty} dE_3 \int_{0}^{E+E_3}  dE_2
   \Theta(E - E_3) f_{\BFB} E_3^2 \left( {E_3 \over 3} - E_1   \right)\;,
 \\
   g_{\BFB,\,12}^{|M_2|^2} &= 
   \frac{1}{2^{8}\pi^6 E^2} \int_{0}^{\infty} dE_3 \int_{0}^{E+E_3}  dE_2
    \Theta(E_3 - E) f_{\BFB} E^2 \left( {E \over 3} - E_2  \right)\;.
\end{align}  
Now we turn to  $g_{\BFB,\,2}^{|M_2|^2}$. The insertion
(\ref{eq:theta-insertion-for-M2-BFB}) yields
  \begin{align}
    g_{\BFB,\,21}^{|M_2|^2} &= - \frac{1}{2^{11}\pi^6 E^2}
    \int_{0}^{\infty} dE_3 \int_{0}^{E+E_3} dE_2 f_{\BFB} \nonumber \\
    &\times
    \int dq\, \Theta(E - E_3)  \Theta(q - E + E_3) \Theta(|2E_2 - E_3  -E| -q) g_{\BFB}^{|M_2|^2}\;,  \label{eq:M2-g21} \\
    g_{\BFB,\,22}^{|M_2|^2} &= - \frac{1}{2^{11}\pi^6 E^2}
    \int_{0}^{\infty} dE_3 \int_{0}^{E+E_3} dE_2 f_{\BFB} \nonumber \\
    &\times \int dq\, \Theta(E_3 - E)\Theta(q + E - E_3) \Theta(|2E_2
    - E_3 -E| -q) g_{\BFB}^{|M_2|^2}\;.   \label{eq:M2-g22}
\end{align}  
We proceed in the same manner as in
(\ref{eq:theta-insertion-for-M2-BFB}) and  insert 
  \begin{align}
  1 =  \Theta(E_3 + E-2E_2 -q)+ \Theta(2E_2 - E_3 -E -q)\;,
\end{align}  
which leads to four integrals
  \begin{align}
  g_{\BFB,\,21}^{|M_2|^2} &=  g_{\BFB,\,211}^{|M_2|^2} +
  g_{\BFB,\,212}^{|M_2|^2} \; , \\
  g_{\BFB,\,22}^{|M_2|^2} & =  g_{\BFB,\,221}^{|M_2|^2}
  +g_{\BFB,\,222}^{|M_2|^2} \; .
\end{align}  
 Note that
\begin{align*}
 \Theta(q- E + E_3 ) \Theta(E + E_3 -2 E_2 -q) & \Rightarrow  E_2 <
 E_3 \;, \\
 \Theta(q- E + E_3 ) \Theta(2 E_2 -E - E_3 -q)  & \Rightarrow  E <
 E_2\; , \\
  \Theta(q + E - E_3 ) \Theta(E + E_3 -2 E_2 -q) & \Rightarrow  E_2 < E \;, \\
 \Theta(q + E - E_3 ) \Theta(2 E_2 -E - E_3 -q)  & \Rightarrow  E_3 < E_2\; ,
\end{align*}
so that we include the corresponding $\Theta$-functions, and then
integrate over $q$, namely,
  \begin{align}
  \label{eq:M2-g211}
  g_{\BFB,\,211}^{|M_2|^2} &= - \frac{1}{2^{11}\pi^6 E^2}
  \int_{0}^{\infty} dE_3 \int_{0}^{E+E_3}  dE_2\,   f_{\BFB}   
  \int_{E-E_3}^{E+E_3-2E_2} dq\,  \Theta(E_3-E_2) \Theta(E - E_3)
    g_{\BFB}^{|M_2|^2}  
  \nonumber \\  &=
   \frac{1}{3}\frac{1}{ 2^{8}\pi^6 E^2}  \int_{0}^{\infty} dE_3
  \int_{0}^{E+E_3}  dE_2 \Theta(E_3-E_2) \Theta(E - E_3)\nonumber \\  &\times
   f_{\BFB}( E_2-E_3  ) \left[ (E_2-E_3)(E_2+2 E_3) - 3(E_2+E_3)E  \right]\;,
\end{align}  
  \begin{align}
  \label{eq:M2-g212}
  g_{\BFB,\,212}^{|M_2|^2} &= - \frac{1}{2^{11}\pi^6 E^2}
  \int_{0}^{\infty} dE_3 \int_{0}^{E+E_3}  dE_2 \,
  f_{\BFB}  
  \int_{E-E_3}^{2E_2-E-E_3} dq\, \Theta(E_2-E) \Theta(E - E_3) g_{\BFB}^{|M_2|^2}  
  \nonumber \\  &=
  - \frac{1}{3} \frac{1}{ 2^{8}\pi^6 E^2}  \int_{0}^{\infty} dE_3
  \int_{0}^{E+E_3}  dE_2 \Theta(E_2-E) \Theta(E - E_3)
   f_{\BFB}(E_2 - E)^3\; ,
\end{align}  
  \begin{align}
  \label{eq:M2-g221}
  g_{\BFB,\,221}^{|M_2|^2} &= - \frac{1}{2^{11}\pi^6 E^2}
  \int_{0}^{\infty} dE_3 \int_{0}^{E+E_3}  dE_2 \,
  f_{\BFB}  
  \int_{E_3-E}^{E+E_3-2E_2} dq\, \Theta(E-E_2) \Theta(E_3 - E) g_{\BFB}^{|M_2|^2}  
  \nonumber \\  &=  \frac{1}{3} \frac{1}{ 2^{8}\pi^6 E^2}  \int_{0}^{\infty} dE_3
  \int_{0}^{E+E_3}  dE_2\Theta(E-E_2) \Theta(E_3 - E)
   f_{\BFB}(E_2 - E)^3\;,
\end{align}  
  \begin{align}
  \label{eq:M2-g222}
  g_{\BFB,\,222}^{|M_2|^2} &= - \frac{1}{2^{11}\pi^6 E^2}
  \int_{0}^{\infty} dE_3 \int_{0}^{E+E_3}  dE_2 \;
  f_{\BFB}  
  \int_{E_3-E}^{2E_2-E-E_3} dq\, \Theta(E_2-E_3) \Theta(E_3 - E)  g_{\BFB}^{|M_2|^2}  
  \nonumber \\  &= - \frac{1}{3}  \frac{1}{ 2^{8}\pi^6 E^2}  \int_{0}^{\infty} dE_3
  \int_{0}^{E+E_3}  dE_2 \Theta(E_2-E_3) \Theta(E_3 - E)\nonumber \\  &\times
   f_{\BFB}( E_2-E_3  ) \left[  (E_2-E_3)(E_2+2 E_3) - 3(E_2+E_3)E  \right],
\end{align}  

\smallskip \noindent\textbf{Result for \boldmath$I_{\BFB}^{|M_2|^2} $}\smallskip

For the contributions from $|M_2|^2$ we thus find
\begin{align}
 \label{eq:I-BFB-M2-final}
 I_{\BFB}^{|M_2|^2} &= g_{\BFB,\,11}^{|M_2|^2} + 
 g_{\BFB,\,12}^{|M_2|^2} + g_{\BFB,\,211}^{|M_2|^2}
 + g_{\BFB,\,212}^{|M_2|^2} +g_{\BFB,\,221}^{|M_2|^2} +g_{\BFB,\,222}^{|M_2|^2}
 \nonumber \\
 &= \frac{1}{256\pi^6} \int_{0}^{\infty} dE_3 \int_{0}^{E+E_3} dE_2
 f_{\BFB}
 \nonumber \\
 & \times \Bigg\{ + \Theta(E - E_3) \frac{ E_3^2}{E^2} \left( {E_3 \over 3} - E_1
 \right)  + \Theta(E_3 - E) 
  \left( {E \over 3} - E_2  \right) \nonumber \\
 &  \hphantom{ \quad\times \Bigg\{ }
 +   \Theta(E_3-E_2) \Theta(E - E_3)
 \frac{ E_2-E_3  }{ 3 E^2} \left[  (E_2-E_3)(E_2+2 E_3) - 3(E_2+E_3)E  \right] \nonumber \\
 &  \hphantom{ \quad\times \Bigg\{ }
 -  \Theta(E_2-E) \Theta(E - E_3) \frac{(E_2 - E)^3}{3 E^2} \nonumber \\
 &  \hphantom{ \quad\times \Bigg\{ }
 +  \Theta(E-E_2) \Theta(E_3 - E) \frac{(E_2 - E)^3}{3 E^2}\nonumber \\
 &  \hphantom{ \quad\times \Bigg\{ }
 -  \Theta(E_2-E_3) \Theta(E_3 - E)
   \frac{ E_2-E_3  }{3 E^2} \left[  (E_2-E_3)(E_2+2 E_3) - 3(E_2+E_3)E
   \right] \Bigg\}\; .
\end{align}

This concludes the calculation for the BFB-processes since there is
no contribution from the matrix element $| M_3 |^2$ because of  $c_{
  \BFB,\, 3 }^{(\alpha)} =0 $.

\section{Production Rate for FFF Processes}
\label{sec:production-rate-FFF}

We now turn to the processes where all in- and outgoing particles for the
$2\rightarrow 2$ scatterings are fermions. We have to compute
  \begin{align}
  \label{eq:hard-prod-rate-FFF-app}
  \gammahard^{(\FFF)} & =   \sum_{\alpha = 1}^3 
  \frac{g_\a^2}{\mplanck^2} \left( 1+\frac{M^2_{ \alpha
      }}{3\mgrav^2}\right)   \sum_{i=1}^3
  \left[ c_{ \FFF,\, i }^{(\alpha)}\, I_{ \FFF}^{|M_i|^2}
    \right].
\end{align}  

\subsection[Contribution from  \texorpdfstring{$| M_1 |^2$}{M1}]{Contribution from \texorpdfstring{\boldmath$| M_1 |^2$}{M1}}
\label{sec:FFF-M1}

Let us first discuss matrix element $|M_1|^2$. The solution of the
integral
  \begin{align}
  \label{eq:I-FFF-1}
  I_{\FFF}^{|M_1|^2} &=  {
    1\over (2\pi)^3 2E}  \int {d\Omega_p \over 4\pi} \int
  \left[\prod_{i=1}^3 \frac{ d^3 p_i }{ (2\pi)^3 2E_i}
  \right] (2\pi)^4 \delta^4( P_1+ P_2- P- P_3)\nonumber \\
  & \times f_{\FFF} |M_1|^2  \Theta( |\mathbf{p}_1 - \mathbf{p}_3 | - k^*  )
\end{align}  
is obtained as in section~\ref{sec:BFB-M1}. Instead of $f_{\BFB}$, we
now have to integrate over the statistical factor $f_{\FFF}$
(\ref{eq:distributions-product}).

The analogous expression to (\ref{eq:BFB-g11-v2}) is 
\begin{align}
 \label{eq:FFF-g11-v2}
 g_{\FFF,\,11}^{|M_1|^2} &=  \frac{1}{48 \pi^6}\ffermi{E}
\int_0^{\infty} dE_3 E_3^2\ffermi{E_3} (1 -\ffermi{E_3})\;.
\end{align}  
and the $E_3$ integration yields
  \begin{align}
  g_{\FFF,\,11}^{|M_1|^2} &= 
  \frac{T^3 \ffermi{E} }{48 \pi^6} 
  \int_0^{\infty} dx \frac{ x^2 e^x }{ (e^x + 1)^2 } 
  \nonumber\\
  &= - \frac{T^3 \ffermi{E} }{48 \pi^6}  \sum_{n=1}^\infty (-1)^n   n
  \int_0^\infty dx x^2 e^{-nx} 
  \nonumber\\
  &= - \frac{T^3 \ffermi{E} }{48 \pi^6} 
  \sum_{n=1}^\infty (-1)^n  \frac{2}{n^2} =  \frac{T^3 \ffermi{E} \zeta(2) }{48 \pi^6}\;.
\end{align}  
Now consider $g_{\FFF,\,12}^{|M_1|^2}$ which we have written in
section~\ref{sec:BFB-M1} as two parts
  \begin{align}
  g_{\FFF,\,12}^{|M_1|^2} & =  g_{\FFF,\,121}^{|M_1|^2} +
  g_{\FFF,\,122}^{|M_1|^2}\; ,
\end{align}  
namely, [cf.~(\ref{eq:BFB-121-M1-version2}), (\ref{eq:BFB-122-M1-version2})],
  \begin{align}
  \label{eq:FFF-121-M1-version2}
  g_{\FFF,\,121}^{|M_1|^2} &= \frac{1}{2^8 \pi^6 E^2}
\int_0^{\infty}dE_3\int_{E_3+k^*}^{E_3+E}dE_1
 f_{\FFF}
\frac{(E_1^2+E_3^2)E_2^2}{E_1-E_3}\;, \\
  \label{eq:FFF-122-M1-version2}
  g_{\FFF,\,122}^{|M_1|^2}  &= -\frac{1}{2^8 \pi^6 E^2}\int_0^{\infty} 
dE_3\int_0^{E_3-k^*} dE_1 f_{\FFF}
\frac{E^2(E_1^2+E_3^2)}{E_1-E_3}\;.
\end{align}  
An integration by parts will exhibit the logarithmic $k^*$ dependence in
the surface term so that we find for (\ref{eq:FFF-121-M1-version2})
  \begin{align}
  g_{\FFF,\,121}^{|M_1|^2} &=  g_{\FFF,\,121}^{|M_1|^2\,\text{surface}} +  g_{
    \FFF, \,121}^{|M_1|^2\,\text{partial}}\; ,
\end{align}  
where the surface term reads
  \begin{align}
  \label{eq:FFF-121-surface}
  g_{\FFF,\,121}^{|M_1|^2\,\text{surface}} &= \frac{1}{2^8 \pi^6 E^2}
  \int_0^{\infty}dE_3 \left[  \ln{\left( \frac{ | E_1 - E_3  |}{ E_3
        } \right)}  f_{\FFF} (E_1^2+E_3^2) E_2^2  \right]_{E_1 = E_3 +
    k^*}^{E_1 = E_3 + E} \nonumber \\
  &= -  \frac{\ffermi{E}  }{2^7 \pi^6 } \int_0^{\infty}dE_3 
   \ln{\left( \frac{ k^* }{ E_3 } \right)} \frac{ E_3^2\, e^{ E_3/T } }{
  (e^{ E_3/T }  +1)^2 }\;.
\end{align}  
Again, $E_2 = E +E_3 -E_1$ is understood and we have set $k^*
\rightarrow 0$ in the distribution functions after evaluating the
borders.  The integral is  solved in an analogous manner as in
(\ref{eq:BFB-121-surface}) so that we find
  \begin{align}
  \label{eq:FFF-g121-surface-final}
   g_{\FFF,\,121}^{|M_1|^2\,\text{surface}} &= \frac{f_F(E)
     T^3}{3\cdot 2^8
     \pi^4}
    \left[  \ln{\left( \frac{ 2T }{ k^* } \right)} 
     + {3 \over 2} - \gamma +
         {\zeta '(2) \over \zeta(2)}  \right]\;.
\end{align}  
Note that in contrast to (\ref{eq:BFB-g121-surface-final}) there is a
factor of two in the nominator of the logarithm. For the remaining part,
we find for $k^* \rightarrow 0$
  \begin{align}
  g_{ \FFF, \,121}^{|M_1|^2\,\text{partial}}
  &= -  \frac{1}{2^8 \pi^6 E^2}    \int_0^{\infty}dE_3
  \int_0^{ E +E_3  }dE_1\,
  \ln\left(\frac{|E_1-E_3|}{E_3}\right)\nonumber \\ &\times
  \Theta(E_1-E_3)
  \frac{d}{dE_1}\left[ f_{\FFF}E_2^2(E_1^2+E_3^2)\right]\;.
\end{align}  
The second integral (\ref{eq:FFF-122-M1-version2}) does not yield any
additional contribution from the lower border as in
(\ref{eq:limit-g122-borders}). Thus,
\begin{align}
  g_{\FFF,\,122}^{|M_1|^2\,\text{rest}} &= 0\;,
\end{align}
and we easily obtain 
  \begin{align}
  \label{eq:FFF-122-surface}
  g_{\FFF,\,122}^{|M_1|^2\,\text{surface}} &=
  g_{\FFF,\,121}^{|M_1|^2\,\text{surface}} \; .
\end{align}  
Furthermore, we find for \mbox{$k^*
\rightarrow 0$ }
  \begin{align}
  g_{ \FFF, \,122}^{|M_1|^2\,\text{partial}}
  &=  \frac{1}{2^8 \pi^6 E^2} \int_0^{\infty} dE_3
  \int_0^{ E +E_3  }dE_1\,
  \ln\left(\frac{|E_1-E_3|}{E_3}\right)\nonumber \\ &\times
  \Theta(E_3-E_1)
  \frac{d}{dE_1}\left[ f_{\FFF} E^2(E_1^2+E_3^2)\right]\;.
\end{align}  
An integration by parts of 
  \begin{align}
  g_{\FFF,\,2}^{|M_1|^2} & = - \frac{1}{2^{8}\pi^6 E^2} \int_0^\infty
  dE_3 \int_0^{E}dE_1 f_{\FFF}  {1\over E_1 - E_3} \left( E_1^2E_2^2
    -E^2 E_3^2   \right)
\end{align}  
 does not yield surface term; see~(\ref{eq:g2-surface-final}). Hence, we obtain
  \begin{align}
  \label{eq:FFF-M1-g2-final}
    g_{\FFF,\,2}^{|M_1|^2} & =  \frac{1}{2^{8}\pi^6 E^2} \int_0^\infty
  dE_3  \int_0^{E+E_3}
  dE_1 \ln{\left(\frac{|E_1-E_3|}{E_3}\right)} 
  \nonumber \\ &\times  \Theta(E - E_1) {d\over dE_1} \left[
     f_{\FFF} (   E_1^2 E_2^2 - E^2 E_3^2 ) \right]\;.
\end{align}  
Again, $ g_{\FFF,\,3}^{|M_1|^2} =0 $, since the argument preceding
(\ref{eq:BFB-g3-final}) is independent of the quantum statistical
distribution functions. 

\smallskip \noindent\textbf{Result for \boldmath$I_{\FFF}^{|M_1|^2} $}\smallskip

For the  FFF processes,  we find in total from the contribution of $|M_1|^2$:
  \begin{align}
 \label{eq:I-FFF-M1-final}
 I_{\FFF}^{|M_1|^2} &= g_{\FFF,\,11}^{|M_1|^2} + 2\cdot
 g_{\FFF,\,121}^{|M_1|^2\,\text{surface}} + g_{ \FFF,
   \,121}^{|M_1|^2\,\text{partial}}+ g_{ \FFF,
   \,122}^{|M_1|^2\,\text{partial}} +
 g_{\BFB,\,2}^{|M_1|^2} \nonumber \\
 & =  \frac{T^3 \ffermi{E}}{384 \p^4} \left[ \ln{\left( \frac{2  T }{ k^* }
     \right)} + {17 \over 6} - \gamma + {\zeta '(2) \over \zeta(2)} \right] \nonumber\\
  & \quad + \frac{1}{256\pi^6 } \int_0^\infty
  dE_3  \int_0^{E+E_3}
  dE_1  \ln{\left(\frac{|E_1-E_3|}{E_3}\right)} \nonumber \\
  & \quad \times \Bigg\{ - \Theta(E_1-E_3)
  \frac{d}{dE_1}\left[ f_{\BFB}\frac{E_2^2}{E^2} (E_1^2+E_3^2)\right] \nonumber \\
  & \; \hphantom{ \quad\times \Bigg\{ } 
    +  \Theta(E_3-E_1)
  \frac{d}{dE_1}\left[ f_{\BFB} (E_1^2+E_3^2)\right] \nonumber \\
  &  \; \hphantom{ \quad\times \Bigg\{ }   +  \Theta(E - E_1)  {d\over dE_1} \left[
     f_{\BFB} (  \frac{ E_1^2 E_2^2}{E^2} -  E_3^2 ) \right] \Bigg\}\; .
\end{align}

Note that $ |M_2|^2  $ does not contribute to the FFF processes because
$ c_{\BFB\,2 }^{(\alpha)} = 0$ so that we
can immediately turn to the integrals containing  $ |M_3|^2  $.

\subsection[Contribution from \texorpdfstring{$| M_3 |^2$}{M3}]{Contribution from \texorpdfstring{\boldmath$| M_3
    |^2$}{M3}}
\label{sec:FFF-M3}

We now encounter for the first time also the
matrix element $|M_3|^2$, namely,
  \begin{align}
  \label{eq:I-FFF-3}
  I_{\FFF}^{|M_3|^2} &=  {
    1\over (2\pi)^3 2E}  \int {d\Omega_p \over 4\pi} \int
  \left[\prod_{i=1}^3 \frac{ d^3 p_i }{ (2\pi)^3 2E_i}
  \right] (2\pi)^4 \delta^4( P_1+ P_2- P_3 - P)\nonumber \\
  & \times  f_{\FFF} |M_3|^2 \Theta( |\mathbf{p}_1 - \mathbf{p}_3 | - k^*  )\; .
\end{align}  
Recall from (\ref{eq:matrixelementsM1toM3}) that $|M_3|^2 = t^2/s$
so that we can set $k^* \rightarrow 0$ and employ the formalism
developed in section~\ref{sec:BFB-M2}. We choose
(\ref{eq:q-frame}) as the frame of reference for the integrations. 

It follows that
\begin{align}
  \label{eq:s-inq-frame}
  s &= (E + E_3)^2 -q^2\; ,
\end{align}
and with $t$ as in (\ref{eq:t-inq-system}) we find for the squared
matrix element
\begin{align}
  \label{eq:M3-in-q-system}
  |M_3|^2 &= \frac{1}{ (E + E_3)^2 -q^2 }
  \left[ -2 E E_2 \left( 1- \cos\theta \cos\tilde\theta -
    \sin\theta \sin\tilde\theta \sin\phi \right)\right]^2\; .
\end{align}
The angular integrations are straightforward to compute and we get
  \begin{align}
  \label{eq:I-FFF-M3-version2}
   I_{\FFF}^{|M_3|^2} &=  \frac{3}{2^{13}\pi^6 E^2} \int  dE_2  dE_3
 dq  f_{\FFF}\,  g_{\FFF}^{|M_3|^2} \, \Omega
\end{align}  
with
  \begin{align}
  \label{eq:BFB-M3-g}
  g_{\BFB}^{|M_3|^2} = &\left( (E_3+E)^2 -q^2 \right) 
  \left[ 1 - \frac{2}{3} { \left(2E_2^2-6E_2E_3 + 3 E_3 +2E_2E- E^2 \right) \over
    q^2 } \right. \nonumber \\
  &+ \left. {(E_3 - E)^2 (E_1 -E_2)^2 \over q^4 }\right]
\end{align}  
and $\Omega$ as in (\ref{eq:omegathetasM2}). The manipulations of the
$\Theta$-functions are the same as for $|M_2|^2$ so we can merely
present the results. Analog to (\ref{eq:M2-g11}), (\ref{eq:M2-g12}),
(\ref{eq:M2-g211}), (\ref{eq:M2-g212}), (\ref{eq:M2-g221}) and
(\ref{eq:M2-g222}), we find
  \begin{align}
  \label{eq:M3-all-g}
  g_{\FFF,\,11}^{|M_3|^2} &= \frac{1}{2^{8}\pi^6 E^2}
  \int_{0}^{\infty}   dE_3  \int_{0}^{E+E_3}  dE_2 
  \Theta(E-E_3) f_{\FFF} \frac{E_1^2 E_3^2}{E+E_3}
\; ,  \\
  g_{\FFF,\,12}^{|M_3|^2} &= \frac{1}{2^{8}\pi^6 E^2}
  \int_{0}^{\infty}   dE_3  \int_{0}^{E+E_3}  dE_2 
  \Theta(E_3 - E) f_{\FFF} \frac{E^2 E_2^2}{E+E_3}
\; ,  \\
  g_{\FFF,\,211}^{|M_3|^2} &= - \frac{1}{2^{8}\pi^6 E^2}
  \int_{0}^{\infty}   dE_3  \int_{0}^{E+E_3}  dE_2 
  \Theta(E - E_3) \Theta(E_3 - E_2) \nonumber \\ &\times  f_{\FFF} (E_2 -E_3) 
  \left[  E_2(E_3 -E) - E_3( E_3+ E)   \right]
\; ,  \\
  g_{\BFB,\,222}^{|M_3|^2} &=   \frac{1}{2^{8}\pi^6 E^2}
  \int_{0}^{\infty}   dE_3  \int_{0}^{E+E_3}  dE_2 
  \Theta(E_3 - E) \Theta(E_2 - E_3) \nonumber \\ &\times  f_{\FFF} (E_2 -E_3) 
  \left[  E_2(E_3 -E) - E_3( E_3+ E)   \right]
\; .
\end{align}  
while
 \begin{align}
  g_{\FFF,\,212}^{|M_3|^2} &= 0
\; ,  \\
  g_{\FFF,\,221}^{|M_3|^2} &= 0\;.
\end{align}

\clearpage
\smallskip \noindent\textbf{Result for \boldmath$I_{\FFF}^{|M_3|^2} $}\smallskip

The contribution of matrix element $|M_3|^2$ yields
\begin{align}
   \label{eq:I-FFF-M3-final}
   I_{\FFF}^{|M_3|^2} & = g_{\FFF,\,11}^{|M_3|^2} +
   g_{\FFF,\,12}^{|M_3|^2} 
   + g_{\FFF,\,211}^{|M_3|^2} +  g_{\BFB,\,222}^{|M_3|^2} \nonumber \\
   & = \frac{1}{256 \p^6} \int_{0}^{\infty}   dE_3  \int_{0}^{E+E_3}  dE_2
   \, f_{\FFF} \nonumber \\ & \times \Bigg\{
        \Theta(E-E_3) \frac{1}{E^2} \frac{E_1^2
     E_3^2}{E+E_3} \nonumber \\
    & \qquad + \Theta(E_3 - E) \frac{1}{E^2} \frac{E^2 E_2^2}{E+E_3}\nonumber \\
    & \qquad - \Theta(E - E_3) \Theta(E_3 - E_2) \frac{E_2 -E_3}{E^2} 
   \left[  E_2(E_3 -E) - E_3( E_3+ E)   \right] \nonumber \\
    & \qquad + \Theta(E_3 - E) \Theta(E_2 - E_3) \frac{E_2 -E_3}{E^2} 
  \left[  E_2(E_3 -E) - E_3( E_3+ E)   \right] \Bigg\}\; .
\end{align}

\section{Production Rate for BBF Processes}
\label{sec:production-rate-bbf}

The BFB processes contain the contributions with two incoming bosons
and the gravitino and another fermion as outgoing particles. In total,
  \begin{align}
  \label{eq:hard-prod-rate-BBF-app}
  \gammahard^{(\BBF)} & =   \sum_{\alpha = 1}^3 
  \frac{g_\a^2}{\mplanck^2} \left( 1+\frac{M^2_{ \alpha
      }}{3\mgrav^2}\right)   \sum_{i=1}^3
  \left[ c_{ \BBF,\, i }^{(\alpha)}\, I_{ \BBF}^{|M_i|^2}
    \right].
\end{align}  
Note that all $c_{ \BBF,\, 1 }^{(\alpha)} = 0 $ so that we are left
with the computation of $ | M_2  |^2 $ and $ | M_3  |^2 $.
\subsection[Contribution from  \texorpdfstring{$| M_2 |^2$}{M2}]{Contribution from
  \texorpdfstring{\boldmath$| M_2 |^2$}{M2}}
\label{sec:BBF-M2}

We have discussed in great detail how to perfom the integrations for the
non-singular matrix elements $| M_2 |^2$ and $| M_3 |^2$ in sections
\ref{sec:BFB-M2} and \ref{sec:FFF-M3}, respectively. Again, all
that changes is a different combination of quantum mechanical
distribution functions,
\begin{align}
  \label{eq:I-BBF-2}
  I_{\BBF}^{|M_2|^2} &=  {
    1\over (2\pi)^3 2E}  \int {d\Omega_p \over 4\pi} \int
  \left[\prod_{i=1}^3 \frac{ d^3 p_i }{ (2\pi)^3 2E_i}
  \right] (2\pi)^4 \delta^4( P_1+ P_2- P_3- P) 
f_{\BBF} |M_2|^2 \; ,
\end{align}
so that we can immediately present the result:
\begin{align}
 \label{eq:I-BBF-M2-final}
 I_{\BBF}^{|M_2|^2} &= 
 \frac{1}{256\pi^6} \int_{0}^{\infty} dE_3 \int_{0}^{E+E_3} dE_2
 f_{\BBF}
 \nonumber \\
 & \times \Bigg\{  \Theta(E - E_3) \frac{ E_3^2}{E^2} \left( {E_3 \over 3} - E_1
 \right)  + \Theta(E_3 - E) 
  \left( {E \over 3} - E_2  \right) \nonumber \\
 &  \hphantom{ \quad\times \Bigg\{ }
 +   \Theta(E_3-E_2) \Theta(E - E_3)
 \frac{ E_2-E_3  }{ 3 E^2} \left[  (E_2-E_3)(E_2+2 E_3) - 3(E_2+E_3)E \right] \nonumber \\
 &  \hphantom{ \quad\times \Bigg\{ }
 -  \Theta(E_2-E) \Theta(E - E_3) \frac{(E_2 - E)^3}{3 E^2} \nonumber \\
 &  \hphantom{ \quad\times \Bigg\{ }
 +  \Theta(E-E_2) \Theta(E_3 - E) \frac{(E_2 - E)^3}{3 E^2}\nonumber \\
 &  \hphantom{ \quad\times \Bigg\{ }
 -  \Theta(E_2-E_3) \Theta(E_3 - E)
   \frac{ E_2-E_3  }{3 E^2} \left[  (E_2-E_3)(E_2+2 E_3) - 3(E_2+E_3)E  \right] \Bigg\}\;.
\end{align}

\subsection[Contribution from \texorpdfstring{$| M_3 |^2$}{M3}]{Contribution from
  \texorpdfstring{\boldmath$| M_3 |^2$}{M3}}
\label{sec:BBF-M3}

For the squared matrix element $|M_3|^2$ we can set as usual $k^*
\rightarrow 0$, i.e.,
\begin{align}
  \label{eq:I-BBF-3}
  I_{\BBF}^{|M_3|^2} &=  {
    1\over (2\pi)^3 2E}  \int {d\Omega_p \over 4\pi} \int
  \left[\prod_{i=1}^3 \frac{ d^3 p_i }{ (2\pi)^3 2E_i}
  \right] (2\pi)^4 \delta^4( P_1+ P_2- P_3 - P) 
f_{\BBF} |M_3|^2 \; ,
\end{align}
and find in complete analogy to section~\ref{sec:FFF-M3}:
\begin{align}
   \label{eq:I-BBF-M3-final}
   I_{\BBF}^{|M_3|^2} & =  \frac{1}{256 \p^6} \int_{0}^{\infty}   dE_3  \int_{0}^{E+E_3}  dE_2
   \, f_{\BBF} \nonumber \\ & \times \Bigg\{
        \Theta(E-E_3) \frac{1}{E^2} \frac{E_1^2
     E_3^2}{E+E_3} \nonumber \\
    & \qquad + \Theta(E_3 - E) \frac{1}{E^2} \frac{E^2 E_2^2}{E+E_3}\nonumber \\
    & \qquad - \Theta(E - E_3) \Theta(E_3 - E_2) \frac{E_2 -E_3}{E^2} 
   \left[  E_2(E_3 -E) - E_3( E_3+ E)   \right] \nonumber \\
    & \qquad + \Theta(E_3 - E) \Theta(E_2 - E_3) \frac{E_2 -E_3}{E^2} 
  \left[  E_2(E_3 -E) - E_3( E_3+ E)   \right] \Bigg\}\; .
\end{align}


\cleardoublepage

\chapter{Published Work from this Thesis}
\label{cha:published-work-from}

The new $\smgroup$ result for the collision term and its implication
on gravitino dark matter scenarios allow for a collider test probing
the viability of thermal leptogenesis. As a summary of this thesis,
the proposed method has been published in
\href{http://link.aps.org/abstract/PRD/V75/E023509}{Physical Review
  D}~\cite{Pradler:2006qh}.
\\

An e-print of the paper is available on the
\href{http://lanl.arxiv.org}{arXiv} server:\\
\url{http://lanl.arxiv.org/abs/hep-ph/0608344}

%
%
%
%
%


\end{appendix}

\cleardoublepage

\phantomsection 
\addcontentsline{toc}{chapter}{References}
\bibliography{biblio}

\providecommand{\href}[2]{#2}\begingroup\raggedright\begin{thebibliography}{10}

\bibitem{Spergel:2006hy}
D.~N. Spergel {\em et.~al.}, {\it Wilkinson microwave anisotropy probe ({WMAP})
  three year results: Implications for cosmology},
  \href{http://arXiv.org/abs/astro-ph/0603449}{{\tt astro-ph/0603449}}.

\bibitem{Bolz:2000fu}
M.~Bolz, A.~Brandenburg and W.~Buchm{\"u}ller, {\it Thermal production of
  gravitinos},  {\em Nucl. Phys.} {\bf B606} (2001) 518--544
  [\href{http://arXiv.org/abs/hep-ph/0012052}{{\tt hep-ph/0012052}}].

\bibitem{Cerdeno:1998hs}
D.~G. Cerde{\~n}o and C.~Mu{\~n}oz, {\it An introduction to supergravity}, .
  Prepared for 6th Hellenic School and Workshop on Elementary Particle
  Physics:, Corfu, Greece, 6-26 Sep 1998.

\bibitem{Nilles:1983ge}
H.~P. Nilles, {\it Supersymmetry, supergravity and particle physics},  {\em
  Phys. Rept.} {\bf 110} (1984) 1.

\bibitem{Yao:2006px}
{\bf Particle Data Group} Collaboration, W.~M. Yao {\em et.~al.}, {\it Review
  of particle physics},  {\em J. Phys.} {\bf G33} (2006) 1--1232.

\bibitem{Wess:1992cp}
J.~Wess and J.~Bagger, {\it Supersymmetry and supergravity}, . Princeton, USA:
  Univ. Pr. (1992) 259 p.

\bibitem{Giudice:1988yz}
G.~F. Giudice and A.~Masiero, {\it A natural solution to the $\mu$ problem in
  supergravity theories},  {\em Phys. Lett.} {\bf B206} (1988) 480--484.

\bibitem{Brignole:1997dp}
A.~Brignole, L.~E. Iba\~nez and C.~Mu\~noz, {\it Soft supersymmetry-breaking
  terms from supergravity and superstring models},
  \href{http://arXiv.org/abs/hep-ph/9707209}{{\tt hep-ph/9707209}}.

\bibitem{Martin:1997ns}
S.~P. Martin, {\it A supersymmetry primer},
  \href{http://arXiv.org/abs/hep-ph/9709356}{{\tt hep-ph/9709356}}.

\bibitem{Drees:2004jm}
M.~Drees, R.~Godbole and P.~Roy, {\it Theory and phenomenology of sparticles:
  An account of four-dimensional {N=1} supersymmetry in high energy physics}, .
  Hackensack, USA: World Scientific (2004) 555 p.

\bibitem{Rarita:1941mf}
W.~Rarita and J.~S. Schwinger, {\it On a theory of particles with half-integral
  spin},  {\em Phys. Rev.} {\bf 60} (1941) 61.

\bibitem{Lee:1998aw}
T.~Lee and G.-H. Wu, {\it Interactions of a single goldstino},  {\em Phys.
  Lett.} {\bf B447} (1999) 83--88
  [\href{http://arXiv.org/abs/hep-ph/9805512}{{\tt hep-ph/9805512}}].

\bibitem{Denner:1992vz}
A.~Denner, H.~Eck, O.~Hahn and J.~Kublbeck, {\it Feynman rules for
  fermion-number violating interactions},  {\em Nucl. Phys.} {\bf B387} (1992)
  467--484.

\bibitem{Braaten:1991dd}
E.~Braaten and T.~C. Yuan, {\it Calculation of screening in a hot plasma},
  {\em Phys. Rev. Lett.} {\bf 66} (1991) 2183--2186.

\bibitem{Weldon:1983jn}
H.~A. Weldon, {\it Simple rules for discontinuities in finite-temperature field
  theory},  {\em Phys. Rev.} {\bf D28} (1983) 2007.

\bibitem{Braaten:1989mz}
E.~Braaten and R.~D. Pisarski, {\it Soft amplitudes in hot gauge theories: A
  general analysis},  {\em Nucl. Phys.} {\bf B337} (1990) 569.

\bibitem{Vermaseren:2000nd}
J.~A.~M. Vermaseren, {\it New features of {FORM}},
  \href{http://arXiv.org/abs/math-ph/0010025}{{\tt math-ph/0010025}}.

\bibitem{Cutler:1977qm}
R.~Cutler and D.~W. Sivers, {\it Quantum chromodynamic gluon contributions to
  large p(t)-reactions},  {\em Phys. Rev.} {\bf D17} (1978) 196.

\bibitem{Pisarski:1988vd}
R.~D. Pisarski, {\it Scattering amplitudes in hot gauge theories},  {\em Phys.
  Rev. Lett.} {\bf 63} (1989) 1129.

\bibitem{Weldon:1982aq}
H.~A. Weldon, {\it Covariant calculations at finite temperature: The
  relativistic plasma},  {\em Phys. Rev.} {\bf D26} (1982) 1394.

\bibitem{Bolz:2000xi}
M.~Bolz, {\em Thermal production of gravitinos}.
\newblock PhD thesis, DESY, 2000.

\bibitem{Pisarski:1989cs}
R.~D. Pisarski, {\it Renormalized gauge propagator in hot gauge theories},
  {\em Physica} {\bf A158} (1989) 146--157.

\bibitem{Kalashnikov:1979cy}
O.~K. Kalashnikov and V.~V. Klimov, {\it Polarization tensor in {QCD} for
  finite temperature and density},  {\em Sov. J. Nucl. Phys.} {\bf 31} (1980)
  699.

\bibitem{Pisarski:1987wc}
R.~D. Pisarski, {\it Computing finite-temperature loops with ease},  {\em Nucl.
  Phys.} {\bf B309} (1988) 476.

\bibitem{Hahn:2004fe}
T.~Hahn, {\it Cuba: A library for multidimensional numerical integration},
  {\em Comput. Phys. Commun.} {\bf 168} (2005) 78--95
  [\href{http://arXiv.org/abs/hep-ph/0404043}{{\tt hep-ph/0404043}}].

\bibitem{Kohri:2005wn}
K.~Kohri, T.~Moroi and A.~Yotsuyanagi, {\it Big-bang nucleosynthesis with
  unstable gravitino and upper bound on the reheating temperature},  {\em Phys.
  Rev.} {\bf D73} (2006) 123511
  [\href{http://arXiv.org/abs/hep-ph/0507245}{{\tt hep-ph/0507245}}].

\bibitem{Minkowski:1977sc}
P.~Minkowski, {\it $\mu$ $\to$ e $\gamma$ at a rate of one out of 1-billion
  muon decays?},  {\em Phys. Lett.} {\bf B67} (1977) 421.

\bibitem{Yanagida:1980xy}
T.~Yanagida, {\it Horizontal gauge-symmetry and masses of neutrinos},  {\em
  Prog. Theor. Phys.} {\bf 64} (1980) 1103.

\bibitem{Fukugita:1986hr}
M.~Fukugita and T.~Yanagida, {\it Baryogenesis without grand unification},
  {\em Phys. Lett.} {\bf B174} (1986) 45.

\bibitem{Buchmuller:2004nz}
W.~Buchm{\"u}ller, P.~Di~Bari and M.~Pl{\"u}macher, {\it Leptogenesis for
  pedestrians},  {\em Ann. Phys.} {\bf 315} (2005) 305--351
  [\href{http://arXiv.org/abs/hep-ph/0401240}{{\tt hep-ph/0401240}}].

\bibitem{Bolz:1998ek}
M.~Bolz, W.~Buchm{\"u}ller and M.~Pl{\"u}macher, {\it Baryon asymmetry and dark
  matter},  {\em Phys. Lett.} {\bf B443} (1998) 209--213
  [\href{http://arXiv.org/abs/hep-ph/9809381}{{\tt hep-ph/9809381}}].

\bibitem{Fujii:2003nr}
M.~Fujii, M.~Ibe and T.~Yanagida, {\it Upper bound on gluino mass from thermal
  leptogenesis},  {\em Phys. Lett.} {\bf B579} (2004) 6--12
  [\href{http://arXiv.org/abs/hep-ph/0310142}{{\tt hep-ph/0310142}}].

\bibitem{Baer:2000gf}
H.~Baer, M.~A. Diaz, P.~Quintana and X.~Tata, {\it Impact of physical
  principles at very high energy scales on the superparticle mass spectrum},
  {\em JHEP} {\bf 04} (2000) 016
  [\href{http://arXiv.org/abs/hep-ph/0002245}{{\tt hep-ph/0002245}}].

\bibitem{Blair:2002pg}
G.~A. Blair, W.~Porod and P.~M. Zerwas, {\it The reconstruction of
  supersymmetric theories at high energy scales},  {\em Eur. Phys. J.} {\bf
  C27} (2003) 263--281 [\href{http://arXiv.org/abs/hep-ph/0210058}{{\tt
  hep-ph/0210058}}].

\bibitem{Lafaye:2004cn}
R.~Lafaye, T.~Plehn and D.~Zerwas, {\it {SFITTER}: {SUSY} parameter analysis at
  {LHC} and {LC}},  \href{http://arXiv.org/abs/hep-ph/0404282}{{\tt
  hep-ph/0404282}}.

\bibitem{Bechtle:2004pc}
P.~Bechtle, K.~Desch and P.~Wienemann, {\it {FITTINO}, a program for
  determining {MSSM} parameters from collider observables using an iterative
  method},  {\em Comput. Phys. Commun.} {\bf 174} (2006) 47--70
  [\href{http://arXiv.org/abs/hep-ph/0412012}{{\tt hep-ph/0412012}}].

\bibitem{Feng:2004mt}
J.~L. Feng, S.~Su and F.~Takayama, {\it Supergravity with a gravitino {LSP}},
  {\em Phys. Rev.} {\bf D70} (2004) 075019
  [\href{http://arXiv.org/abs/hep-ph/0404231}{{\tt hep-ph/0404231}}].

\bibitem{Steffen:2006hw}
F.~D. Steffen, {\it Gravitino dark matter and cosmological constraints},  {\em
  JCAP} {\bf 0609} (2006) 001 [\href{http://arXiv.org/abs/hep-ph/0605306}{{\tt
  hep-ph/0605306}}].

\bibitem{Cerdeno:2005eu}
D.~G. Cerde{\~n}o, K.-Y. Choi, K.~Jedamzik, L.~Roszkowski and R.~Ruiz~de
  Austri, {\it Gravitino dark matter in the {CMSSM} with improved constraints
  from {BBN}},  {\em JCAP} {\bf 0606} (2006) 005
  [\href{http://arXiv.org/abs/hep-ph/0509275}{{\tt hep-ph/0509275}}].

\bibitem{Buchmuller:2004rq}
W.~Buchm{\"u}ller, K.~Hamaguchi, M.~Ratz and T.~Yanagida, {\it Supergravity at
  colliders},  {\em Phys. Lett.} {\bf B588} (2004) 90--98
  [\href{http://arXiv.org/abs/hep-ph/0402179}{{\tt hep-ph/0402179}}].

\bibitem{Brandenburg:2005he}
A.~Brandenburg, L.~Covi, K.~Hamaguchi, L.~Roszkowski and F.~D. Steffen, {\it
  Signatures of axinos and gravitinos at colliders},  {\em Phys. Lett.} {\bf
  B617} (2005) 99--111 [\href{http://arXiv.org/abs/hep-ph/0501287}{{\tt
  hep-ph/0501287}}].

\bibitem{Martyn:2006as}
H.~U. Martyn, {\it Detecting metastable staus and gravitinos at the {ILC}},
  \href{http://arXiv.org/abs/hep-ph/0605257}{{\tt hep-ph/0605257}}.

\bibitem{Steffen:2005cn}
F.~D. Steffen, {\it Collider signatures of axino and gravitino dark matter},
  \href{http://arXiv.org/abs/hep-ph/0507003}{{\tt hep-ph/0507003}}.

\bibitem{Djouadi:2002ze}
A.~Djouadi, J.-L. Kneur and G.~Moultaka, {\it {SuSpect}: A fortran code for the
  supersymmetric and higgs particle spectrum in the {MSSM}},
  \href{http://arXiv.org/abs/hep-ph/0211331}{{\tt hep-ph/0211331}}.

\bibitem{Belanger:2001fz}
G.~Belanger, F.~Boudjema, A.~Pukhov and A.~Semenov, {\it {micrOMEGAs}: A
  program for calculating the relic density in the {MSSM}},  {\em Comput. Phys.
  Commun.} {\bf 149} (2002) 103--120
  [\href{http://arXiv.org/abs/hep-ph/0112278}{{\tt hep-ph/0112278}}].

\bibitem{Wheelon:1953}
A.~D. Wheelon, {\it On the summation of infinite series in closed form},  {\em
  Journal of Applied Physics} {\bf 25} (1954) 113.

\bibitem{Pradler:2006qh}
J.~Pradler and F.~D. Steffen, {\it Thermal gravitino production and collider
  tests of leptogenesis},  {\em Phys. Rev.} {\bf D75} (2007) 023509
  [\href{http://arXiv.org/abs/hep-ph/0608344}{{\tt hep-ph/0608344}}].

\end{thebibliography}\endgroup

\cleardoublepage

\chapter*{Acknowledgements}
\phantomsection
\addcontentsline{toc}{chapter}{Acknowledgements}

I would like to thank Professor~Alfred~Bartl for strong support and for the
possibility to conduct my diploma thesis abroad at the
Max-Planck-Institute in Munich. Frank Daniel Steffen deserves my
special thanks for proposing the interesting topic which resulted
in a close collaboration. It is a great pleasure to work under his
responsible and encouraging supervision. Furthermore, I want to thank
Georg Raffelt for his friendly advice.

I am thankful to Marco Zagermann for helpful discussions on
supergravity and to \mbox{Professor}~Wilfried Buchm\"uller and Arnd Brandenburg
for correspondence on the gravitino production rate. For brief but
inspiring discussions, I would like to thank Professor Antonio Masiero.

I want to thank my office mates Andreas Biffar, Florian Hahn-Woernle,
Max Huber, Alexander Laschka, Felix Rust, and Tobias Schl\"uter for
having spent a most pleasant and diverse time with them.

Thanks to Simon Gr\"oblacher, Erik H\"ortnagl, Ulrich Matt, and Robert
Prevedel for friendship and not least for  providing me with
accommodation during my visits in \mbox{Vienna.}

Last but not least I wish to thank my family for their trust in me. I
deeply thank my parents for their support.

\fancyhead{} 

\cleardoublepage

\renewcommand{\headrulewidth}{0pt}
\phantomsection
\addcontentsline{toc}{chapter}{Revision History}

Revision History:\\[20pt]
  March 15, 2007 
    \begin{itemize}\itemsep=-0.2cm
      \item Cover sheet: date of hand in of the thesis added
      \item Appendix~A: Footnote added 
      \item Appendix~D: Reference updated
    \end{itemize}
  June 01, 2007 
    \begin{itemize}\itemsep=-0.2cm
    \item Typos corrected in Eqs.~(\ref{eq:Lsugra}),
      (\ref{eq:killing-vector-fields}), and (\ref{susytrafos})
    \end{itemize}

\end{document}